\documentclass[aps,prb,twocolumn,showpacs,tightenlines]{revtex4}
\usepackage{epsfig}
\begin{document}

\def \cT {{\cal T}}
\def \cI {{\cal I}}
\def \cf {{\cal f}}
\def \cG {{\cal G}}
\def \cD {{\cal D}}
\def \cU {{\cal U}}
\def \cV {{\cal V}}
\def \cF {{\cal F}}
\def \cT {{\cal T}}
\def \cH {{\cal H}}
\def \cA {{\cal A}}
\def \cL {{\cal L}}
\def \cR {{\cal R}}
\def \cN {{\cal N}}
\def \cC {{\cal C}}
\def \cS {{\cal S}}
\def \cP {{\cal P}}
\def \cE {{\cal E}}
\def \cM {{\cal M}}
\title{Electronic Transport in Hybrid Mesoscopic Structures: A Nonequilibrium Green Function Approach}

\author{Zhao Yang Zeng,$^{1,2}$ Baowen Li,$^1$ and F. Claro$^3$}

\affiliation {$^1$ Department of Physics, National University of
Singapore, 117542, Singapore\\
$^2$ Department of Physics, Hunan Normal University, Changsha
410081, China \\
 $^3$ Facultad de F\'isica, Pontificia Universidad Cat\'olica
de Chile, Casilla 306, Santiago 22, Chile }

\date{\today}

\begin{abstract}
 We present a unified transport theory of  hybrid structures, in
which a confined normal state ($N$) sample is sandwiched between
two leads each of which can be either a ferromagnet ($F$) or
 a superconductor ($S$) via tunnel barriers. By
introducing a four-dimensional Nambu-spinor space, a general
current formula is derived within the Keldysh nonequilibrium Green
function formalism, which can be applied to various kinds of
hybrid mesoscopic systems with strong correlations even in the
nonequilibrium situation. Such a formula is gauge invariant.   We
also demonstrate analytically for some quantities, such as the
difference between chemical potentials, superconductor order
parameter phases and ferromagnetic magnetization orientations,
that only their relative value appears explicitly in the current
expression. When applied to specific structures, the formula
becomes of the Meir-Wingreen-type favoring strong correlation
effects, and reduces to the Landauer-B\"uttiker-type in
noninteracting systems such as the double-barrier resonant
structures, which we study in detail beyond the wide-band
approximation. We find that the spin-dependent density of states
of the ferromagnetic lead(s) is reflected in the resonant peak and
resonant shoulder structure of the I-V characteristics of
$F/I/N/I/F$ structures with large level spacing. The tunnel
magnetoresistance that exhibits complex behaviors as a function of
the bias voltage, can be either positive or negative, suppressed
or enhanced within the resonant peak region(s), depending on the
couplings to the leads. The Andreev current spectrum of
$F/I/N/I/S$ structures consists of a series of resonant peaks as a
function of the gate voltage, of which the number and amplitude
are strongly dependent on the bias voltage, degree of spin
polarization of the ferromagnetic lead, energy gap of the
superconducting lead,  and the level configuration of the central
region. In $S/I/N/I/S$ resonant structures  with asymmetric
superconducting energy gaps, the Josephson current through a
single resonant level is slightly enhanced in contrast to the
significant enhancement of the Josephson current in $S/N/S$
Junctions. The current-phase relation is relevant to the level
position and the couplings to the superconducting leads.

\end{abstract}


\maketitle

\section{Introduction}
Electronic transport in mesoscopic systems or nanoscale structures
has received extensive theoretical and experimental
attention.\cite{Kouwenhoven}  In mesoscopic systems the sample
size is smaller than the phase coherent length,  and electrons
retain their phase when travelling through the sample. In the
ballistic limit, i.e., when the dimensions of the sample are
smaller than the mean free path, electrons can traverse the system
without any scattering. In contrast to macroscopic systems, the
conductance of mesoscopic systems is sample-specific, since
electron wavefunctions are strongly dependent on the form of the
boundary of the sample and the configuration of scatterers located
within the sample.

To calculate the conductance of mesoscopic systems, one should
first consider the wave nature of electrons.
 The classical Boltzmann transport
equation\cite{Kittel} is obviously inappropriate, since the
assumption that electrons can be viewed as classical particles
does not hold at a mesoscopic scale due to the Heinsenberg
uncertainty limitation.  Linear-response theory \cite{Kubo} is
restricted to the weak perturbation regime and apparently can not
be applied to the nonlinear or nonequilibrium situation.
Electronic transport through a mesoscopic medium is in effect a
wave transmitting process of electrons, which can be associated
with a scattering matrix. In measuring the conductance one always
connects the sample to electron reservoirs through perfect leads.
 \cite{Datta}  In a two-terminal
setup ($\cL$=left, $\cR$=right), the Landauer-B\"uttiker
formula\cite{Buttiker}
 states that the current $\cI$   can be expressed as a convolution of the transmission
probability $\cT$ and the Fermi distribution function $f_{\alpha}$
($ \alpha=\cL,\cR$), i.e.,  $\cI= \frac{2e}{h}\int
\cT(\epsilon)[f_\cL(\epsilon)-f_\cR(\epsilon)]d\epsilon$. The
conductance $\cG$ in the linear-response regime is
$\cG=\frac{2e^2}{h}\int \cT(\epsilon)(-\frac{\partial f}{\partial
\epsilon})d\epsilon$. Such a formulation seems more appealing
since the transport properties are encoded in the corresponding
transmission probability, which can be calculated by various
methods.

 The nonequilibrium Green function (NEGF) approach \cite {Kadanoff,Rammer,Mahan,Haug} has
proven to be a powerful technique to investigate transport
problems in the many-body systems and mesoscopic systems. The
equation of motion for the NEGF $G^<$,  the quantum Boltzmann
equation (QBE),\cite{Kadanoff}  serves a starting point for many
transport calculations in the many-body
problems,\cite{Rammer,Mahan}  where a four-variable distribution
function is required to incorporate the quantum effect due to the
uncertainty principle. The Keldysh formalism of the
NEGF,\cite{Keldysh, Haug} due to its integral form, becomes a
popular method in the  formulation, calculation and simulation of
recent mesoscopic transport problems. Caroli at al. were the first
to employ the Keldysh NEGF technique to study the tunneling
problems of a biased (nonequilibrium) metal-insulator-metal
junction.\cite{Caroli}  Meir and Wingreen \cite{Meir} in 1992
derived a useful formula for the current through an interacting
region with normal leads and applied it to investigate the
transport properties of a quantum dot in Kondo and fractional
quantum Hall regimes.  Later the Keldysh NEGF formalism was used
to analyze the I-V characteristics of
superconductor-superconductor point contacts and the transport
problem in a quantum dot with superconductor leads in Nambu
space.\cite{Cuevas,Fazio,Sun}  By introducing Green functions in
the spinor space, the Keldysh NEGF approach has been also employed
to study a quantum dot connected to two ferromagnetic
electrodes.\cite{Wang} Therefore, incorporating both Nambu and
spinor spaces is a convenient device in order to investigate
transport problems in the presence of both superconductors and
ferromagnets within the Keldysh NEGF formalism. It is the purpose
of this work to present a unified theory of electronic transport
through an interacting region connected to either bulk
ferromagnetic or superconducting leads. In such a formalism,
resonant transmission due to single particle interference,
correlation effects arising from to strong electron-electron
interactions, ferromagnetism and superconductivity proximity
effect in the presence of ferromagnets and superconductors can be
treated in a systematic way. We noticed that there exists a
circuit theory for mesoscopic systems developed by Nazarov et al.
\cite{Nazarov} based on the kinetic equations of quasiclassical
Green functions,\cite{kopnin} which provides an alternative way to
investigate the transport properties of hybrid structures with
arbitrary connections.\cite{Nazarov} However, such a formalism is
not favorable to the systems of strong correlation,  and
apparently inapplicable to the cases where  the single particle
interference effect (for example, in resonant-tunneling
structures) is prominent since the dependence on the relative
coordinate of the quasiclassical Green functions is integrated
out.

Thanks to recent advances in nanofabrication and material growth
technologies, several kinds of  hybrid mesoscopic structures have
been realized experimentally. These nanoscale structures include
mesoscopic junctions such as
normal-metal/superconductor\cite{Poirier} ($N/S$) and
ferromagnet/superconductor\cite{Upadhyay} ($F/S$) contacts,
superconductor/insulator/superconductor\cite{Morpurgo} ($S/I/S$)
and superconductor/ferromagnet/superconductor\cite{Lawrence}
($S/F/S$) junctions, and certain kinds of resonant structures such
as superconductor/quantum-dot/superconductor\cite{Tuominen}
($S-QD-S$), normal-metal/superconducting
quantum-dot/normal-metal\cite{Eiles} ($N-SQD-N$),
normal-metal/ferromagnetic-quantum-dot/normal-metal\cite{Gueron}
($N-FQD-N$) transistors. In a normal-metal/superconductor ($N/S$)
junction, Andreev reflection\cite{Andreev, NS} dominates the
transport process at low bias voltages, in which an electron in
the normal metal slightly above the chemical potential of the
superconductor is reflected as a hole slightly below the chemical
potential at the interface between the normal metal and
superconductor with an electron pair moving into the
superconductor, and vice versa. When two superconductor components
are coupled together through an insulator or a normal metal,
electron pairs can move coherently from one superconductor to the
other, yielding a nonzero current even in the zero bias limit-the
well known DC Josephson effect, and an oscillating current at
finite bias-the AC Josephson effect.\cite{Tinkham}  The $S/N/S$
Josephson junction must be a mesoscopic system, with the length
smaller than the phase coherent length of electrons in the normal
region, to ensure electron (hole)'s coherent motion inside the
normal part. Then a reflected electron (hole) can interfere
constructively with itself, a process that produces a set of
decoupled forward or backward `Andreev energy levels' carrying
positive or negative Josephson current.\cite{Kulik, Bagwell}   An
impurity inside the normal region couples the Andreev energy
levels, and thus modifies the quasi-particle energy spectrum and
other quantities. In the presence of a ferromagnetic metal, a
spin-polarized current may be induced due to the imbalance of the
spin populations  at the chemical potential.\cite{Prinz} Spin
imbalance also introduces a net magnetic moment-the magnetization
of ferromagnets. When two ferromagnets participate in a transport
experiment, the relative orientation of the magnetizations of
these two ferromagnets will play an  important role in the
transport properties, and the spin-valve effect
arises.\cite{Julliere} Combining  ferromagnets and
superconductors, one may expect some new transport features, since
there is no complete Andreev reflection at the $F/S$ interface.
The conductance of a $F/S$ junction can  be either smaller or
larger than the $N/S$ case, depending on the degree of spin
polarization of the ferromagnet.\cite{Jong}  When ferromagnets,
superconductors and confined (interacting) normal metals are
integrated together, the interplay between ferromagnetism,
superconductivity and electron-electron interaction is anticipated
to lead to
 more interesting and more complicated transport properties.
 Despite the basic interest in the fundamental theory as mentioned
 above, hybrid mesoscopic  systems also boast potential
 applications in future electronic devices which employ both the charge and spin
 degree of freedom of electrons.

Starting from a microscopic Hamiltonian, we derive in this paper a
 general current formula within the Keldysh NEGF formalism for hybrid mesoscopic systems in
 which a central nanoscale interacting normal region is weakly connected
 to two leads, each of which is either a ferromagnet or a superconductor,
 thus providing a unified theory of electron transport in
 general hybrid structures, which incorporates resonant tunneling, strong correlation, ferromagnetism and
 superconductivity proximity effect. Such a formula can be also applied to the
 nonequilibrium situation. Rather than from the original
 mean-field Stoner ferromagnet\cite{Jong} and BCS
 superconductor Hamiltonian,\cite{Tinkham} we calculate the
 current from their diagonalized forms after appropriate Bogoliubov transformations,
 with which  the  ferromagnetism and superconducting proximity as well as the
 chemical potentials of the system are embodied in the tunneling parts of the
 system Hamiltonian. Such a procedure is found to be a crucial
 step in the analysis of our transport problem, and facilitates the
 applications of the general formula to the specific forms for
given structures, which are Meir-Wingreen-type formulae.
 \cite{Meir}
  Employing such a procedure it is very easy to check
 whether such a theory satisfies the condition of
 gauge invariance, a requirement of all transport theories.
Moreover, the energy-dependence and  bias-voltage-dependence of
the level-width functions and the distribution functions in the
current formula are derived in a strict and natural way, while
this has been done somewhat phenomenologically in the other
formalisms.\cite{Meir, Haug, Sun, Wang, Zhu} This merit allows us
to investigate the I-V characteristics of hybrid mesoscopic
systems with a much more broad bias region.  In addition we
demonstrate that only their relative value for some physical
quantities appears in the current formula after some unitary
transformations. These quantities include the chemical potential,
magnetization orientation of ferromagnet and the phase of the
superconductor order parameter. Physically only their relative
value can be measured in a transport experiment for these physical
quantities, thus justifying the ad hoc assumption that one of them
can be always set to zero.\cite{Cuevas, Fazio,Sun,Wang} Such a
formalism can be directly extended to the cases with more than two
external leads, which can be either ferromagnetic or
superconducting. A shorter paper which summarizes the formulation
has been reported elsewhere.\cite{Zeng}

In order to illustrate the validity and versatility of our
formulation, we apply the derived formulae to a non-interacting
double-barrier resonant structure (DBTS) beyond the wide-band
approximation which is usually used in the Keldysh NEGF
formalisms.\cite{Haug, Meir,Fazio,Sun,Wang}   We neglect the
interaction effects, since in a regime where these interactions
are not important, we can then see more clearly how ferromagnetism
and superconductivity influence the transport properties of a
normal metal resonant structure coupled to ferromagnetic and/or
superconducting leads. As demonstrated in Section III, we derive
the final current formula based on our formulation in a more
systematic and  economic way than others.\cite{Sun, Zhu} Some
unexpected and novel transport features are found. When the level
spacing of the central normal region is comparable to the
bandwidth of the ferromagnetic lead(s), the I-V curves show
resonant peaks plus resonant shoulders, reflecting directly the
profile of the density of states (DOS) of the Stoner ferromagnet.
This observation provides an alternative way to measure the degree
of spin polarization of the system. The tunnel magnetoresistance
(TMR) decreases non monotonically, as well as oscillates, as a
function of the applied bias voltage between the ferromagnetic
leads. It is enhanced or suppressed within the resonant regions
depending on the couplings to the two sides. We also find negative
TMR at some bias voltages in the strong coupling limit. These
features tell us that there is richer physics in the TMR of a
resonant structure. In the presence of ferromagnetic and
superconucting leads, a series of peaks emerges in the Andreev
current whenever the resonant Andreev reflection condition at the
$N/S$ interface is satisfied as the gate voltage applied to the
central part varies. The number and height of these Andreev
current peaks are strongly dependent on the bias voltage and the
degree of spin polarization of the ferromagnet lead.  Interesting
step and peak structures are observed in the I-V characteristics,
which may be used to determine the DOS of both ferromagnetic and
superconducting leads. Finally we investigate the DC Josephson
current in $S/I/N/I/S$ structures. It is shown that the DC
Josephson current is slightly enhanced if the energy gaps of
superconductors becomes asymmetric, in contrast to the $S/N/S$
systems. The current-phase relation is also weakly dependent on
the asymmetry of the superconductor energy gaps.

The rest of this paper is organized as follows: In Section II the
full Hamiltonian of an interacting normal metal placed in between
either ferromagnetic or superconducting bulk leads is given. We
first express the current in terms of the nonequilibrium Green
functions in the Nambu-spinor space in general cases, and then
present current expressions for the specific structures. The
 gauge invariance is proven to hold as well.
Section III is devoted to the applications of the current formulae
derived in Section II to non-interacting double-barrier
structures, with a detailed analysis based on the analytical
results and numerical demonstrations. Concluding remarks are given
in Section IV. An appendix is included to present the expressions
of the self-energy matrices and level-width matrices due to the
elastic couplings to the leads.

\section{Formulation of the Problem}

We consider electron motion along the longitudinal direction $x$
in a hybrid sandwich structure schematically shown in Fig. $1$.
The central part is assumed to be in the normal state, connected
via tunnel barriers (insulators or point contacts, etc.) to two
bulk materials acting as leads, each of which can be either a
ferromagnet, or a conventional BCS superconductor. We adopt the
Stoner model\cite{Jong,Slonczewski} for the ferromagnet and the
BCS Hamiltonian\cite{Tinkham} for the superconductor. The Stoner
model Hamiltonian is characterized by a mean-field exchange
magnetization ${\bf h}$, and can be written as
\begin{figure}
\epsfig{file=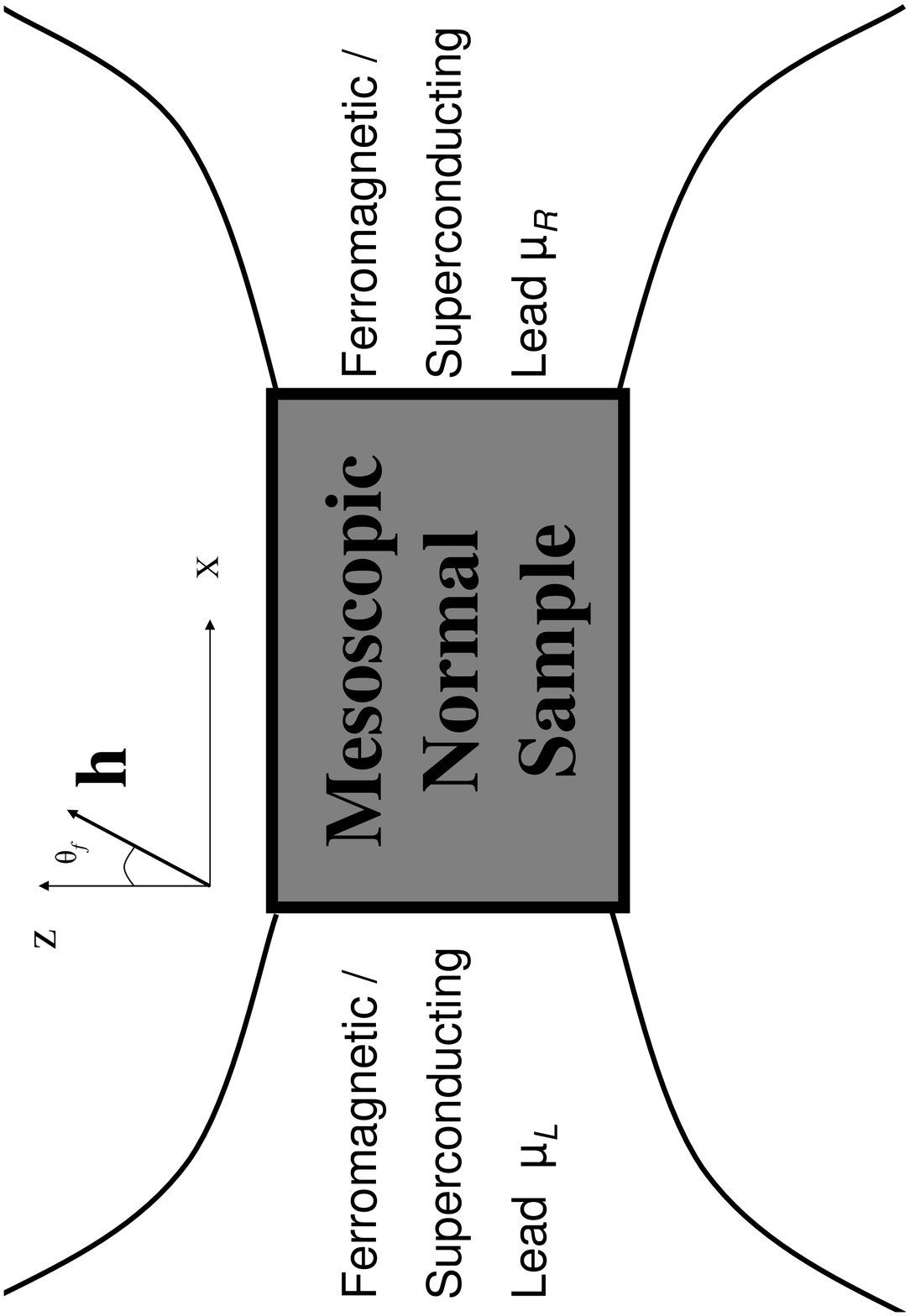, width=8cm} \caption{A schematic diagram
of a two-terminal hybrid mesoscopic structure. A mesoscopic normal
region is attached to either ferromagnetic or superconducting
leads with chemical potential $\mu_\cL$ and $\mu_\cR$. In the
former case the magnetization ${\bf h}$ makes an angle $\theta_f$
relative to the normal $z$ axis. The current is assumed to flow
along the longitudinal $x$ axis from the higher-chemical-potential
lead to the lower-chemical-potential one.}
\end{figure}
\begin{equation}
\label{eq:Hamiltonian-fx} H^F=\int dx \Psi^\dagger(x
)\Big(-\frac{\hbar^2}{2m^*}\nabla^2_x-\hat{\bf \sigma}\cdot {\bf
h}-\mu\Big)\Psi(x),
\end{equation}
where $m^*$ is the electron effective mass, $\hat{\bf
\sigma}=(\hat { \sigma}_x,\hat{ \sigma}_y, \hat{\sigma}_z)$ is the
Pauli spin operator, $\mu$ is the chemical potential and
$\Psi^\dagger=(\psi^\dagger_\uparrow,\psi^\dagger_\downarrow)$ is
the spinor field. In what follows we assume that the magnetization
${\bf h}$ makes an angle $\theta_f$ relative to the $z$ axis,
while we ignore the orientation with respect to the plane
perpendicular to the transport direction being not relevant to the
transport properties.\cite{Slonczewski}

 Within the mean-field approximation, the BCS
Hamiltonian takes the form
\begin{eqnarray}
\label{eq:Hamiltonian-sx}
 &&H^S=\sum\limits_{\sigma}\int dx
 \Psi^\dagger_\sigma(x)\Big(-\frac{\hbar^2}{2m^*}\nabla^2_x-\mu\Big)\Psi_\sigma(x
)\nonumber \\
&&+\int dx \big[\Delta(x)\Psi^\dagger_\uparrow(x
)\Psi^\dagger_\downarrow(x)- \Delta^*(x)\Psi_\downarrow(x
)\Psi_\uparrow(x)\big].
\end{eqnarray}
In Eq. (\ref{eq:Hamiltonian-sx})  $\Psi_\sigma$ is the field
operator of electrons with spin $\sigma=\uparrow, \downarrow$,
$\Delta(x)=U\langle\Psi_\uparrow(x)\Psi_\downarrow(x)\rangle$ is
the off-diagonal pair potential, with $U$ a negative constant
characterizing the electron-electron attraction. In general the
pair potential $\Delta$ needs to be determined self-consistently,
and in this work will be assumed to be position and energy
independent for simplicity.

Since we are concerning about transport properties of electrons
rather than their motion in real space it is more convenient to
deal with the lead Hamiltonian in $k$-space. Expanding the
electron field operator in terms of the eigenfunctions of the
momentum operator in the longitudinal(tunneling) direction as
$\Psi_\sigma(x)=\sum f_{k\sigma }e^{i kx}(s_{k\sigma }e^{i kx}) $
we cast the ferromagnet and superconductor Hamiltonian
(\ref{eq:Hamiltonian-fx}) and (\ref{eq:Hamiltonian-sx}) into the
following forms in $k$-space (subscript $\gamma=\cL$ and $\cR$ are
added to denote which side of the structure the ferromagnet or
superconductor is located at)
\begin{eqnarray}
\label{eq:Hamiltonian-fk}
 \cH^{(F)}_\gamma &=&\sum_{k\sigma
}[\varepsilon _{\gamma k }-sgn(\sigma) h_{\gamma} \cos
\theta_{\gamma f}-\mu_\gamma ]f_{\gamma k\sigma }^{\dagger }f_{\gamma k\sigma }\nonumber \\
&& \hspace{2cm} +\sum_{k\sigma }h_{\gamma}\sin \theta_{\gamma f}
f_{\gamma k\sigma }^{\dagger}f_{\gamma k\stackrel{-}{\sigma }},   \\
\label{eq:Hamiltonian-sk}
 \cH^{(S)}_\gamma &=&\sum_{k\sigma
}(\varepsilon_{\gamma k
}-\mu_\gamma)s_{\gamma k\sigma }^{\dagger }s_{\gamma k\sigma }+\nonumber \\
&& \hspace{0.5cm} \sum_{k}\left[ \Delta_\gamma s_{\gamma k\uparrow
}^{\dagger }s_{\gamma -k\downarrow }^{\dagger }+\Delta^*_\gamma
s_{\gamma -k\downarrow } s_{\gamma k\uparrow }\right],
\end{eqnarray}
where $\varepsilon_{\gamma k}=\hbar^2 k^2/2m^*$,
$\stackrel{-}\sigma$ stands for the opposite to $\sigma$. Here the
order parameter $\Delta_\gamma$ is characterized by its magnitude
and phase: $|\Delta_\gamma|e^{i\varphi_\gamma}$, and, as we will
show, $|\Delta_\gamma|$ opens an energy gap in the excitation
spectrum of the superconductor. $f_{\gamma k\sigma
}(f^\dagger_{\gamma k\sigma })$ and $s_{\gamma k\sigma
}(s^\dagger_{\gamma k\sigma })$ are the electron destruction
(creation) operators of spin $\sigma$ in state $k$ in the
ferromagnet and superconductor, respectively.
 Henceforth,  the notations
$\sigma=\uparrow, \downarrow$ and $\sigma=\pm$ are used
interchangeably.   In what follows physical quantities such as the
particle operator $\psi$, particle energy $\varepsilon$, and the
chemical potential $\mu$ of the different parts are labelled by
subscript $\gamma=\cL, \cR; \cC$, whenever convenient. In
addition, ferromagnetic or superconducting characteristics of the
leads is stressed by adding a subscript or a superscript $f$ or
$s$ to some quantities.

The Hamiltonian of the central region
 $\cH_{\cC}$ in momentum space
can be modeled by
\begin{eqnarray}
\label{Hamiltonian:central} \cH_{\cC} &=&\sum_{n\sigma
}(\varepsilon _{n\sigma
}-\mu_\cC)\psi_{cn\sigma }^{\dagger }\psi_{cn\sigma }+\nonumber \\
&& \hspace{1.5cm} \cH_{int}(\{\psi_{cn\sigma
}^{\dagger}\},\{\psi_{cn\sigma }\}),
\end{eqnarray}
where $\psi_{cn\sigma }^{\dagger }(\psi_{cn\sigma })$ creates
(destroys) an electron of spin $\sigma$ in state $n$, and
$\cH_{int}$ represents the interaction terms in the central
region. It may include the electron-electron Coulomb interaction
\begin{equation}
\cH_{int}^{el-el}=\sum\limits_{\matrix{n,m,\sigma, \sigma'\cr
n\sigma\neq m\sigma'}}U_{n\sigma,m\sigma'}\psi_{cn\sigma
}^{\dagger }\psi_{cn\sigma }\psi_{cm\sigma' }^{\dagger
}\psi_{cm\sigma' },
\end{equation}
or the electron-phonon interaction
\begin{equation}
\cH_{int}^{el-ph}=\sum\limits_{q}\hbar \omega_q \zeta_q^\dagger
\zeta_q+\sum\limits_{n,\sigma; q}U_{n\sigma,q}\psi_{cn\sigma
}^{\dagger }\psi_{cn\sigma }(\zeta_q^\dagger+\zeta_{-q}).
\end{equation}
Here the first term is the free-phonon Hamiltonian, while the
second represents the electron-phonon interaction, with
interaction matrix element $U_{n\sigma,q}$.
$\zeta^\dagger_q(\zeta_q)$ is the phonon creation (destruction)
operator in mode $q$.

The couplings between the leads and the central region can be
modeled by tunneling Hamiltonian, no matter how the leads are
coupled to the central region, provided the couplings are not
strong enough. Certainly the coupling strengths depend on the
detailed configuration of the setup and should be determined in a
self-consistent manner. However, for simplicity they are assumed
known and can be written as
\begin{eqnarray}
\cH_{\cT}^{\gamma (F)} &=&\sum_{kn;\sigma }\left[ V^{\gamma
f}_{kn;\sigma }f_{\gamma k\sigma }^{\dagger }\psi_{cn\sigma
}+V_{kn;\sigma }^{\gamma f*}\psi_{cn\sigma }^{\dagger }f_{\gamma
k\sigma}\right], \\
\cH_{\cT}^{\gamma (S)} &=&\sum_{kn;\sigma }\left[ V^{\gamma
s}_{kn;\sigma }s_{\gamma k\sigma }^{\dagger }\psi_{cn\sigma
}+V_{kn;\sigma }^{\gamma s*}\psi_{cn\sigma }^{\dagger }s_{\gamma
k\sigma }\right].
\end{eqnarray}

To see tunneling processes more clearly, and, more importantly, to
facilitate the analysis of gauge invariance and the simplification
of the general current formula (\ref{eq:current}) to the forms of
specific systems, we first diagonalize the Hamiltonian of the
leads by Bogoliubov transformations. For the ferromagnetic lead
one has
\begin{equation}
\label {bogoliubov-f}
 f_{\gamma k\sigma}=\cos (\theta_{\gamma
f}/2) \psi_{\gamma fk\sigma}-sgn(\sigma) \sin(\theta_{\gamma f}/2)
\psi_{\gamma fk \stackrel{-}{\sigma}},
\end{equation}
and for the superconducting lead,
\begin{equation}
\label {bogoliubov-s}
 e^{-i\varphi_\gamma/2}s_{\gamma
k\sigma}=\cos \theta_{\gamma sk} \psi_{\gamma
sk\sigma}+sgn(\sigma) \sin \theta_{\gamma sk}\cP \psi_{\gamma
sk\stackrel{-}\sigma}^{\dagger}.
\end{equation}
In Eqs. (\ref{bogoliubov-f},\ref{bogoliubov-s}), $\cP^\dagger
(\cP)$ is the pair creation (destruction) operator guaranteeing
particle conservation, which transforms a given N-particle state
into an (N+2)-particle ((N-2)-particle) state, i.e.,
$\cP^\dagger/\cP|N\rangle=|N+2\rangle/|N-2\rangle$, and
\begin{equation}
\theta_{\gamma sk}=\arctan\left(\frac{\varepsilon_{\gamma k}+
\sqrt{\varepsilon_{\gamma
k}^2+|\Delta_\gamma|^2}}{\varepsilon_{\gamma k}-
\sqrt{\varepsilon_{\gamma k}^2+|\Delta_\gamma|^2}}\right)^{1/2}.
\end{equation}

Substituting the Bogoliubov transformations (\ref{bogoliubov-f})
and (\ref{bogoliubov-s}) into the lead Hamiltonian
(\ref{eq:Hamiltonian-fk}) and (\ref{eq:Hamiltonian-sk}),  we get
the following diagonalized forms for the ferromagnetic and
superconducting leads, respectively,
\begin{eqnarray}
\label{eq:Hamiltonian-f}
 \cH_\gamma^{(F)}
&=&\sum_{k\sigma}[\varepsilon_{\gamma k}-sgn(\sigma)
h_{\gamma}-\mu_\gamma]
\psi_{\gamma fk\sigma}^{\dagger }\psi_{\gamma fk\sigma}\nonumber \\
&=&\sum_{k\sigma}\varepsilon_{\gamma fk\sigma}
\psi_{\gamma fk\sigma}^{\dagger }\psi_{\gamma fk\sigma}, \\
\label{eq:Hamiltonian-s}
\cH_\gamma^{(S)}&=&\sum_{k\sigma}(\sqrt{\varepsilon_{\gamma
k}^2+|\Delta_\gamma|^2}-\mu_\gamma)
 \psi^{\dagger}_{\gamma sk\sigma}\psi_{\gamma sk\sigma} +\nonumber \\
 & & \hspace{5cm}Constant\nonumber \\ & &
 =
 \sum_{k\sigma}\varepsilon_{\gamma sk}\psi^{\dagger}_{\gamma sk\sigma}\psi_{\gamma
 sk\sigma}+Constant
 \end{eqnarray}
Now the particle number operator commutes with the corresponding
lead Hamiltonian. After the Bogoliubov transformations, the
diagonalized lead Hamiltonian describe the excitation (quasi
particle) properties. The minimum energy of the excitations in a
superconductor is $|\Delta_\gamma|$, implying an energy gap in the
excitation spectrum.

With the Bogoliubov transformations (\ref{bogoliubov-f}) and
(\ref{bogoliubov-s}) we turn the tunneling Hamiltonian into
 \begin{widetext}
 \begin{eqnarray}
\label{eq:tunnelh-f}
 \cH_\cT^{\gamma (F)} &=&\sum_{kn;\sigma
}\left(V^{\gamma f}_{kn;\sigma }\big[\cos (\theta_{\gamma f}/2)
\psi_{\gamma fk\sigma}^\dagger-sgn(\sigma) \sin(\theta_{\gamma
f}/2) \psi_{\gamma
f k \stackrel{-}{\sigma}}^\dagger\big]\psi_{cn\sigma }\right. +\nonumber \\
 &&\hspace{5cm}\left.
V^{\gamma f*}_{kn;\sigma }\psi_{cn\sigma }^\dagger\big[\cos
(\theta_{\gamma f}/2) \psi_{\gamma fk\sigma}-sgn(\sigma)
\sin(\theta_{\gamma f}/2) \psi_{\gamma f k \stackrel{-}{\sigma}}
\big]\right),\\
\label{eq:tunnelh-s}
 \cH_\cT^{\gamma (S)} &=&\sum_{kn;\sigma
}\left( V^{\gamma s}_{kn;\sigma }\big[ \cos\theta_{\gamma sk}
\psi_{\gamma sk\sigma}^{\dagger}+sgn(\sigma)\sin\theta_{\gamma sk}
\psi_{\gamma
sk\stackrel{-}\sigma}\cP^\dagger\big]e^{i\varphi_\gamma/2}\psi_{cn\sigma
}
\right.+\nonumber \\
 &&\hspace{5cm}\left.V^{\gamma
s*}_{kn;\sigma }\psi_{cn\sigma }^\dagger
e^{-i\varphi_\gamma/2}\big[\cos\theta_{\gamma sk} \psi_{\gamma
sk\sigma}+sgn(\sigma) \sin(\theta_{\gamma sk})\cP \psi_{\gamma
sk\stackrel{-}\sigma}^\dagger \big]\right).
\end{eqnarray}
\end{widetext}
The associated physical processes are more obvious and clear in
the semiconductor model:\cite{Bardeen,Tinkham} an electron of spin
$\sigma$ in the central regime can tunnel  into either the spin
$\sigma$ band or $\stackrel{-}{\sigma}$ band of the ferromagnetic
lead, or tunnel into a spin $\sigma$ state or condensate into an
electron pair with a hole state of opposite spin being created,
and vice versa.

 In superconductors, correlation between two
creation or annihilation quasi-particle operators with opposite
spins are very important, relating to the Andreev reflection in
transport processes. When ferromagnets are introduced, the
correlation between a creation and an annihilation quasi-particle
operator with opposite spins needs to be considered. To
incorporate these two kinds of correlations in a unified way and
consider the ferromagnet and superconductor on the same footing,
we here introduce a generalized Nambu-spin representation spanning
a 4-dimensional spin-orientated particle-hole space ${\bf \Psi}_x=
\left(\matrix{ \psi_{x\uparrow }^{\dagger } & \psi_{x \downarrow }
& \psi_{x\downarrow }^{\dagger } & \psi_{x\uparrow} \cr }
 \right)^\dagger.$
Within the Keldysh NEGF formalism,  Green functions are defined as
\begin{equation}
{\bf G}_{x,y}(t_1,t_2)=i \langle T_C[{\bf \Psi}_x(t_1) \otimes {\bf
\Psi}_y^\dagger(t_2)]\rangle,
\end{equation}
where $T_C$ is the time-ordering operator along the closed time
path.\cite{Haug} The usual retarded/advanced and lesser/greater
Green functions then take the form
\begin{eqnarray*}
{\bf G}_{\alpha, \beta}^{r/a}(t_1,t_2)
&=&\sum\limits_{i,j}{\bf G}_{\alpha i, \beta j}^{r/a}(t_1,t_2)\\
\nonumber &=& \mp i \vartheta(\pm t_1 \mp
t_2)\sum\limits_{i,j}\langle[{\bf \Psi}_{\alpha i}(t_1)\otimes\nonumber
\\ & & \hspace{-0.2cm} {\bf
\Psi}_{\beta j}^\dagger(t_2)+{\bf \Psi}_{\beta j}^\dagger(t_2)\otimes{\bf \Psi}_{\alpha i}(t_1)]\rangle, \\
{\bf G}_{\alpha , \beta }^{</>}(t_1,t_2)&=& \sum\limits_{i,j}{\bf
G}_{\alpha i, \beta j}^{</>}(t_1,t_2)
\\&=&\pm i\sum\limits_{i,j}\langle{\bf \Psi}_{\beta j}^\dagger(t_2)/{\bf
\Psi}_{\alpha i}(t_1) \otimes \nonumber \\ && \hspace{1cm} {\bf
\Psi}_{\alpha i}(t_1) /{\bf \Psi}_{\beta j}^\dagger(t_2)\rangle,
\end{eqnarray*}
where $\alpha, \beta=\gamma f,\gamma s,c$ and $i,j=k,n$.

In this 4-dimensional Nambu-spinor space the total Hamiltonian can
be rewritten in the following compact form
\begin{equation}
\label {Hamiltonian-T}
 \cH=\cH_\cC+\cH_\cL+\cH_\cR+\cH_\cT^\cL+\cH_\cT^\cR,
 \end{equation}
 where
 \begin{eqnarray}
 \label{Hamiltonian-C}
 \cH_\cC&=&\sum\limits_{n}{\bf \Psi}_{cn}^\dagger
{\bf E}_{cn}{\bf \Psi}_{cn}+ \nonumber\\
 && \hspace{2cm}
  \cH_{int}(\{{\bf
\Psi}_{cn}^\dagger,{\bf \Psi}_{cn}\}), \\
 \label{Hamiltonian-LR}
\cH_{\gamma}^{(F/S)}&=&\sum\limits_{k}{\bf \Psi}_{\gamma
f/sk}^\dagger
{\bf E}_{\gamma f/s k}{\bf \Psi}_{\gamma f/sk}, \\
 \label{Hamiltonian-Tu}
\cH^{\gamma {(F/S)}}_\cT &=&\sum_{kn}\big[{\bf \Psi}_{\gamma f/s
k}^\dagger {\bf V}^{\gamma f/s}_{kn}(t){\bf \Psi}_{cn}+H.c.\big].
\end{eqnarray}
In writing down Eq.
(\ref{Hamiltonian-C},\ref{Hamiltonian-LR},\ref{Hamiltonian-Tu}),
we have introduced the energy matrices
\begin{equation}
{\bf E}_{\alpha}=\left(\matrix { \epsilon_{\alpha \uparrow} &0
&0&0\cr 0 &- \epsilon_{\alpha \downarrow}&0&0\cr
0&0&\epsilon_{\alpha \downarrow}&0\cr 0&0&0&-\epsilon_{\alpha
\uparrow}\cr }\right), \hspace{0.5cm} \alpha=cn, \gamma f/s k
\end{equation}
and the tunneling matrices
\begin{eqnarray}
\label{tunnel matrix-f} {\bf V}^{\gamma f}_{kn}(t)&=&{\bf
R}^f(\frac{\theta_{\gamma f}}{2}){\bf
V}^{\gamma f}_{kn}{\bf P}(\mu_{\gamma \cC} t), \\
 \label{tunnel matrix-s}
{\bf V}^{\gamma s}_{kn}(t)&=&{\bf R}^s(\theta_{\gamma sk}){\bf
V}^{\gamma s}_{kn}{\bf P}(\mu_{\gamma \cC}
t+\frac{\varphi_\gamma}{2}), \\
\mu_{\gamma \cC}&=&\mu_\gamma-\mu_\cC \nonumber
\end{eqnarray}
in which
\begin{eqnarray*}
{\bf V}^{\gamma f/s}_{kn}
 &=&\left (\matrix{ V^{\gamma f/s}_{kn} &
 0 & 0
  & 0 \cr
 0 &-V^{\gamma f/s*}_{kn}& 0 &
  0\cr
 0 & 0
 &V^{\gamma f/s}_{kn} & 0 \cr
 0 & 0 & 0 &
  -V^{\gamma f/s*}_{kn} \cr }\right),\\
  {\bf R}^{ f}(x)
 &=&\left (\matrix{ \cos x &
 0 & \sin x
  & 0 \cr
 0 &\cos x& 0 &
  -\sin x\cr
 -\sin x& 0
 &\cos x& 0 \cr
 0 & \sin x & 0 &
  \cos x\cr }\right), \\
{\bf R}^{ s}(x)
 &=&\left (\matrix{ \cos x &
  -\cP\sin x &0
  & 0 \cr
 \cP^*\sin x  &
 \cos x& 0 & 0\cr
 0& 0&\cos x &\cP\sin x \cr
 0 &0 & -\cP^*\sin x  &
\cos x\cr }\right), \\
 {\bf P}(x) &=& \left(\matrix{ e^{ix/\hbar} & 0 & 0 &0\cr
 0& e^{-ix/\hbar}&0&0\cr
 0 & 0&e^{ix/\hbar} & 0 \cr
 0 & 0 & 0& e^{-ix/\hbar}\cr }\right).
\end{eqnarray*}
are the {\it coupling, rotation} and {\it phase} matrices,
respectively. Note that we have performed a gauge transformation
\cite{Rogovin} to get the above Hamiltonian. Chemical potential is
now incorporated in the phase operator ${\bf P}(\mu_{\gamma
\cC}t)$ in the tunneling matrices Eqs.(\ref{tunnel
matrix-f},\ref{tunnel matrix-s}), which along with the {\it
rotation} operators is very useful to demonstrate gauge invariance
for our system as shown below.

The current flowing from lead $\gamma=\cL,\cR$ to the central
region can be defined as the rate of change of the electron number
$N_\gamma=\sum_{k\sigma} f_{\gamma k\sigma}^{\dagger}f_{\gamma
k\sigma}(  s_{\gamma k\sigma}^{\dagger}s_{\gamma k\sigma})
=\sum_{k\sigma} \psi_{\gamma f k\sigma}^{\dagger}\psi_{\gamma f
k\sigma}(\psi_{\gamma s k\sigma}^{\dagger}\psi_{\gamma s
k\sigma})$ in the lead. Within the Keldysh NEGF formalism, the
current is expressed as
\begin{eqnarray}
\label{eq:current-gamma}
\cI_{\gamma}(t)&=&-e<\dot{N_\gamma}>=\frac{ie}{\hbar}<[N_\gamma,H_\cT]>
 \nonumber\\
&=&- \frac{e}{\hbar}\sum\limits_{i=1,3}\sum\limits_{nk}\left([{\bf
G}_{cn,\gamma f/sk}^<(t,t){\bf V}^{\gamma
f/s}_{kn}(t)-\right.\nonumber\\
&&\hspace{2cm} \left.{\bf V}^{\gamma f/s\dagger}_{kn}(t){\bf G}_{
\gamma f/s
k,cn}^<(t,t)]\right)_{ii} \nonumber \\
&=&\frac{2e}{\hbar}\sum\limits_{nk}^{i=1,3}{\rm
Re}\left\{\Big({\bf V}^{\gamma f/s\dagger }_{kn}(t){\bf G}_{\gamma
f/sk,cn}^<(t,t)\Big)_{ii}\right \}.\nonumber \\
\end{eqnarray}

  Since the Hamiltonian of  lead $\gamma$ is of the form
$\psi_{\gamma f/s k\sigma}^{\dagger}\psi_{\gamma f/s k\sigma}$,
the equations of motion for $ {\bf G}_{\gamma f/sk, cn}^<$ along
with the Langreth analytic continuation \cite{Langreth} yield the
following Dyson equations
\begin{eqnarray}
\label{eq:dyson}
 {\bf G}_{\gamma f/sk,
cn}^<(t,t')&=&\sum\limits_{m}\int dt_1 \Big[{\bf g}_{\gamma f/sk,
\gamma f/sk}^r(t,t_1) {\bf V}^{\gamma f/s}_{km}(t_1)\nonumber
\\&&\hspace{-1cm}{\bf G}_{cm, cn}^<(t_1,t') +{\bf g}_{\gamma f/sk, \gamma
f/sk}^<(t,t_1)\nonumber\\ && \hspace{1cm}
{\bf V}^{\gamma f/s}_{km}(t_1){\bf G}_{cm, cn}^a(t_1,t')\Big], \\
{\bf G}_{cn,\gamma f/sk}^<(t,t')&=&\sum\limits_{m} \int
dt_1\Big[{\bf G}_{cn, cm}^<(t,t_1){\bf V}^{\gamma f/s\dagger
}_{km}(t_1)\nonumber\\&&\hspace{-1cm} {\bf g}_{\gamma f/sk,\gamma
f/sk}^r(t_1,t')+{\bf G}_{cn,
cm}^r(t,t_1)\nonumber\\&&\hspace{0.5cm}{\bf V}^{\gamma f/s\dagger
}_{km}(t_1){\bf g}_{\gamma f/sk, \gamma f/sk}^<(t_1,t')\Big],
\end{eqnarray}
 in which the unperturbed retarded/advanced Green function $
 {\bf g}_{\gamma f/s, \gamma f/s}^{r/a}$ of lead $\gamma$ can be readily
 obtained from the Hamiltonian $(\ref{eq:Hamiltonian-f}, \ref{eq:Hamiltonian-s})$ as  diagonal
 matrices
 \begin{eqnarray}
 \label{eq:unper-green}
{\bf g}_{\gamma f/sk, \gamma f/sk}^{r/a}(t,t')
 &=&\mp i\vartheta(\pm t \mp t')\nonumber \\
 &&\hspace{-2cm} \left (\matrix{g^{\uparrow-}_{\gamma f/sk} & 0 & 0
  & 0 \cr
 0 &g^{\downarrow+}_{\gamma f/sk}& 0 & 0 \cr
 0 & 0&g^{\downarrow-}_{\gamma f/sk} & 0 \cr
 0 & 0 & 0 &g^{\uparrow+}_{\gamma f/sk}
   \cr }\right), \\
g^{\sigma\mp}_{\gamma f/sk}(t,t')&=& e^{ \mp i\varepsilon_{\gamma
f/s k\sigma}(t-t')/\hbar},
\end{eqnarray}
and the lesser(greater) Green functions are related to the
retarded(advanced) Green functions by ${\bf g}_{\gamma f/sk,
\gamma f/sk}^{</>}(t,t') = \big[ {\bf f}_\gamma
(\varepsilon_{\gamma f/sk})-\frac12{\bf 1}\pm \frac12{\bf
1}\big]\big[{\bf g}_{\gamma f/sk, \gamma f/sk}^a(t,t')-{\bf
g}_{\gamma f/sk, \gamma f/sk}^r(t,t')\big]$.  The Fermi
distribution matrix  ${\bf f}_\gamma(\varepsilon_{\gamma f/sk})$
reads
\begin{eqnarray}
{\bf f}_\gamma(\varepsilon_{\gamma f/sk})
 &=&\nonumber\\
 &&\hspace{-2cm}\left (\matrix{ f(\varepsilon_{\gamma
f/sk\uparrow}) & 0 & 0
  & 0 \cr
 0 &f(-\varepsilon_{\gamma
f/sk\downarrow}) & 0 & 0 \cr
 0 & 0&f(\varepsilon_{\gamma
f/sk\downarrow})  & 0 \cr
 0 & 0 & 0 & f(-\varepsilon_{\gamma
f/sk\uparrow})  \cr}\right),\nonumber
\end{eqnarray}
where $f(x)=(1+e^{x/k_BT})^{-1}$ and we have used the relation
$f(-x)=1-f(x)$.

 Substituting Eq. (\ref{eq:dyson}) into (\ref{eq:current-gamma}),
we obtain
\begin{eqnarray}
\label{current-gamma}
 \cI_{\gamma}(t)&=&
\frac{2e}{\hbar}
 \sum\limits^{i=1,3}_{nm}\int_{-\infty}^{t}dt_1 {\rm Re}\Big\{\Big({\bf
\Sigma}^r_{\gamma f/s; nm}(t,t_1)\nonumber\\ & & \hspace{1cm} {\bf
G}_{cm,cn}^{<}(t_1,t)+{\bf \Sigma}^<_{\gamma f/s;
nm}(t,t_1)\nonumber\\ & &  \hspace{3cm} {\bf
G}_{cm,cn}^{a}(t_1,t)\Big)_{ii}\Big\},
\end{eqnarray}
where
\begin{eqnarray}
{\bf \Sigma}^{r,a, </>}_{\gamma f/s;nm}(t_1,t_2)&=&\sum_{k}{\bf
V}^{\gamma f/s\dagger}_{kn}(t_1) \nonumber\\
 &&{\bf g}^{r/a,
</>}_{\gamma f/sk,\gamma f/sk}(t_1,t_2){\bf V}^{\gamma
f/s}_{km}(t_2)\nonumber
\end{eqnarray}
is the self-energy matrix (see Appendix A) arising from electron
tunneling between the central region and lead $\gamma$.

For steady transport, no charge piles up in the central normal
region. One then has $\cI_\cL(t)=-\cI_\cR(t)$.\cite{Meir}  After
symmetrizing the current formula (\ref{current-gamma}), we finally
get by setting $\cI(t)=[\cI_\cL(t)-\cI_\cR(t)]/2$
\begin{eqnarray}
\label{eq:current}
 \cI(t)&=&
\frac{e}{\hbar}\sum\limits^{i=1,3}\int\limits_{-\infty}^{t}dt_1
{\rm Re} {\rm Tr} \Big\{\Big( \big[{\bf \Sigma}^r_{\cL
f/s}(t,t_1)- \nonumber\\&&\hspace{1cm}{\bf \Sigma}^r_{\cR
f/s}(t,t_1)\big]{\bf G}_{c,c}^{<}(t_1,t)+\big[{\bf \Sigma}^<_{\cL
f/s}(t,t_1)-\nonumber\\&&\hspace{2cm}{\bf \Sigma}^<_{\cR
f/s}(t,t_1)\big]{\bf G}_{c,c}^{a}(t_1,t) \Big)_{ii}\Big\} ,
\end{eqnarray}
where the trace is over the level indices of the central region.
Eq. $(\ref{eq:current})$ along with the self-energy matrices given
in Appendix A is the central result of this work. The current is
expressed in terms of the local properties ($ {\bf G}^{r/a}$) and
the occupation ($ {\bf G}^{</>})$ of the central interacting
 region and the equilibrium properties (${\bf
\Sigma}^{</>}$) of the leads. It is emphasized that the current is
usually independent of time except the presence of two
superconductor leads with nonzero bias voltage. Formula
(\ref{eq:current}) can be employed to investigate both equilibrium
and nonequilibrium electronic transport in various kinds of hybrid
mesoscopic systems, including $F/I/N/I/F$, $F/I/N/I/S$,
$F/I/N/I/N$, $S/I/N/I/S$, $S/I/N/I/N$, and $F-QD-F$, $F-QD-S$,
$F-QD-N$, $S-QD-S$, $S-QD-N$ structures as well, in which
arbitrary interactions are allowed in the central part of the
structure.

It is not difficult to check that Eq. (\ref{eq:current}) is gauge
invariant, i.e., the current $\cI(t)$ remains unchanged under a
global energy shift in the whole region. This can be achieved
through a gauge transformation for the Hamiltonian of the system
\begin{eqnarray*}
 \hat{f}(\epsilon_0 t)&=&\exp\Big \{\frac{i}{\hbar}\epsilon_0
t \big(\sum\limits_{n\sigma}\psi^\dagger_{cn\sigma
}\psi_{cn\sigma} \nonumber
\\  && \hspace{2cm} +\sum\limits_{\gamma=\cL, \cR; k\sigma}\psi^\dagger_{\gamma f/s k\sigma}\psi_{\gamma f/s
k\sigma} \big) \Big\},
\end{eqnarray*}
where $\epsilon_0$ is just the energy shift. Such a gauge
transformation gives rise, in turn, to the {\it phase}
transformations of all the terms in the right-hand side of Eq.
(\ref{eq:current})
\begin{eqnarray*}
{\bf \Sigma}^{r/<}_{\gamma f/s}(t,t_1)&\rightarrow& {\bf
P}(\varepsilon_0t){\bf \Sigma}^{r/<}_{\gamma
f/s}(t,t_1){\bf P}^\dagger(\varepsilon_0t_1),\nonumber \\
 {\bf G}^{a/<}_{c,c}(t_1,t)&\rightarrow&
{\bf P}^\dagger(\varepsilon_0t_1){\bf G}^{a/<}_{c,c }(t_1,t){\bf
P}(\varepsilon_0t).
\end{eqnarray*}
The above procedures, equivalent to applying a {\it phase}
transformation to the current operator: ${\bf
P}^\dagger(\varepsilon_0t)\cI(t){\bf P}(\varepsilon_0t)$, ensure
that the current remains the same under such a transformation.
Therefore, the current formula (\ref{eq:current}) is  gauge
invariant.

Now we check whether the current becomes zero if we take the zero
bias limit $\mu_\cL=\mu_\cR=\mu_0$. We first perform a {\it phase}
operation ${\bf P}(\mu_{0\cC }t+\varphi_\cR/2)$ corresponding to
the gauge transformation $\hat{f}(\mu_{\cC 0}t)$,
$\mu_{0\cC}=\mu_0-\mu_\cC$ to Eq.(\ref{eq:current}), obtaining
\begin{eqnarray}
 \cI(t) &=&-\frac{e}{\hbar}\sum\limits^{i=1,3}\int\frac{d\varepsilon}{2\pi}
{\rm Im} {\rm Tr}\Big\{\Big(\frac12\big[\tilde{{\bf \Gamma}}^{\cL
f/s}_{/\varrho}(\varepsilon)-{\bf \Gamma}^{\cR
f/s}_{/\varrho}(\varepsilon)\big]\nonumber \\
&& \hspace{0.5cm}\widetilde{{\bf G}}_{c,c}^{<}(\varepsilon)-
\big[\tilde{{\bf \Gamma}}^{\cL f/s}_{/\varrho}(\varepsilon){\bf
f}_\cL(\varepsilon)-{\bf \Gamma}^{\cR
f/s}_{/\varrho}(\varepsilon){\bf
f}_\cR(\varepsilon)\big]\nonumber\\ &&
\hspace{2.2cm}\widetilde{{\bf G}}_{c,c}^{a}(\varepsilon)
\Big)_{ii}\Big\},
\end{eqnarray}
where
\begin{eqnarray*}
\tilde{{\bf \Gamma}}^{\cL f/s}_{/\varrho}(\varepsilon)&=&{\bf
P}^\dagger(\frac{\varphi_s}{2}){\bf \Gamma}^{\cL
f/s}_{/\varrho}(\varepsilon){\bf P}(\frac{\varphi_s}{2}),\\
\widetilde{{\bf G}}_{c,c}^{r,a/<}(\varepsilon)&=&\int
d(t-t')e^{i\varepsilon(t-t')/\hbar}{\bf P}(\mu_{0\cC}
t+\frac{\varphi_\cR}{2})\nonumber\\
&& \hspace{0.9cm}{\bf G}^{r,a/<}_{c,c}(t,t'){\bf
P}^\dagger(\mu_{0\cC}t'+\frac{\varphi_\cR}{2}),
\end{eqnarray*}
with $\varphi_s=\varphi_\cL-\varphi_\cR$. From the
fluctuation-dissipation theorem  $\widetilde{{\bf G}}_{c,c}^{</>}=
[{\bf f}_{eq}(\varepsilon)-\frac12{\bf 1}\pm \frac12{\bf
1}](\widetilde{{\bf G}}_{c,c}^{a}-\widetilde{{\bf G}}_{c,c}^{r})$
$({\bf f}_\cL={\bf f}_\cR={\bf f}_{eq})$, one can readily verify
that the current is zero except in the presence of two
superconductor leads with different superconducting order
parameter phases.
 In this case, there still exists a DC Josephson current in the zero bias limit due to the coherent
tunneling of {\it quasi-particle pairs}. This can be seen more
clearly in the expressions  of the current in the specific systems
(\ref{eq:current-fnf},\ref{eq:current-fns},\ref{eq:current-sns}).

Up to now we have obtained the expression for the current in a
general case in which each of the two leads can be either a
ferromagnet or a superconductor. Next we apply this general result
(\ref{eq:current}) to the specific structures we are interested
in.  We first consider the case in which two leads are
ferromagnetic. Inserting the expressions of the self-energy
matrices ${\bf \Sigma}_{\gamma f}(t_1,t_2)$ (Appendix A) into
Eq.(\ref{eq:current}), we get the current in a $F/I/N/I/F$ or
$F-QD-F$ structure after a {\it rotation} transformation and a
{\it phase} transformation
\begin{eqnarray}
\label{eq:current-fnf}
 \cI_{fnf}&=&\frac{ie}{2\hbar}\sum\limits^{i=1,3}\int\frac{d\varepsilon}{2\pi}
{\rm Tr}\Big\{\Big(\big[\hat{{\bf \Gamma}}^{\cL f}(\varepsilon\mp
eV)-{\bf \Gamma}^{\cR f}(\varepsilon)\big]
\nonumber \\
&& \hspace{0.2cm} \widehat{{\bf G}}_{c,c}^{<}(\varepsilon)+
\big[\hat{{\bf \Gamma}}^{\cL f}(\varepsilon\mp eV){\bf
f}_\cL(\varepsilon\mp
eV)-\nonumber\\
&& \hspace{0.5cm}{\bf \Gamma}^{\cR f}(\varepsilon){\bf
f}_\cR(\varepsilon)\big] \big[\widehat{{\bf
G}}_{c,c}^{r}(\varepsilon)-\widehat{{\bf
G}}_{c,c}^{a}(\varepsilon)\big]\Big)_{ii} \Big\},
\end{eqnarray}
where  $\hat {\bf \Gamma}^{\cL f}={\bf
R}^{f\dagger}(\frac{\theta_{ f}}{2}) {\bf \Gamma}^{\cL f}{\bf
R}^f(\frac{\theta_{f}}{2})$, $ \theta_f=\theta_{\cL f}-\theta_{\cR
f}$, and
\begin{eqnarray*}
\widehat {\bf G}^{r,a/<}_{c,c}(\varepsilon)&=&\int
d(t-t')e^{i\varepsilon(t-t')/\hbar}{\bf P}(\mu_{\cR \cC} t){\bf
R}^f(\frac{\theta_{\cR f}}{2}) \nonumber\\
&&\hspace{1cm} {\bf G}^{r,a/<}_{c,c}(t,t'){\bf
R}^{f\dagger}(\frac{\theta_{\cR f}}{2}){\bf P}^\dagger(\mu_{\cR
\cC} t'), \nonumber \\
{\bf f}_{\gamma}(\varepsilon\mp c)
 &=& \nonumber\\
 && \hspace{-1cm}
 \left (\matrix{ f(\varepsilon-c) & 0 & 0
  & 0 \cr
 0 &f(\varepsilon+c) & 0 & 0 \cr
 0 & 0&f(\varepsilon-c)  & 0 \cr
 0 & 0 & 0 & f(\varepsilon+c)  \cr}\right).\nonumber
\end{eqnarray*}
The expression of tunneling current (\ref {eq:current-fnf})
resembles formally the current formula derived by Meir and
Wingreen \cite{Meir} for a confined region coupled to two normal
electrodes. The difference lies in that the coupling matrices and
Green functions in (\ref {eq:current-fnf}) are spanned in the
Nambu-spinor space, which reflects  the  dependence of the current
on the spin polarization of the ferromagnetic leads and the
relative orientation of the magnetic moments. When we set to zero
the magnetic moments of the two leads, Eq. (\ref {eq:current-fnf})
reduces to Equation $(5)$ in the paper of Meir and
Wingreen,\cite{Meir} since in this case the ferromagnetic leads
become normal metals. As seen from Eq. (\ref {eq:current-fnf}),
current is only dependent on the relative orientation of the
magnetizations of two leads, although there is an apparent
$\theta_{\cR f}$ dependence in the expression for $\widehat {\bf
G}^{r,a/<}$. Nevertheless, this dependence of the Green functions
$\widehat {\bf G}^{r,a/<}$ on the orientation of the ferromagnet
magnetization comes from the self-energy matrices
$\widehat{\Sigma}^{\cL f}$ and $\widehat{\Sigma}^{\cR f}$ after
the {\it rotation} operation ${\bf R}^f(\frac{\theta_{\cR
f}}{2})$, hence they only depend on the relative orientation as
can be seen more clearly in the non-interacting model.

If one lead ($\cL$) is ferromagnetic and the other ($\cR$) is
superconducting it is expected that Andreev reflection process,
dependent on the spin polarization of the ferromagnet, will
dominate the current at low bias voltages. Applying a {\it phase}
and a {\it rotation} transformations to Eq. (\ref{eq:current}),
simple integration gives
\begin{eqnarray}
\label{eq:current-fns}
 \cI_{fns}&=&\frac{ie}{2\hbar}\sum\limits^{i=1,3}\int\frac{d\varepsilon}{2\pi}
{\rm Tr}\Big\{\Big(\big[{\bf \Gamma}^{\cL f}(\varepsilon\mp
eV)-{\bf \Gamma}^{\cR s}_\rho(\varepsilon)\big]\nonumber \\
&& \hspace{0.2cm}\breve{{\bf G}}_{c,c}^{<}(\varepsilon)+ \big[{\bf
\Gamma}^{\cL f}(\varepsilon\mp eV){\bf f}_\cL(\varepsilon\mp
eV)-\nonumber\\
&& \hspace{0.5cm}{\bf \Gamma}^{\cR s}_\rho(\varepsilon){\bf
f}_\cR(\varepsilon)\big] \big[\breve{{\bf
G}}_{c,c}^{r}(\varepsilon)-\breve{{\bf
G}}_{c,c}^{a}(\varepsilon)\big]\Big)_{ii} \Big\},
\end{eqnarray}
in which the full Green functions are
\begin{eqnarray*}
\breve {\bf G}^{r,a/<}_{c,c}(\varepsilon)&=&\int
d(t-t')e^{i\varepsilon(t-t')/\hbar}{\bf P}(\mu_{\cR \cC}
t+\frac{\varphi_\cR}{2}) \nonumber\\
&&\hspace{-1.1cm} {\bf R}^f(\frac{\theta_{\cL f}}{2}){\bf
G}^{r,a/<}_{c,c}(t,t'){\bf R}^{f\dagger}(\frac{\theta_{\cL
f}}{2}){\bf P}^\dagger(\mu_{\cR \cC} t'+\frac{\varphi_\cR}{2}).
\end{eqnarray*}
In Eq. (\ref {eq:current-fns}) the current does not depend on the
orientation of the magnetization of the ferromagnetic lead and the
phase of  the order parameter of the superconductor lead. This can
be clearly demonstrated by expanding the full Green functions of
the central part perturbatively, as we will show below in the
non-interacting case. In addition, one can divide the current into
several parts implying the contributions from different physical
processes such as normal particle tunneling and Andreev
reflection, after expanding the right hand side of Eq. (\ref
{eq:current-fns}). We will show it later in the non-interaction
case.

When two leads are superconducting, the situation becomes much
more complicated. As did in the previous examples, we derive the
following current formula for $S/I/N/I/S$ or $S-QD-S$ systems
\begin{eqnarray}
\label{eq:current-sns}
 \cI_{sns}(t) &=&-\frac{e}{\hbar}\sum\limits^{i=1,3}\int\frac{d\varepsilon}{2\pi}
{\rm Im}{\rm Tr}\Big\{\Big(\frac12\big[\tilde{{\bf \Gamma}}^{\cL
s}_\varrho(\varepsilon\mp
eV;t)-\nonumber \\
&& \hspace{1cm}{\bf \Gamma}^{\cR
s}_\varrho(\varepsilon)\big]\widetilde{{\bf
G}}_{c,c}^{<}(\varepsilon;t)- \big[\tilde{{\bf \Gamma}}^{\cL
s}_\rho(\varepsilon\mp eV;t)\nonumber\\
&&  \hspace{-0.3cm}{\bf f}_\cL(\varepsilon\mp eV)-{\bf
\Gamma}^{\cR s}_\rho(\varepsilon){\bf
f}_\cR(\varepsilon)\big]\widetilde{{\bf
G}}_{c,c}^{a}(\varepsilon;t) \Big)_{ii}\Big\},
\end{eqnarray}
where
\begin{eqnarray*}
\tilde{{\bf \Gamma}}^{\cL s}_{\varrho/\rho}(\varepsilon\mp
eV;t)&=&{\bf P}^\dagger(eVt+\frac{\varphi_s}{2}){\bf \Gamma}^{\cL
s}_{\varrho/\rho}(\varepsilon\mp eV)\nonumber\\
 && \hspace{3cm} {\bf P}(eVt+\frac{\varphi_s}{2}),\\
\widetilde{{\bf G}}_{c,c}^{r,a/<}(\varepsilon;t)&=&\int
d(t-t')e^{i\varepsilon(t-t')/\hbar}{\bf P}(\mu_{\cR \cC}
t+\frac{\varphi_\cR}{2})\nonumber\\
&& \hspace{1cm}{\bf G}^{r,a/<}_{c,c}(t,t'){\bf P}^\dagger(\mu_{\cR
\cC} t'+\frac{\varphi_\cR}{2}),
\end{eqnarray*}
with $\varphi_s=\varphi_\cL-\varphi_\cR$. One may  wonder why we
use the notation $\widetilde{{\bf
G}}_{c,c}^{r,a/<}(\varepsilon;t)$ with the additional variable $t$
other than $\widehat{{\bf G}}_{c,c}^{r,a/<}(\varepsilon)$. The
reason is that
 the full Green functions ${\bf G}^{r,a/<}_{c,c}$ should be calculated
 in the presence of tunneling between the central part and
 the two sides, as well as the interactions in the
 central region. In the present case, the $t$-dependence can not
 be avoided in the self-energy matrices, while it can be
 removed by a unitary {\it phase} operation when only one
 superconductor is involved. The current through a confined interacting
 region connected to  two superconductor leads is generally
 time dependent, as in the case of biased weak Josephson
 links.\cite{Tinkham}  However, in the limiting case of zero bias,
 the current is a time-independent nonzero quantity, as can be
 seen from Eq. (\ref {eq:current-sns}). In other theoretical treatments,
 \cite{Arnold,Cuevas} a double-time Fourier
 transformation is usually taken as $X(t,t')=\frac{1}{2\pi} \sum_n\int d\omega
 e^{-i \omega t}e^{i(\omega+n\omega_0/2)t'} X(\omega,
 \omega+n\omega_0/2)$, where $\omega_0=2eV/\hbar$, and the
 current yields a harmonic expansion of the
 fundamental frequency $\cI(t)=\sum_n I_n e^{in\omega_0 t}$.
 In fact, the Green functions $\widetilde{\bf G}^{r,a/<}_{d,d}$ in (\ref{eq:current-sns}) can be expanded
 in powers
 of the fundamental frequency $\omega_0$, i.e., $\widetilde {\bf
 G}^{r,a/<}_{d,d}(\varepsilon,t)=\sum_m \widetilde {\bf
 G}^{r,a/<}_{d,d}(\varepsilon,\varepsilon+m\omega_0/2)e^{im\omega_0
 t/2}$, which with the expression for the Green function $\widetilde {\bf G}^{r,a/<}_{d,d}$
 below Eq.(\ref{eq:current-sns}) is
 exactly of the form of  the double-energy
 transformation.\cite{Arnold}
 However, we show here that one  can  obtain {\it in principle} the time
 dependence of the current, as long as one can derive the full Green functions of the central
 part, which need further investigation.

 So far we have presented a general formulation to calculate the current
 through a confined normal region connected to two leads being
 either ferromagnetic or  superconducting, and the
 Meir-Wingreen-type
 formulae in the specific cases. Although the formalism can not
 applied to the strong coupling of the central normal part to the
 outer world as the circuit theory of the hybrid mesoscopic
 transport,\cite{Nazarov} it permits us to investigate the effects
 of the single particle interference and strong electron-electron
 interaction on the transport properties of  hybrid mesoscopic
 systems, which is ignored in the circuit theory.
 Compared to the other formalisms
 based on the similar Keldysh NEGF technique,\cite{Meir, Haug, Sun, Wang, Zhu, Lin} our formalism is
 more systematic and more general. In the present formalism,  one do not need to make
 additional ad hoc assumptions as mentioned in the introduction.
 One can also judge what quantities can be measured in experiment
 in a much more complicated structure, by observing  simply the
 energy-independent arguments in the exponential functions or the
 triangle functions in the unitary matrices of the tunneling parts of
 the Hamiltonian after the Bogoliubov transformations. As we will
 demonstrate, we obtain the final current formula after simple
 matrix algebra rather than difficult mathematical techniques.\cite{Sun, Wang, Zhu, Lin}
 This kind of mathematical simplicity makes our formalism more appearing than
 others. More importantly, an explicit energy- and bias-voltage
 dependence of the level-width functions and distribution
 functions allows us to investigate the I-V characteristics in a
 much more wider range of bias voltage.

\section{Applications to the noninteracting model in the central region}

In this section we use the formulae developed above to study
transport properties of various kinds of hybrid mesoscopic systems
in which, for simplicity and convenience of comparison with other
theories, the interaction effects in the central confined region
are not considered. The absence of the interactions permits an
analysis of the genuine physical influence of ferromagnetism and
superconductor proximity on the transport properties in hybrid
structures. One of the best candidates for a non-interaction
confined region is double-barrier resonant structures (DBTSs) with
quantized discrete energy levels.\cite{Datta} Therefore we adopt
the double-barrier model with the emitter and collector replaced
by either a ferromagnet or a superconductor. Throughout the
following calculations we use the following approximations: (i)
the level shift is omitted, (ii) the coupling coefficients are
real constants, independent of spin and energy such that
$V^{\gamma f/s}_{kn;\sigma}=V^{\gamma f/s}=V^{\gamma f/s *}$.
However, we will abandon the usual\cite{Haug, Meir,Fazio,Sun,Wang}
wide-band approximation of the level-width functions which is
reasonable in the low-voltage transport, since in this paper we
also deal with high bias voltage situation. As we will show, this
permits us to investigate the current within a much wider region
of bias voltage and find interesting transport features of the
same resonant structures which can  not be found in the other
formalisms based on the Keldysh NEGF technique.\cite{Sun, Wang,
Zhu}

In the absence of interactions within the intermediate normal
metal, the full retarded/advanced Green function can be solved
from Dyson  equation
\begin{eqnarray}
\label{eq:dyson-gra}
 {\bf G}_{c,c}^{r/a}(t,t')&=&{\bf
G}_{c,c}^{0r/a}(t,t')+\int
dt_1\int dt_2{\bf G}_{c,c}^{0r/a}(t,t_1)\nonumber \\
&&\hspace{1cm}{\bf \Sigma}^{r/a}(t_1,t_2){\bf
G}_{c,c}^{r/a}(t_2,t')\nonumber\\
&=&{\bf G}_{c,c}^{0r/a}(t,t')+\int
dt_1\int dt_2{\bf G}_{c,c}^{r/a}(t,t_1)\nonumber \\
&&\hspace{1cm}{\bf \Sigma}^{r/a}(t_1,t_2){\bf
G}_{c,c}^{0r/a}(t_2,t'),
\end{eqnarray}
in which ${\bf G}_{c,c}^{0r/a}$ is the decoupled Green function,
which becomes when the central region is isolated from the outside
world
\begin{widetext}
\begin{eqnarray}
\label{eq:green-dd}
 {\bf g}_{c,c}^{r/a}(t,t')=\mp i\vartheta(\pm
t_1\mp t_2) \sum\limits_{n}\left(
\matrix{e^{-i(\varepsilon_{n\uparrow}-\mu_\cC)(t_1-t_2)/\hbar}
&0&0&0\cr 0&
e^{i(\varepsilon_{n\downarrow}-\mu_\cC)(t_1-t_2)/\hbar}&0&0\cr
0&0&e^{-i(\varepsilon_{n\downarrow}-\mu_\cC)(t_1-t_2)/\hbar}&0\cr
0&0&0&
e^{i(\varepsilon_{n\uparrow}-\mu_\cC)(t_1-t_2}/\hbar)\cr}\right).\nonumber\\
\end{eqnarray}
\end{widetext}

The lesser/greater Green function of the central region is
calculated via Keldysh equation
\begin{eqnarray}
\label{eq:keldysh}
 {\bf G}_{c,c}^{</>}(t,t')&=&\int dt_1\int
dt_2\int dt_3\int dt_4\nonumber\\
&&\left[{\bf 1}+{\bf G}_{c,c}^{r}(t,t_1){\bf
\Sigma}^{r}(t_1,t_2)\right]\nonumber\\
&&{\bf G}_{c,c}^{0</>}(t_2,t_3)\nonumber\\
&&\left[{\bf 1}+{\bf \Sigma}^{a}(t_3,t_4){\bf
G}_{c,c}^{a}(t_4,t')\right]+\nonumber\\
&& \int dt_1\int dt_2{\bf
G}_{c,c}^{r}(t,t_1)\nonumber\\
&&\hspace{1cm}{\bf \Sigma}^{</>}(t_1,t_2){\bf
G}_{c,c}^{a}(t_2,t').
\end{eqnarray}

Once the full Green functions in the central region are known, we
then have the complete knowledge to investigate tunneling
processes in the specific structures. In the following we study
electron tunneling in three typical hybrid structures: (a)
$F/I/N/I/F$ magnetic DBTSs, (b) $F/I/N/I/S$ DBTSs, and (c)
$S/I/N/I/S$ DBTSs.

\subsection{F/I/N/I/F structures}

\begin{figure}
\epsfig{file=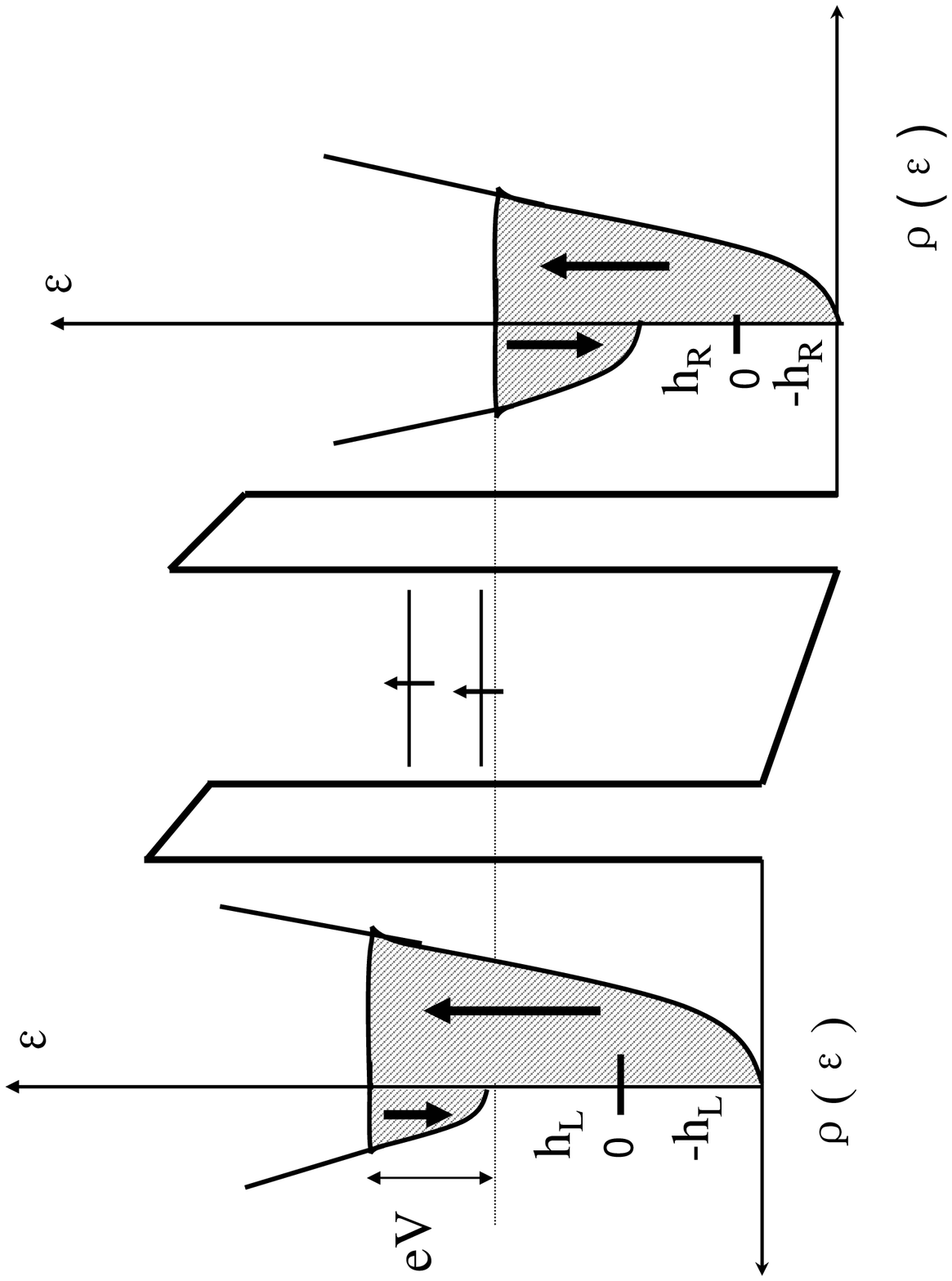, width=8cm} \caption{A schematic potential
profile for a biased magnetic double-barrier structure with two
ferromagnetic electrodes characterized by their respective
magnetizations $h_{\cL}$ and $h_{\cR}$. The hatched regions denote
the states occupied by electrons.}
\end{figure}

When two ferromagnets are separated by a thin nonmagnetic barrier,
two kinds of physical effects arise. The first is the {\it spin
valve} effect,\cite{Julliere, Slonczewski}  showing a
$(1+\varepsilon\cos\theta)$ dependence of the tunnel conductance
on the relative orientation $\theta$ between the involved two
magnetizations. The other is tunnel magnetoresistance
(TMR),\cite{Julliere} defined by $\Delta R/R=(R_a-R_p)/R_a$, where
$R_p$ and $R_a$ are the resistances when two magnetizations are
parallel and antiparallel, respectively. The {\it spin valve} and
TMR are due to the spin polarization induced by an exchanging
coupling between electron spins and the internal
magnetization,\cite{Julliere, Slonczewski}  and the relative
orientation of the magnetizations can be adjusted by applying a
magnetic field. A TMR up to  $11.8\% $ at room temperatures, and
$24\% $ at $4.2 K$, was reported in $CoFe/Al_2O_3/Co$ planar
magnetic junctions.\cite{Moodera}  It is observed that the TMR
decreases with increasing bias voltage.\cite{Moodera}

Recently, double-barrier magnetic resonant structures has
attracted much experimental \cite{Brehmer} and
theoretical\cite{Wang, Zhang,Tanamoto, Sheng, Barnas,Petukhov}
attention. The theoretical results show that the TMR of resonant
magnetic structures is enhanced compared to the single magnetic
junctions due to resonant tunneling.\cite{Zhang,Tanamoto, Sheng,
Barnas,Petukhov}  However, there appears a controversial issue
related to the TMR enhancement, on whether it is for
peak\cite{Zhang} or valley current.\cite{Tanamoto} In addition,
Sheng et al.\cite{Sheng} found both positive and negative TMR in
$F/I/F/I/F$ double junctions. To clarify these issues, we
re-examine this problem using our formula (\ref{eq:current-fnf})
in terms
 of the two-band free-electron spin-polarization model\cite{Slonczewski} for the
ferromagnetic leads. A typical double-barrier magnetic structure
is schematically plotted in Fig. 2.

We take two steps to calculate the full retarded/advanced Green
function ${\bf G}^{r/a}_{c,c}(t,t')$ of the central region. First
we decouple the system from the left ferromagnet, denoting the
corresponding retarded/advanced Green function by ${\bf
\cG}^{r/a}_{c,c}(t,t')$, then we couple the central region to the
left ferromagnet and calculate the full Green function ${\bf G}$
from ${\bf \cG}$. From the Dyson equation (\ref{eq:dyson-gra}) we
have
\begin{eqnarray}
\label{eq:dyson-full-fnf}
 \widehat{{\bf
G}}_{c,c}^{r/a}(t,t')&=&\widehat{{\bf \cG}}_{c,c}^{r/a}(t,t')+\int
dt_1\int dt_2\widehat{{\bf \cG}}_{c,c}^{r/a}(t,t_1)\nonumber \\
&&\hspace{1cm}\widehat{{\bf \Sigma}}^{r/a}_{\cL
f}(t_1,t_2)\widehat{{\bf
G}}_{c,c}^{r/a}(t_2,t'), \\
\label{eq:dyson-part-fnf} \widehat{{\bf
\cG}}_{c,c}^{r/a}(t,t')&=&\widehat{{\bf g}}_{c,c}^{r/a}(t,t')+\int
dt_1\int dt_2\widehat{{\bf g}}_{c,c}^{r/a}(t,t_1)\nonumber \\
&&\hspace{1cm}\widehat{{\bf \Sigma}}^{r/a}_{\cR
f}(t_1,t_2)\widehat{{\bf \cG}}_{c,c}^{r/a}(t_2,t'),
\end{eqnarray}
where (${\bf X}={\bf G}^{r,a/<},{\bf \cG}^{r,a/<},{\bf g}^{r,a/<}$
and ${\bf \Sigma}^{r,a/<}$)
\begin{eqnarray*}\widehat{\bf X}(t,t')={\bf P}(\mu_{\cR \cC}
t){\bf R}^f(\frac{\theta_{\cR f}}{2}) {\bf X}(t,t'){\bf
R}^{f\dagger}(\frac{\theta_{\cR f}}{2}){\bf P}^\dagger(\mu_{\cR
\cC} t').\end{eqnarray*} Substituting the self-energy matrices
${\bf \Sigma}^{r/a}_{\cR f}$ (Appendix A) into the Dyson equation
(\ref{eq:dyson-part-fnf}), one has
\begin{eqnarray}
\widehat{{\bf
\cG}}_{c,c}^{r/a}(t,t')&=&\int\frac{d\varepsilon}{2\pi}
e^{-i\varepsilon(t-t')/\hbar}\Big[\widehat{{\bf
g}}^{r/a-1}_{c,c}(\varepsilon)\pm\frac{i}{2} {\bf
\Gamma}^{\cR f}(\varepsilon)\Big]^{-1}\nonumber \\
&=&\int\frac{d\varepsilon}{2\pi}
e^{-i\varepsilon(t-t')/\hbar}\widehat{{\bf
\cG}}_{c,c}^{r/a}(\varepsilon),
\end{eqnarray}
where the retarded/advanced Green function for the isolated
central region is
\begin{widetext}
\begin{eqnarray}
\widehat{{\bf g}}_{c,c}^{r/a}(\varepsilon)&=&
\left(\matrix{(\sum\limits_{n}
\frac{1}{\varepsilon-\epsilon'_{n\uparrow}\pm i0^+})^{-1}
&0&0&0\cr 0&
(\sum\limits_{n}\frac{1}{\varepsilon+\epsilon'_{n\downarrow}\pm
i0^+})^{-1}&0&0\cr
0&0&(\sum\limits_{n}\frac{1}{\varepsilon-\epsilon'_{n\downarrow}\pm
i0^+})^{-1}&0\cr 0&0&0&
(\sum\limits_{n}\frac{1}{\varepsilon+\epsilon'_{n\uparrow}\pm
i0^+})^{-1}\cr}\right),
\end{eqnarray}
\end{widetext}
with
$\epsilon'_{n\sigma}=\varepsilon_{n\sigma}-\mu_{\cC}+\mu_{\cR}$.
The full retarded/advanced Green function is obtained in the same
way
\begin{eqnarray}
\label{eq:ragreenfnfno}
 \widehat{{\bf
G}}_{c,c}^{r/a}(t,t')&=&\int\frac{d\varepsilon}{2\pi}
e^{-i\varepsilon(t-t')/\hbar}
 \Big[\widehat{ {\bf
\cG}}^{r/a-1}_{c,c}(\varepsilon)\pm \nonumber
\\&&
\hspace{3cm}\frac{i}{2} \hat{{\bf \Gamma}}^{\cL f}(\varepsilon\mp eV) \Big]^{-1}\nonumber \\
&=&\int\frac{d\varepsilon}{2\pi} e^{-i\varepsilon(t-t')/\hbar}
 \Big[ \widehat{{\bf
g}}^{r/a-1}_{c,c}(\varepsilon)\pm\frac{i}{2}\nonumber
\\&&\hspace{1.5cm}{\bf
\Gamma}^{\cR f}(\varepsilon) \pm\frac{i}{2}\hat{ {\bf
\Gamma}}^{\cL f}(\varepsilon\mp eV)) \Big]^{-1}\nonumber \\
&=&\int\frac{d\varepsilon}{2\pi}
e^{-i\varepsilon(t-t')/\hbar}\widehat{{\bf
G}}_{c,c}^{r/a}(\varepsilon).
\end{eqnarray}

The lesser Green function is associated with the retarded and
advanced Green functions via the Keldysh equation
(\ref{eq:keldysh}) where $\hat {\bf g}^<(t,t')=0$
\begin{eqnarray}
\label{eq:lessgreenfnfno}
 \widehat{\bf G}_{c,c}^{<}(t,t')&=&\int
dt_1\int dt_2\widehat{\bf
G}_{c,c}^{r}(t,t_1)\Big[\widehat{\bf \Sigma}^{</>}_{\cL f}(t_1,t_2)\nonumber\\
&&\hspace{2cm}+\widehat{\bf \Sigma}^{</>}_{\cR f}
(t_1,t_2)\Big]\widehat{\bf
G}_{c,c}^{a}(t_2,t')\nonumber \\
&=&\int\frac{d\varepsilon}{2\pi}
e^{-i\varepsilon(t-t')/\hbar}\widehat{{\bf
G}}_{c,c}^{r}(\varepsilon)\big[\hat{{\bf \Gamma}}^{\cL
f}(\varepsilon\mp eV)\nonumber\\
&&\hspace{1cm}{\bf f}_\cL(\varepsilon \mp eV)-{\bf \Gamma}^{\cR
f}(\varepsilon){\bf f}_\cR(\varepsilon)\big]\widehat{{\bf G}}_{c,c}^{a}(\varepsilon)\nonumber\\
&=&\int\frac{d\varepsilon}{2\pi}
e^{-i\varepsilon(t-t')/\hbar}\widehat{{\bf
G}}_{c,c}^{<}(\varepsilon).
\end{eqnarray}

Substituting the advanced  and lesser Green functions
(\ref{eq:ragreenfnfno}) and  (\ref{eq:lessgreenfnfno}) into Eq.
(\ref{eq:current-fnf}) or (\ref{eq:current}), one gets the
following Landauer-B\"uttiker-type\cite{Buttiker} formula for the
current through a non-interacting $F/I/N/I/F$ magnetic structure,
\begin{equation}
\label{eq:current-fnf-non}
 \cI_{fnf}(\theta_f)=\frac{2e}{h}\int d\varepsilon
[f_\cL(\varepsilon-eV)-f_\cR(\varepsilon)]\cT_{fnf}(\varepsilon,\theta_f),\nonumber\\
\end{equation}
where the transmission coefficient $\cT_{fnf}$ is cast into the
following compact form
\begin{eqnarray}
\label{eq:transmission-fnf}
&&\cT_{fnf}(\varepsilon,\theta_f) \nonumber\\
&=&\frac12 \sum\limits_{i=1,3} \left(\hat{{\bf \Gamma}}^{\cL
f}(\varepsilon\mp eV)\widehat{{\bf G}}_{c,c}^{r}(\varepsilon) {\bf
\Gamma}^{\cR f}(\varepsilon)\widehat{{\bf
G}}_{c,c}^{a}(\varepsilon)\right)_{ii} \nonumber\\
&=&\frac12\Big[(\cos^2\frac{\theta_f}{2}\hat{\Gamma}^{\cL
f}_{\uparrow}\Gamma^{\cR
f}_{\uparrow}+\sin^2\frac{\theta_f}{2}\hat{\Gamma}^{\cL
f}_{\downarrow}\Gamma^{\cR
f}_{\uparrow})|\widehat{G}^{r}_{c,c;11}|^2 \nonumber\\ &&
-(2\Gamma^{\cR f}_{\uparrow}\Gamma^{\cR
f}_{\downarrow}+\cos^2\frac{\theta_f}{2}\hat{\Gamma}^{\cL
f}_{\downarrow}\Gamma^{\cR
f}_{\uparrow}+\sin^2\frac{\theta_f}{2}\hat{\Gamma}^{\cL
f}_{\uparrow}\Gamma^{\cR
f}_{\uparrow})\nonumber\\
&& |\widehat{G}^{r}_{c,c;13}|^2 -(2\Gamma^{\cR
f}_{\uparrow}\Gamma^{\cR
f}_{\downarrow}+\cos^2\frac{\theta_f}{2}\hat{\Gamma}^{\cL
f}_{\uparrow}\Gamma^{\cR f}_{\downarrow}+\nonumber \\
 &&\sin^2\frac{\theta_f}{2}
\hat{\Gamma}^{\cL f}_{\downarrow}\Gamma^{\cR
f}_{\downarrow})|\widehat{G}^{r}_{c,c;31}|^2+
(\cos^2\frac{\theta_f}{2}\hat{\Gamma}^{\cL
f}_{\downarrow}\Gamma^{\cR
f}_{\downarrow}+\nonumber \\
 && \sin^2\frac{\theta_f}{2}\hat{\Gamma}^{\cL
f}_{\uparrow}\Gamma^{\cR
f}_{\downarrow})|\widehat{G}^{r}_{c,c;33}|^2 \Big].
\end{eqnarray}
Notice that we have dropped the arguments $\varepsilon-eV$  in
$\hat{\Gamma}^{\cL f}_{\sigma}$ and $\varepsilon$  in $\Gamma^{\cR
f}_{\sigma}$ for brevity. The full retarded Green function in Eq.
(\ref{eq:transmission-fnf}) is determined by the matrix inversion
$\widehat{\bf G}^r_{c,c}=[\widehat{\bf g}^{r-1}_{c,c}+\frac i 2
\hat{\bf \Gamma}^{\cL f}+\frac i 2 {\bf \Gamma}^{\cR f}]^{-1}$,
and we have
\begin{eqnarray*}
\widehat{G}^r_{c,c;11}&=&
 \widehat{G}^{-1}_{ff}(\varepsilon)\Big[(\sum\limits_{n}
\frac{1}{\varepsilon-\epsilon'_{n\downarrow}+i0^+})^{-1}+\nonumber\\
&&\frac{i}{2} (\cos^2\frac{\theta_f}{2}\hat{\Gamma}^{\cL
f}_{\downarrow}+\sin^2\frac{\theta_f}{2}\hat{\Gamma}^{\cL
f}_{\uparrow}+\Gamma^{\cR f}_{\downarrow})\Big],  \\
 \widehat{G}^r_{c,c;13}&=&\widehat{G}^r_{c,c;31}= \frac{-\frac{i}{4}\sin\theta_f(\hat{\Gamma}^{\cL
f}_{\uparrow}-\hat{\Gamma}^{\cL f}_{\downarrow})}{
 \widehat{G}_{ff}(\varepsilon)},  \\
\widehat{G}^r_{c,c;33}&=&\widehat{G}^{-1}_{ff}(\varepsilon)\Big[
(\sum\limits_{n}
\frac{1}{\varepsilon-\epsilon'_{n\uparrow}+i0^+})^{-1}+\nonumber\\
&&  \frac{i}{2}(\cos ^2\frac{\theta_f}{2}\hat{\Gamma}^{\cL
f}_{\uparrow}+\sin^2\frac{\theta_f}{2}\hat{\Gamma}^{\cL
f}_{\downarrow}+\Gamma^{\cR f}_{\uparrow})\Big]
,\\
\widehat{G}_{ff}(\varepsilon)&=&\widehat{G}^r_{c,c;11}\widehat{G}^r_{c,c;33}-\widehat{G}^r_{c,c;13}
\widehat{G}^r_{c,c;31}.
\end{eqnarray*}
To obtain the last equality of Eq. (\ref{eq:transmission-fnf}), we
have used
\begin{eqnarray}
\big(\widehat{g}^{r/a-1}_{c,c;11}\pm \frac
{i}{2}\Gamma^{f}_{11}\big)\widehat{G}^{r/a}_{c,c;13}
&=&\pm\frac {i}{2}\Gamma^{f}_{13}\widehat{G}^{r/a}_{c,c;33}\nonumber \\
\big(\widehat{g}^{r/a-1}_{c,c;33}\pm \frac i2\Gamma^{
f}_{33}\big)\widehat{G}^{r/a}_{c,c;31}&=&\pm\frac i2\Gamma^{
f}_{31}\widehat{G} ^{r/a}_{c,c;11}, \nonumber
\end{eqnarray}
where  ${\bf \Gamma}^{f}=\hat{\bf \Gamma}^{\cL f}+{\bf
\Gamma}^{\cR f}$.

One sees from Eqs. (\ref{eq:current-fnf-non}) and
(\ref{eq:transmission-fnf}) that the current has a generic
dependence on the relative orientation $\theta_f$ between the two
magnetizations. By observing the current expression and
scrutinizing the structure of the Green functions, it is not
difficult to find that the tunneling current through the magnetic
structure is generally maximized at $\theta_f=0$ (parallel
magnetization) and minimized at $\theta_f=\pi$ (antiparallel
magnetization), a typical {\it spin valve} effect also in magnetic
resonant tunneling devices (data not shown here). The
ferromagnetism is reflected in the $\theta_f$- and $\Gamma^{\gamma
f}_{\sigma}$-dependence of the full Green functions of the central
part $\hat{G}_{c,c}$, as well the transmission function $\cT$.
When at least one lead is nonmagnetic, the $\theta_f$-dependence
can be removed with the help of a {\it rotation} operation ${\bf
R}^f$. If we set $ \Gamma^{\cL f}_\uparrow=\Gamma^{\cL
f}_\downarrow$ and $ \Gamma^{\cR f}_\uparrow=\Gamma^{\cR
f}_\downarrow$, the current formula (\ref{eq:current-fnf-non})
will  recover the usual Landauer-B\"uttiker formula and the
transmission is finally simplified to the Breit-Wigner type in the
single level case. We notice that the current formula
(\ref{eq:current-fnf-non}) is formally similar to the results of
Wang et al.\cite{Wang} and Zhu et al.,\cite{Zhu} however, the
discrepancy is nontrivial. The current formula
(\ref{eq:current-fnf-non}) allows us to calculate the I-V curves
in a much wider region of bias voltage yielding rich physics,
while according to the theoretical treatments of Wang et
al.\cite{Wang} and Zhu et al.,\cite{Zhu} the current formula is
restricted to the low bias voltage case  where the level-width
functions can be viewed as energy-independent constants, and thus
may result in even wrong consequences in the large bias voltage
limit.

\begin{figure}
\epsfig{file=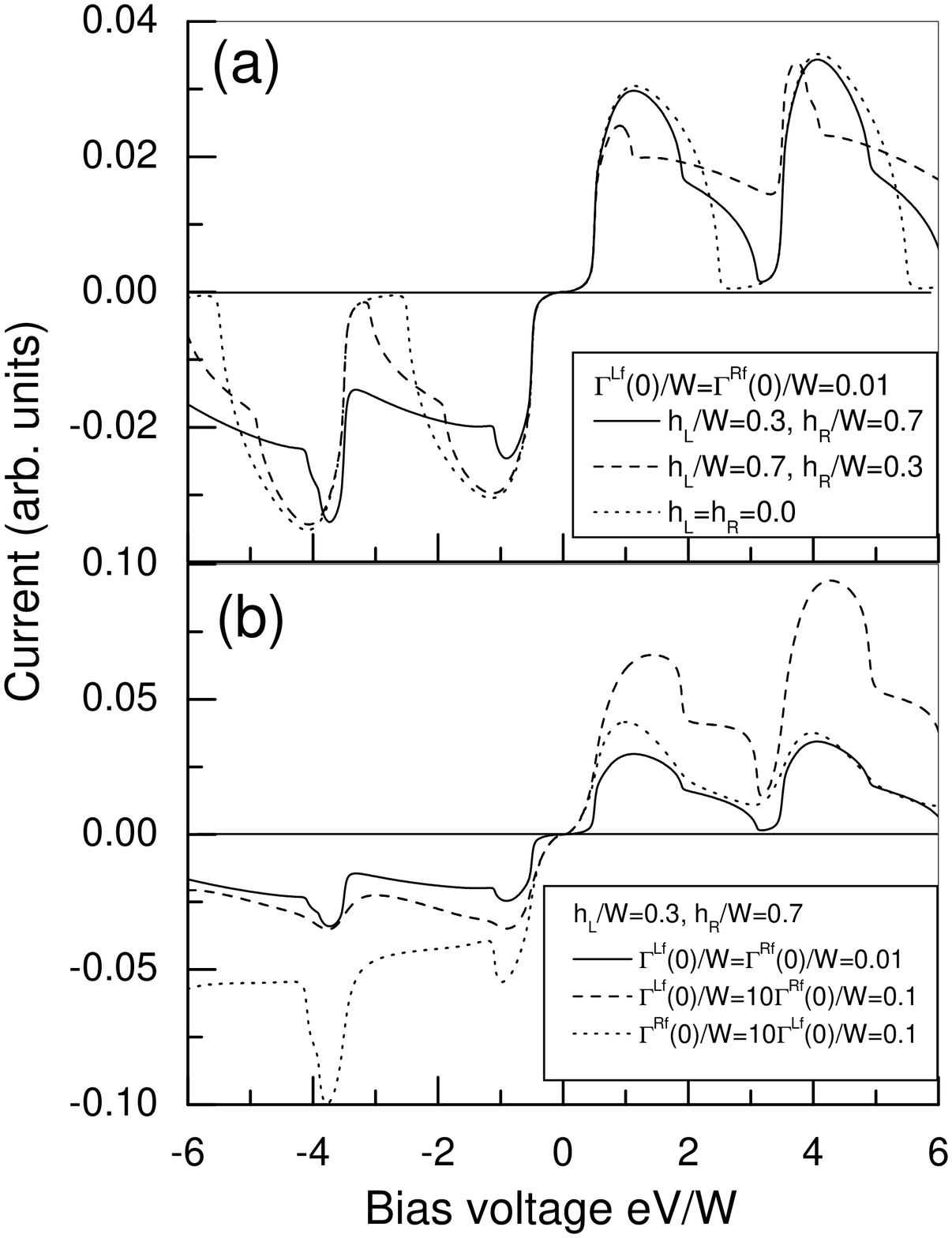, width=8cm} \caption{I-V characteristics
 of a magnetic double-barrier structure for (a) different magnetizations $h_\cL$ and $
h_\cR$ and (b) different couplings $\Gamma^{\cL f}(0)$ and
$\Gamma^{\cR f}(0)$ at temperature $k_BT=0.1W$. The level spacing
is $1.5W$, larger than the bandwidth $W$.}
\end{figure}

Of particular interest is the current-voltage characteristics of
double-barrier structures. In subsequent calculations, we
approximate the density of states of the ferromagnetic leads by
that of the two-band free-electron spin-polarization
model\cite{Slonczewski, Tanamoto} and take into consideration the
finite width of these two bands. In this model the dimensionless
DOS of the spin bands is $\rho^{\gamma
f}_\sigma(\varepsilon)\propto\sqrt{(\varepsilon+\sigma
h_{\gamma}+W)/W}$, where $W$ is the bandwidth measured from the
band bottom to the Fermi level. For some ferromagnetic metals,
this is quite an appropriate approximation.\cite{Slonczewski,
Sterns, Bratkovsky} In the absence of a magnetic field in the
central region,
$\varepsilon_{n\uparrow}=\varepsilon_{n\downarrow}=\varepsilon_n$.
To model the bias voltage drop inside the well we take
$\mu_\cC=\mu_\cR-eV/2$, since the bias potential is assumed
 distributing uniformly across the double-barrier structure. Without
any loss of generality we consider two quasi-stationary levels in
the well, of which the energy of the lowest one is $0.25W$ when
$V=0$ and the level spacing is chosen as $1.5W$, larger than the
bandwidth of the ferromagnets. This assumption is, in practice,
quite reasonable for the narrow-band ferromagnetic metals and
quantum wells with very large level spacing or small quantum dots
with very large charging energy. The I-V curves are shown in Fig.
$3$. The dotted line in Fig. $3(a)$, the well-known I-V
characteristics of usual double-barrier structures, is given for
comparison with the ferromagnetic case. In the presence of
ferromagnets, the structure of resonant shoulders neighboring to
resonant peaks is celebrated in the I-V plots. The ratio of the
peak width to the shoulder width is about
$(W-h_\gamma)/2h_\gamma$. Moreover, the valley current in a normal
resonant tunneling structure is lifted when the leads become
ferromagnetic. These surprising results, unexpected within the
wide-band approximation, can be understood from the potential
profile of this kind of magnetic double junction structure, shown
in Fig. $2$. It is well known that the current through a usual
double-barrier structure is resonantly enhanced when one of the
well levels falls into the region of a Fermi sea, i.e.,
$eV<\varepsilon_n+eV/2<W+eV$.\cite{Liu}  In the ferromagnetic
situation, the Fermi sea is distorted and comprises two distinct
parts: one with both spin-up and spin-down bands, and the other
with only a spin-up or spin-down band, as displayed in Fig. $2$.
We thus have two types of Fermi sea for ferromagnets, one is
represented by $h_\gamma+eV<\varepsilon_n+eV/2<W+eV$ and the other
is $-h_\gamma+eV<\varepsilon_n+eV/2<h_\gamma+eV$. It is obvious
that the resonant current through one of the well levels being
within the sea of the first type is larger than that of the second
type, which is clearly reflected in the I-V characteristics in
Fig. $3(a)$.

Next we investigate the influence of coupling asymmetry on the
tunneling current. The results are presented in Fig. $3(b)$. The
magnitude of resonant current is significantly enhanced when one
of the couplings becomes $10$ times as large. The coupling
asymmetry induces a more significant enhancement of the tunneling
current if it is the coupling to the higher-voltage lead (emitter)
that is stronger, consistent with the tight-binding numerical
result in the usual DBTS.\cite{Kim}  The ferromagnetic I-V
characteristics (peaks plus shoulders) in the reverse-bias case is
blurred, also due to the same coupling asymmetry effect. These
features can be understood in a similar way. The resonant current
is roughly proportional to the ratio $\Gamma^{\cL f}(0) \rho^{\cL
f}\Gamma^{\cR f}(0)\rho^{\cR f} /(\Gamma^{\cL f}(0)\rho^{\cL
f}+\Gamma^{\cR f}(0)\rho^{\cR f})^2$, which becomes larger when
one of the couplings $\Gamma^{\cL f}(0)$ or $\Gamma^{\cR f}(0)$ is
enhanced. However, the magnitude of this enhancement also depends
on the DOS of the ferromagnet lead $\rho^{\gamma f}$. If one
strengthens the coupling to the lower-voltage lead (collector),
the tunneling current is slightly enhanced since the DOS  of the
collector is comparatively large.

Following Sheng et al.,\cite{Sheng} we define the tunnel
magnetoresistance (TMR)  as $\Delta
R/R=[\cI_{fnf}(\pi)-\cI_{fnf}(0)]/{\rm
max}(\cI_{fnf}(\pi),\cI_{fnf}(0))$. In Fig. $4$ we give the TMR as
a function of the bias voltage for some typical couplings, and the
$I-V$ curves in the cases of parallel and antiparallel alignments
of magnetizations for the convenience of comparison and analysis.
In contrary to the monotonous decay with the bias
voltage,\cite{Moodera}  the TMR in magnetic DBTSs displays complex
dependence on the bias voltage no matter what the values of
couplings are, which arises from the resonant tunneling of
electrons.
 This feature reveals that there is  richer
physics in the TMR  of magnetic resonant structures.  We notice
also that the bias voltage dependence of the TMR can be
comparatively simple if the collector ferromagnet is of low degree
of spin polarization $h_\gamma/W$, as shown in the negative bias
domains of Fig. $4$. This phenomenon can be ascribed to the weak
perturbation of the spin-up and spin-down DOS of the ferromagnet
with small $h_\gamma/W$ by an external magnetic field. In
addition, the peculiar behavior of the TMR also depends  on the
strengths and symmetry of the elastic couplings. In the strong
coupling case (Fig. $4(d)$), the TMR shows a resonant behavior
similar to that in the tunneling current,\cite{Sheng, Zhang}  and
can even be negative at some bias voltages.\cite{Sheng} In other
cases (Fig. $4(a)$-$(c)$), the TMR drops within the resonant peak
region and then develops peaks at the boundaries between the
current peaks and shoulders, similar to the result of a
non-interacting quantum dot coupled to two magnetic
leads.\cite{Tanamoto}   The different oscillatory behaviors of the
TMR with the bias voltage for different couplings  imply that the
analysis on the TMR in the resonant structure\cite{Tanamoto} may
not stand. It is worth noticing that the TMR will eventually decay
to zero in the large bias voltage limit, due to the trivial
dependence on the interchange of the spin-up and spin-down DOS of
the lead at lower voltage.  It is interesting to notice that the
TMR reaches a maximum of $18\% $ for asymmetric couplings. The
maximum would increase further as the couplings become more
asymmetric. The TMR ratio given by our simple model is consistent
with the estimation for a resonant structure with $Fe$
electrodes\cite{Bratkovsky} and that of the Coulomb-blockade-free
double junction model.\cite{Brataas} In magnetic resonant
structures,  TMR depends not only on the DOS of two electrodes as
in the single junction case, but also on the spectral density of
the central well associated with $\widehat{G}^r_{c,c}$, and thus
exhibits complicated dependence on the bias voltage. As for the
coupling dependence of peaks and valleys in the TMR curve, it is
associated with the sensitivity  to the distortion in the spin-up
and spin-down DOS of the leads. Such a sensitivity strongly
depends on the coupling strengths and which type of the Fermi sea
the well levels fall into. In general, electrons in a one-band
Fermi sea in the weak coupling case can detect much better the
change in the DOS of the other ferromagnetic lead, and so the TMR
develops a peak at the boundary between the two distinct types of
Fermi sea.

\begin{figure}
\epsfig{file=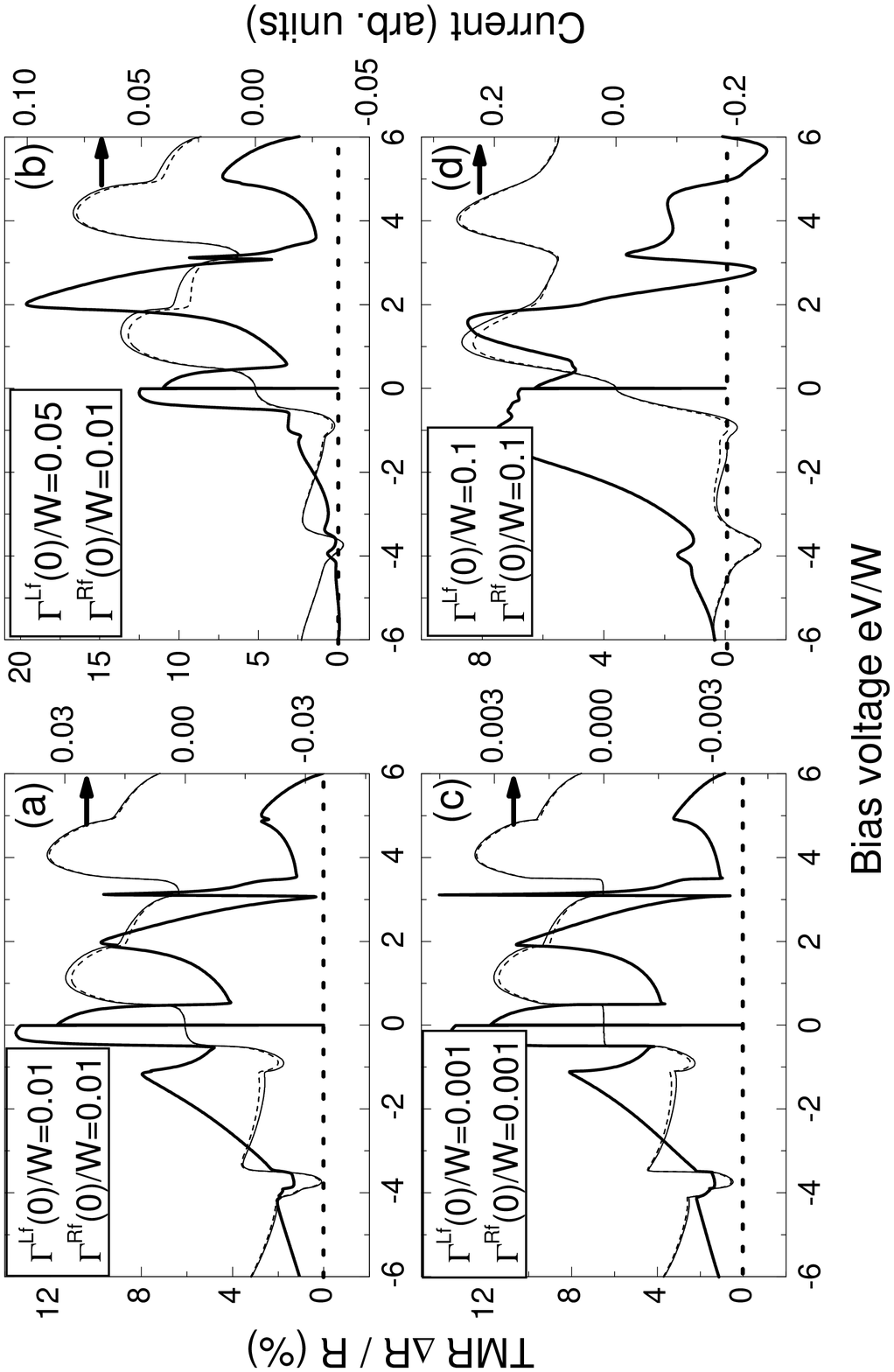, width=9cm}  \caption{TMR versus bias voltage
of a magnetic double-barrier structure for different couplings at
temperature $k_BT=0.1W$. The thick lines are the results for TMR,
and the thin solid and dashed lines represent the tunneling
current with parallel and antiparallel magnetizations,
respectively. The other parameters are the same as in Fig. $3$.}
\end{figure}

To summarize this subsection, we have studied the I-V
characteristics and TMR behavior in a double-barrier magnetic
structure. It is found that both a peak and a shoulder emerge
within the resonant region, manifesting directly the
  DOS profile of the ferromagnets. This finding may provide a new
  way to measure the degree of spin polarization of a
  ferromagnet. The TMR of resonant structures
  exhibits complex dependence on the bias voltage. It is
  either enhanced or suppressed, depending on the
  strengths and symmetry of the elastic couplings of the central region to
  the magnetic leads.

\subsection{F/I/N/I/S structures}

At an $N/S$ interface a dissipative current in the normal metal
can be converted into a dissipationless  supercurrent in the
superconductor via the Andreev reflection
process.\cite{Andreev,NS} Owing to the spin imbalance in the
ferromagnet, the Andreev current is suppressed in a $F/S$
contact.\cite{Jong}  Blonder, Tinkham and Klapwijk presented a
one-dimensional model based on the Bogoliubov-de-Gennes equation
to analyze the transport processes at an $N/S$ interface in terms
of normal electron transmission and Andreev reflection
probability, known as the BTK theory.\cite{Blonder}  Cuevas et al.
in 1996 also uncovered some kinds of electron tunneling processes
in the $N/S$ quantum point contacts within the Keldysh NEGF
formalism starting from a microscopic Hamiltonian.\cite{Cuevas}
The scattering matrix theory\cite{Beenakker, Claughton}
 and Keldysh NEGF formalism \cite{Sun, Fazio} of electronic transport in
 $N-QD-S$
 systems were also presented. Quite recently Zhu et al\cite{Zhu}
 investigated a $2F-QD-S$ structure using the Keldysh NEGF method,
 obtaining some interesting results. However, in the Keldysh NEGF
 treatment to the $N-QD-S$\cite{Sun, Fazio} or
 $2F-QD-S$,\cite{Zhu}
 they always made some assumptions that the  ferromagnetic
 magnetization is along the $z$ axis and the superconductor order parameter is a real
 quantity,
  and take the wide-band limit.
 Here we use the current formula (\ref{eq:current-fns}) for
 $F/I/N/I/S$
 systems to investigate the resonant Andreev current and I-V characteristics of a
 genuine non-interacting hybrid structure (see Fig. $5$) beyond the wide-band
 limit.

\begin{figure}
\epsfig{file=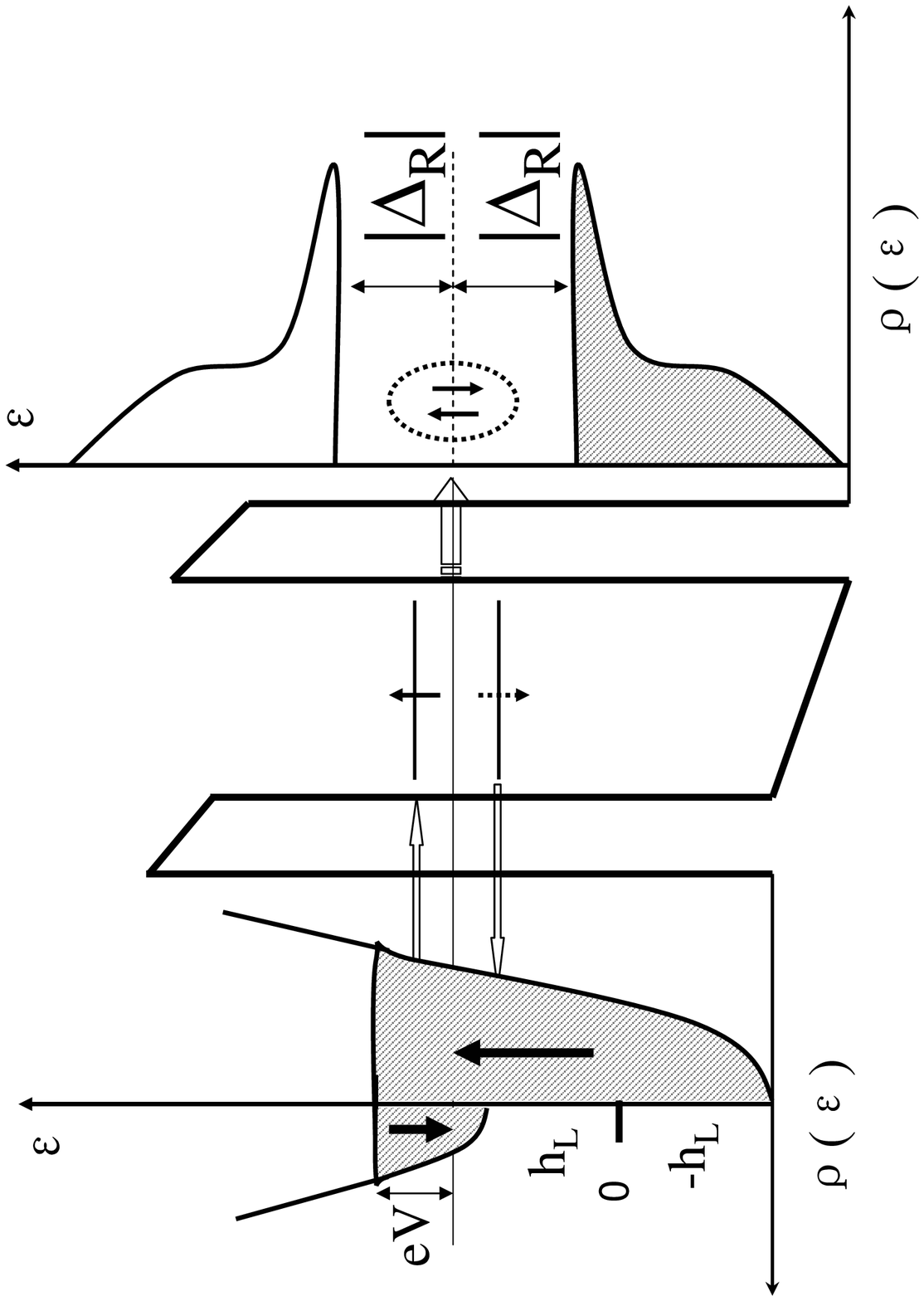, width=8cm}
 \caption{A schematic potential profile
for a biased DBTS connected to a ferromagnetic and a
superconducting leads. The magnetization of the ferromagnetic lead
is $h_{\cL}$ and the energy gap of the superconductor is
$|\Delta_{\cR}|$. The hatched regions represent occupied electron
states. A typical Andreev reflection process is shown: a spin-up
electron above the chemical potential of the superconductor is
reflected as a spin-down hole below the chemical potential at the
$NS$ interface, and finally into the ferromagnetic lead.}
\end{figure}

Following similar procedures as in the last subsection, we derive
the various kinds of full Green functions of the normal region for
a noninteracting $F/I/N/I/S$  resonant structure as
\begin{eqnarray}
\breve{\bf G}_{c,c}^{r,a/<}(t,t')&=&\int\frac{d\varepsilon}{2\pi}
e^{-i\varepsilon(t-t')/\hbar}\breve{{\bf
G}}_{c,c}^{r,a/<}(\varepsilon),
\end{eqnarray}
where (${\bf X}={\bf G}^{r,a/<},{\bf g}^{r,a/<}$ and
${\bf\Sigma}^{r,a/<}$)
\begin{eqnarray*}
\breve{\bf X}(t,t')&=& {\bf P}(\mu_{\cR \cC}
t+\frac{\varphi_{\cR}}{2}){\bf R}^f(\frac{\theta_{\cR f}}{2}) {\bf
X}(t,t')\nonumber\\
&&\hspace{2cm}{\bf R}^{f\dagger}(\frac{\theta_{\cR f}}{2}){\bf
P}^\dagger(\mu_{\cR \cC} t'+\frac{\varphi_{\cR}}{2}) \nonumber
\end{eqnarray*}
 and
\begin{eqnarray}
\label{eq:rgreen-fns}
 \breve{{\bf
G}}_{c,c}^{r/a}(\varepsilon)&=&\Big[\breve{\bf
g}^{r/a-1}_{c,c}(\varepsilon)\pm\frac{i}{2}{\bf \Gamma}^{\cR
s}_\varrho(\varepsilon)\pm\nonumber\\
&&\hspace{3.0cm} \frac{i}{2}{\bf \Gamma}^{\cL
f}(\varepsilon\mp eV)  \Big]^{-1}, \\
\breve{\bf G}_{c,c}^{<}(\varepsilon)&=&\breve{\bf
G}_{c,c}^{r}(\varepsilon)\big[{\bf \Gamma}^{\cL
f}(\varepsilon\mp eV){\bf f}_\cL(\varepsilon \mp eV)\nonumber\\
&&\hspace{3cm}-{\bf \Gamma}^{\cR s}_\rho(\varepsilon){\bf
f}_\cR(\varepsilon)\big]\breve{\bf
G}_{c,c}^{a}(\varepsilon).\nonumber
\end{eqnarray}

Substituting the above Green functions into Eq.
(\ref{eq:current-fns}), we obtain
\begin{eqnarray}
\label{eq:current-fns-non}
\cI_{fns}&=&\cI^A_{fns}+\cI^{N}_{fns}\nonumber\\
&=&\frac{2e}{h}\int d\varepsilon
 [f_\cL(\varepsilon-eV)-f_\cL(\varepsilon+eV)]\cT_{fns}^A(\varepsilon)\nonumber\\
 &&+\frac{2e}{h}\int d\varepsilon
 [f_\cL(\varepsilon-eV)-f_\cR(\varepsilon)]\cT_{fns}^N(\varepsilon),
\end{eqnarray}
where
\begin{eqnarray}
\label{eq:andreev-transmission}
 \cT^A_{fns}&=&  \frac12\sum\limits_{i=1,3}\left({\bf \Gamma}^{\cL
f}(\varepsilon\mp eV)\breve{\bf
G}_{c,c}^r(\varepsilon)\right)_{ii+1}\nonumber\\&&\hspace{2cm}\left({\bf
\Gamma}^{\cL f}(\varepsilon\mp eV)
      \breve{\bf G}_{c,c}^a(\varepsilon)
      \right)_{i+1i}\nonumber\\
&=&\frac12\Big[\Gamma^{\cL f}_{\uparrow}(\varepsilon-e
V)\Gamma^{\cL f}_{\downarrow}(\varepsilon+e V)
|\breve{G}_{c,c;12}^r|^2+\nonumber\\ && \hspace{1cm} \Gamma^{\cL
f}_{\downarrow}(\varepsilon-e V)\Gamma^{\cL
f}_{\uparrow}(\varepsilon+e
V)|\breve{G}_{c,c;34}^r|^2\Big],\\
\label{eq:normal-transmission}
 \cT^{N}_{fns}&=& \frac12 \sum\limits_{i=1,3} \left( {\bf \Gamma}^{\cL
 f}(\varepsilon\mp eV)\breve{\bf G}_{c,c}^r(\varepsilon) {\bf \Gamma}^{\cR s}_\rho
(\varepsilon)\breve{\bf G}_{c,c}^a(\varepsilon)\right)_{ii} \nonumber\\
&=& \frac12 \rho^{\cR s}(\varepsilon) \Gamma^{\cR s}
\Big\{ \Gamma^{\cL
 f}_{\uparrow}(\varepsilon-eV)\big[|\breve{G}_{c,c;11}^r|^2+|\breve{G}_{c,c;21}^r|^2\nonumber\\
 &&  -2\frac{|\Delta_\cR|}{\varepsilon}
      {\rm Re}\{\breve{G}^r_{c,c;11}\breve{G}_{c,c;21}^a\}\big]+\Gamma^{\cL f}_{\downarrow}(\varepsilon-eV)
      \nonumber \\
      &&\hspace{1cm}\big[|\breve{G}_{c,c;33}^r|^2+|\breve{G}_{c,c;43}^r|^2+\nonumber\\&&
      \hspace{1.5cm}2\frac{|\Delta_\cR|}{\varepsilon}
      {\rm Re}\{\breve{G}^r_{c,c;33}\breve{G}_{c,c;43}^a\}\big] \Big\}.
  \end{eqnarray}
The elements of the Green function matrix
$\breve{G}^{r/a}_{c,c;11}$,
$\breve{G}^{r/a}_{c,c;12}$,$\breve{G}^{r/a}_{c,c;21}$,
$\breve{G}^{r/a}_{c,c;33}$, $\breve{G}^{r/a}_{c,c;34}$  and
$\breve{G}^{r/a}_{c,c;43}$ are derived from Eq.
(\ref{eq:rgreen-fns}) as
\begin{eqnarray*}
\breve{G}^{r}_{c,c;11}&=& \frac{(\sum\limits_{n}
\frac{1}{\varepsilon+\epsilon'_{n\downarrow}+i0^+})^{-1}+
\frac{i}{2}\big[\Gamma^{\cL
f}_{\downarrow}(\varepsilon+eV)+\Gamma^{\cR s}\varrho^{\cR
s}(\varepsilon)\big]}{
 \breve{G}_{fs1}(\varepsilon)},  \\
 \breve{G}^{r}_{c,c;12}&=& \breve{G}^{r}_{c,c;21}=
\frac{\frac i2\Gamma^{\cR s}\varrho^{\cR
s}(\varepsilon)\frac{|\Delta_\cR|}{\varepsilon}}{
 \breve{G}_{fs1}(\varepsilon)},  \\
 \breve{G}^{r}_{c,c;33}&=& \frac{(\sum\limits_{n}
\frac{1}{\varepsilon+\epsilon'_{n\downarrow}+i0^+})^{-1}+
\frac{i}{2}\big[\Gamma^{\cL
f}_{\uparrow}(\varepsilon+eV)+\Gamma^{\cR s}\varrho^{\cR
s}(\varepsilon)\big]}{
 \breve{G}_{fs2}(\varepsilon)},  \\
\breve{G}^{r}_{c,c;34}&=&\breve{G}^{r}_{c,c;43}=- \frac{\frac
i2\Gamma^{\cR s}\varrho^{\cR
s}(\varepsilon)\frac{|\Delta_\cR|}{\varepsilon}}{
\breve{G}_{fs2}(\varepsilon)},
\end{eqnarray*}
in which
\begin{eqnarray}
\breve{G}_{fs1}(\varepsilon)&=&\Big\{\big(\sum\limits_{n}
\frac{1}{\varepsilon-\epsilon'_{n\uparrow}+i0^+}\big)^{-1}+
\frac{i}{2}\big[\Gamma^{\cL
f}_{\uparrow}(\varepsilon-eV)\nonumber\\&& +\Gamma^{\cR
s}\varrho^{\cR
s}(\varepsilon)\big]\Big\}\Big\{\big(\sum\limits_{n}
\frac{1}{\varepsilon+\epsilon'_{n\downarrow}+i0^+}\big)^{-1}+\nonumber\\&&
\hspace{1cm}\frac{i}{2}\big[\Gamma^{\cL
f}_{\downarrow}(\varepsilon+eV)+\Gamma^{\cR s}\varrho^{\cR
s}(\varepsilon)\big]\Big\}+\nonumber \\
&&\hspace{2.5cm}\frac 14\Big[\Gamma^{\cR s}\varrho^{\cR
s}(\varepsilon)\frac{|\Delta_\cR|}{\varepsilon}\Big]^2,\\
\breve{G}_{fs2}(\varepsilon)&=&\Big\{\big(\sum\limits_{n}
\frac{1}{\varepsilon-\epsilon'_{n\downarrow}+i0^+}\big)^{-1}+
\frac{i}{2}\big[\Gamma^{\cL
f}_{\downarrow}(\varepsilon-eV)\nonumber\\&& +\Gamma^{\cR
s}\varrho^{\cR
s}(\varepsilon)\big]\Big\}\Big\{\big(\sum\limits_{n}
\frac{1}{\varepsilon+\epsilon'_{n\uparrow}+i0^+}\big)^{-1}+\nonumber\\&&
\hspace{1cm}\frac{i}{2}\big[\Gamma^{\cL
f}_{\uparrow}(\varepsilon+eV)+\Gamma^{\cR s}\varrho^{\cR
s}(\varepsilon)\big]\Big\}+\nonumber \\
&&\hspace{2.5cm}\frac 14\Big[\Gamma^{\cR s}\varrho^{\cR
s}(\varepsilon)\frac{|\Delta_\cR|}{\varepsilon}\Big]^2.
\end{eqnarray}

Compared to the work for the $N-QD-S$ system from the similar
Keldysh formalism,\cite{Sun} the derivation of the final current
formula (\ref{eq:current-fns-non}),(\ref{eq:andreev-transmission})
and (\ref{eq:normal-transmission}) from the formalism we developed
is more direct, simple and systematic. What we need to do is just
some simple matrix algebra, while complicated mathematical
techniques are needed in the derivation of the Green functions in
the formalism of  Sun et al..\cite{Sun} Also the current formula
permits us to investigate the I-V characteristics within a much
wider bias voltage region in $F/I/N/I/S$ DBTSs.

The ferromagnetism and superconductor proximity are manifested in
the dependence on the magnetization $h_\cL$ and the magnitude of
the superconducting order parameter $|\Delta_\cR|$ of the full
Green functions through self-energy matrices.  From expressions
(\ref{eq:current-fns-non},\ref{eq:andreev-transmission},\ref{eq:normal-transmission}),
one observes that the current through a $F/I/N/I/S$ resonant
structure results from  different
contributions.\cite{Blonder,Cuevas}
 $\cI^A_{fns}$ is the Andreev
reflection current: a spin-up/down electron/hole associated with
spectral weight $\Gamma^{\cL f}_\uparrow(\varepsilon-eV)$/
$\Gamma^{\cL f}_\downarrow(\varepsilon-eV)$ incident from the
 ferromagnetic lead is reflected as a spin-down/up hole/electron
 with  spectral weight $\Gamma^{\cL
 f}_\downarrow(\varepsilon+eV)$/
 $\Gamma^{\cL f}_\uparrow(\varepsilon+eV)$ backward into the
 original lead, and at the same time two electrons in the normal
 region are removed into the superconductor as an electron pair
 with probability  $|\breve{G}^r_{c,c;12}|^2/|\breve{G}^r_{c,c;34}|^2$.
 $\cI^N_{fns}$ comes from three kinds of physical processes. The
 first and fourth terms in (\ref{eq:normal-transmission}) represent
 the contribution from normal electron transmission, a
 spin-up/down electron/hole tunnels into the superconductor with
 probability
 $|\breve{G}^r_{c,c;11}|^2/|\breve{G}^r_{c,c;33}|^2$; the
 second and fifth terms describe the `branch-crossing' process in
 the BTK theory,\cite{Blonder}  a spin-up/down electron/hole in the
 ferromagnetic lead is converted into a spin-down/up hole/electron in the
 superconductor side, with particle pairs of opposite spins
 created in the normal region $|\breve{G}^r_{c,c;21}|^2/|\breve{G}^r_{c,c;43}|^2$.
 The terms left correspond to the net transfer of electrons/holes,
 along with the creation/annihilation of particle pairs inside the well and the
 annihilation/creation of pairs into the superconductor lead with
 probability proportional to
 ${\rm Re}\{\breve{G}^r_{c,c;11}\breve{G}_{c,c;21}^a\}/{\rm Re}\{\breve{G}^r_{c,c;33}\breve{G}_{c,c;43}^a\}$.
 At absolute zero temperature, the only contribution to the
 current is $\cI^A_{fns}$ for $eV<|\Delta_\cR|$, since in this case $\rho^{\cR
 s}(\varepsilon)$ in $\cT^N_{fns}$ becomes zero and then
 $I^N_{fns}=0$.
 When $eV>|\Delta_\cR|$ all processes contribute to the
 current. If one sets $h_\cL=0$ and assumes the wide-band approximation, the current formula will reduce to the
 result obtained by Sun et al\cite{Sun} in the $N-QD-S$
 case.

 Assuming a single active level $\varepsilon_0$ in the well, we get the
 following linear-response conductance of the  $F/I/N/I/S$ system
 \begin{eqnarray}
 \label{eq:conductance-fns1}
 \cG_{fns}(\varepsilon_0)&=& \nonumber \\&&
 \hspace{-2cm} \frac{4e^2}{h}\frac{\Gamma^{\cL
 f}_\uparrow  \Gamma^{\cL
 f}_\downarrow(\Gamma^{\cR
 s})^2/4}{\displaystyle \Big[\varepsilon_0^2+\frac{\Gamma^{\cL f}_\uparrow \Gamma^{\cL f}_\downarrow
 +(\Gamma^{\cR s})^2}{\displaystyle 4}\Big]^2+\frac{\varepsilon_0^2(\Gamma^{\cL
 f}_\uparrow-\Gamma^{\cL
 f}_\downarrow )^2}{\displaystyle 4}}.
 \end{eqnarray}
 For the completely polarized ferromagnetic lead, i.e.,
 $h_\cL/W=1$, and $\Gamma^{\cL f}_\downarrow=0$, the linear
 conductance turns out to be zero, since there is no state available for the Andreev
 reflected spin-down holes.  If the magnetization
 $h_\cL$ is zero,  $\Gamma^{\cL f}_\uparrow=\Gamma^{\cL f}_\downarrow=
 \Gamma^{\cL}(0)$, the ferromagnetic lead becomes a normal metal, and the conductance is reduced to
  \begin{equation}
  \label{eq:conductance-nns}
  \cG_{nns}(\varepsilon_0)=
  \frac{4e^2}{h}\Big(\frac{2\Gamma^{\cL
  }(0)\Gamma^{\cR
  s}}{4\varepsilon_0^2+ [\Gamma^{\cL }(0)]^2
  +(\Gamma^{\cR s})^2}\Big)^2,
  \end{equation}
 which is the same as the result obtained by Beenakker from the
 scattering matrix approach.\cite{Beenakker} In contrast to a single $F/S$
 junction,\cite{Jong}
 the conductance of an $N/I/N/I/S$ resonant structure is always
 not
 less than that of the $F/I/N/I/S$ structure, regardless of the value
 of the magnetization $h_\cL$. At
 $\varepsilon_0=0$ the conductance in the $N/I/N/I/S$ structure is maximal for
 symmetric couplings $\Gamma^{\cL}(0)=\Gamma^{\cR s}$, equaling to $4e^2/h$ twice that in the
 $N/I/N/I/N$ case. Moreover the line shapes of the linear-response conductances (\ref{eq:conductance-fns1}) and
 (\ref{eq:conductance-nns}) which decay as $\varepsilon_0^{-4}$
 are not of the simple Lorentzian form
 $\Gamma^\cL\Gamma^\cR/[\varepsilon_0^2+(\Gamma^\cL+\Gamma^\cR)^2/4]$.

Let us analyze further the spin polarization dependence of
$\cG_{fns}$ at resonance  $\varepsilon_0=0$ for
 different couplings. Setting $\Gamma^{\cR s}=\lambda \Gamma^{\cL
 f}(0)$, Eq. (\ref{eq:conductance-fns1}) evolves into
\begin{equation}
\label{eq:conductance-fns2} \cG_{fns}=\frac{4e^2}{h}\frac{4 \kappa
\lambda^2}{(\kappa+\lambda^2)^2},
\end{equation}
where $\kappa=\sqrt{1-h_\cL^2/W^2}$ is a quantity characterizing
the degree of spin polarization of the ferromagnetic lead:
$\kappa=1$ for normal metals and $\kappa=0$ for completely
polarized ferromagnets. When $\lambda\ge1$, the conductance
increases with increasing $\kappa$, implying that the conductance
decreases when the degree of spin polarization rises. If
$\lambda<1$, the conductance first increases with increasing the
spin polarization and then decreases rapidly after it reaches its
maximum value $4e^2/h$. The critical  value of $\kappa$ is given
by $\kappa=\lambda^2$. This interesting result is also obtained by
Zhu et al.,\cite{Zhu} shown in  Fig. $2$ of their paper.

\begin{figure}
\epsfig{file=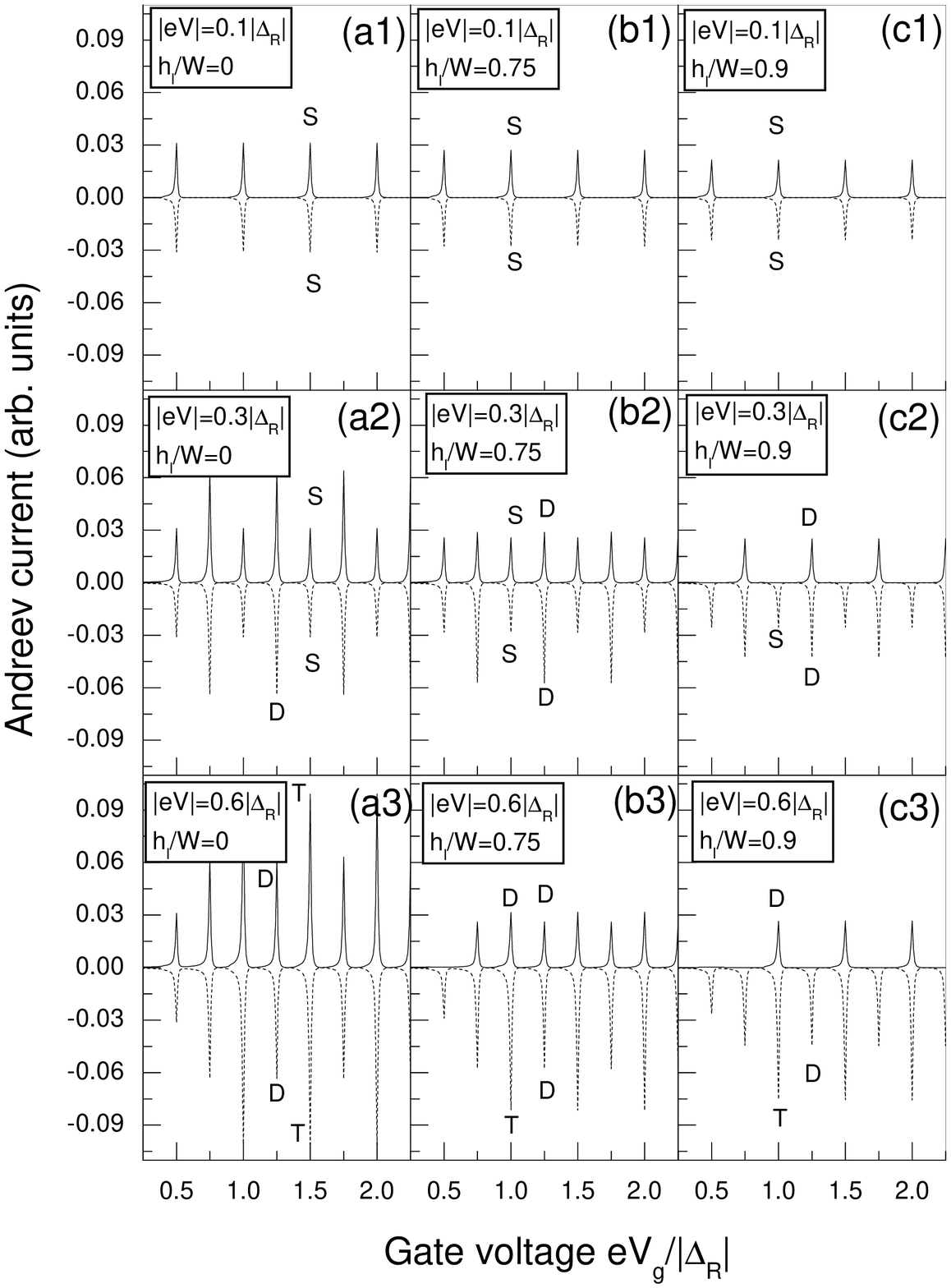, width=9cm} \caption{The Andreev current
spectra at zero temperature and fixed bias voltage for different
spin polarizations  (a) $h_\cL/W=0$ , (b) $h_\cL/W=0.75$ and (c)
$h_\cL/W=0.9$, where $W=2|\Delta_\cR|$. The full lines correspond
to the results when $eV>0$ and the  dashed  when $eV<0$. Labels
$S$, $D$ and $T$ are used to denote the current peaks arising from
the resonant Andreev reflections involving single, double or
triple levels, respectively. Here we assume $20$ levels with
identical level separation $0.5|\Delta_\cR|$, and the first level
aligns with the chemical potential of the superconductor lead when
$V_g=0$. The other parameters are $\Gamma^{\cL f}(0)=\Gamma^{\cR
s}=0.01|\Delta_\cR|$.}
\end{figure}

Next we explore the dependence of the Andreev current spectrum on
the degree of spin polarization of the ferromagnetic lead. From
the schematic view of the resonant Andreev reflection processes in
$F/I/N/I/S$ structures, one immediately becomes aware that the
resonant Andreev current is determined mainly by the applied bias
voltage $V$, the ratio of the strength of ferromagnetic
magnetization to the bandwidth $h_\cL/W$, and the level separation
$\delta \varepsilon_n$.   We choose a special level separation
$\Delta \varepsilon_n=0.501|\Delta_\cR|$, a case in which
  at most three levels are allowed to fall into the energy gap of
    the superconductor lead. For simplicity we assume identical
    level separations and do not consider the influence of the bias
    voltage on the level shift for convenience of comparison.
    At fixed bias voltage smaller than the energy gap $|\Delta_\cR|$,
    {\it resonant Andreev reflection takes place whenever
    the chemical potential of the superconductor lies just
    in between two levels and there are  states available for the reflected
    electrons/holes}. The energy levels $\epsilon_n-eV_g$ can be shifted up and down by
    tuning continuously the gate voltage $V_g$. Therefore, one can expects a series of peaks in the Andreev current as a
    function of $V_g$.

    In Fig. 6 we
    present numerical results of the Andreev current as a
    function of the gate voltage $eV_g$ for different spin
    polarizations $h_\cL/W$ and different bias voltages $eV$.
    The cases of positive and
    negative bias voltage are considered to compare the Andreev current
    contributed from electron and hole transmission.
    Let us first inspect the Andreev current spectra in the
    $N/I/N/I/S$ case (Fig. 6(a1)-(6a3)).\cite{Sun}
    At a small positive bias voltage $eV=0.1|\Delta_\cR|$(Fig. 6(a1)),
    a series of peaks labelled by $S$ with the same separation as
    the level spacing
    is observed. These peaks come from the resonant Andreev reflection
    processes by electron tunneling through a single level
    aligned with the chemical potential of the superconductor. When
    the bias voltage is small, there is no possibility for two levels to satisfy
    the resonant Andreev reflection condition.  As the bias increases to
    $eV=0.3|\Delta_\cR|$ (Fig. 6(a2)), there is possibility for two neighboring levels to lie equally above and
    below the chemical potential of the superconductor. The condition of the resonant Andreev reflection
    involving two levels can be satisfied and two-level Andreev reflection also
    contributes to the Andreev current.
    As a result another series of resonant Andreev
    current peaks labelled by $D$ is observed neighboring to the original series
    from single level contributions.  These $D$ peaks,
    with the same spacing as the $S$ series and
    $0.25|\Delta_\cR|$ away from it, stand out for their double
    height compared to the $S$ ones.  This is because for the $D$ peaks
    two neighboring levels are involved in the corresponding
    resonant Andreev reflection processes, so the probability
    is doubled as compared to the single level
    situation. At a still higher bias voltage $eV=0.6|\Delta_\cR|$ (Fig. 6(a3))
    the Andreev current spectrum can be
    understood similarly, the $S$-type peaks are now replaced by
    $T$ ones with tripled amplitudes,  resulting from three-level
    contributions. Since at $eV=0.6|\Delta_\cR|$, two additional
    levels near the middle one in alignment with the
    chemical potential also contribute to the Andreev current, so there are
    three neighboring levels taking part in resonant Andreev reflections,
    making the height of the $T$ series three times as
    that of the $S$ ones. When the bias is reversed, the Andreev
    current becomes negative, implying that the Andreev reflection
    is induced by  hole transmission. However, the Andreev
    current spectra remains unchanged.  Since in normal metals
    the DOS is spin degenerate,
    complete resonant Andreev reflections are guaranteed for electrons as
    well as their hole counterparts. Therefore, except for the sign
    the Andreev current spectra are the same for
    electron ($V>0$) and hole transmission ($V<0$).
     The above phenomena can
    also be understood from the intuitive diagrams in Fig. $7$, with
     different spin-up and spin-down bands of the ferromagnets replaced by
    identical ones of the normal metals.

    \begin{figure}
\epsfig{file=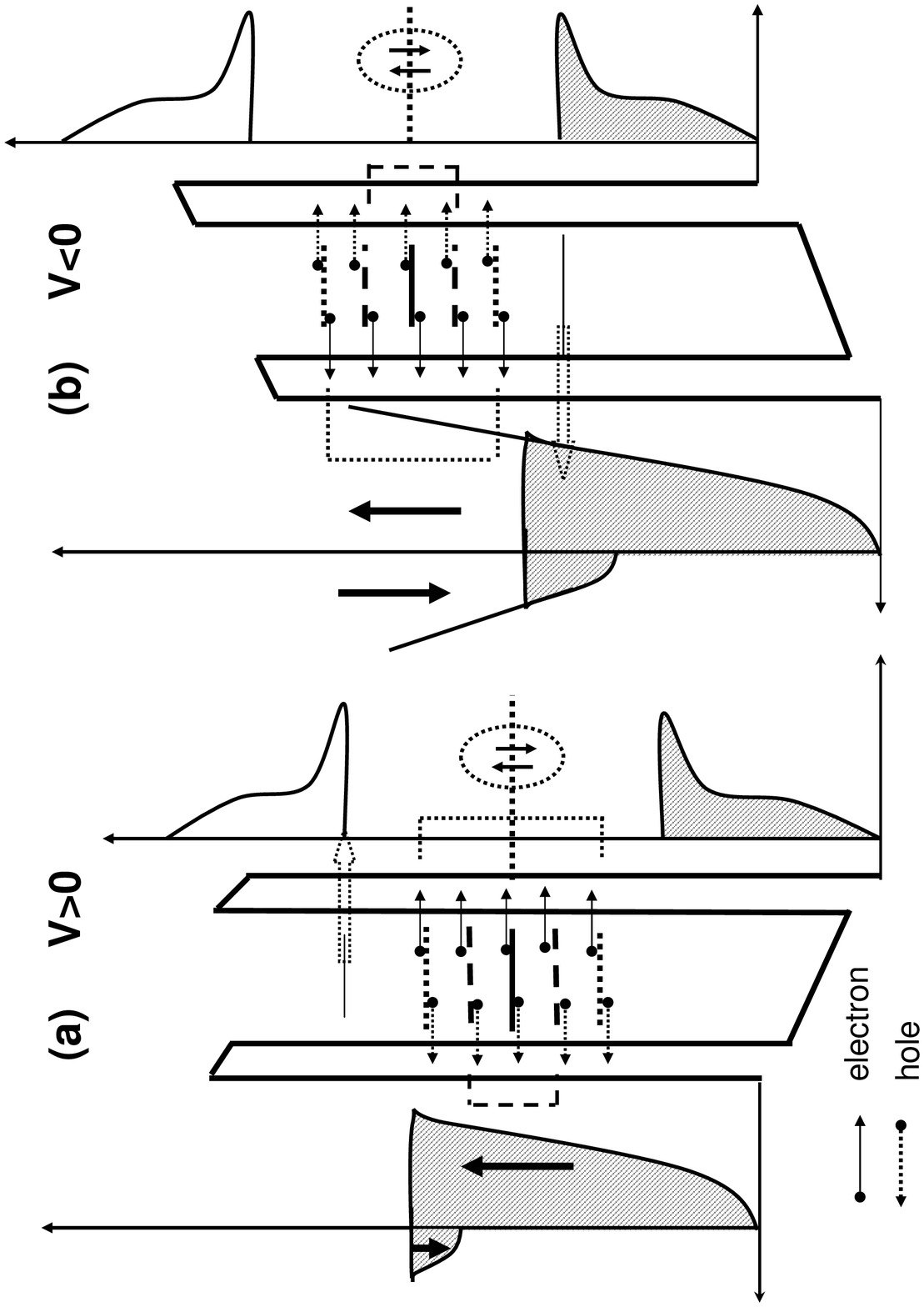, width=9cm} \caption{Schematic views of the
resonant Andreev reflection processes arising from the electron
(a)  and hole (b) transmission from the ferromagnetic lead. The
hatched region represents states that are occupied by electrons.
The Andreev current peaks labelled by $S$ in Fig. 6 is originated
from the Andreev reflection process involving a {\it single} level
represented by the solid line, $D$ {\it double} levels by dashed
lines and $T$ {\it triple} levels by dotted lines. The block
arrows stand for the process at the edges of the superconducting
band involving electron tunneling through the level located at
$\varepsilon_n=|\Delta_\cR|$, which results in sharp peaks in the
I-V characteristics.}
\end{figure}

    In a $F/I/N/I/S$ resonant structure the Andreev current
    depends not only on the position of the quantum well levels as in the normal case, but
    also on whether there are available states for the backward
    reflecting
    holes. The Andreev spectra for electron
    and hole transmission appear to be different if the
    ferromagnetic lead is of large spin polarization. For the
    completely polarized ferromagnetic lead, no Andreev current
    is expected due to the absence of empty states for the
    returning holes. We therefore choose the spin polarizations
    $h_\cL/W=0.75$ and $h_\cL/W=0.9$ for our purpose, where
    $W=2|\Delta_\cR|$, and the results are given in
    Fig. 6 (b1-b3) and (c1-c3).  At a low bias voltage
    $eV=0.1|\Delta_\cR|$ there are still available states for the
    reflecting holes and the Andreev current exhibits the same
    resonant spectrum as in the normal case, with slightly suppressed peak amplitude.
    If the bias voltage
    is $eV=0.3|\Delta_\cR|$,  the spin-down band of polarization $h_\cL/W=0.9$ moves above the
    chemical potential of the superconducting lead, leaving only the possibility for a spin-up
    electron to be transmitted through a level below the chemical potential
    and then reflected backwards through the neighboring level above the
    chemical potential
    to the spin-down band. Hence we can only observe the $D$-type
    peaks with amplitude half that in the normal case in the spectrum (Fig. 6(c2)).
    As $h_\cL/W=0.75$,  the $D$-type Andreev current peaks with half the amplitude of the normal case
     are
    originated from the same resonant Andreev reflection processes by spin-up
    electrons going through the states below the chemical potential as in the case $h_\cL/W=0.9$.
    This is the reason why we observe the amplitudes of the $S$ and $D$-type
    peaks to be nearly the same (Fig. 6(b2)).
    As the bias voltage $eV$ is
    further increased to $0.6|\Delta_\cR|$, the spin-down bands for
    $h_\cL/W=0.75$ and $h_\cL/W=0.9$ shift above the chemical
    potential of the superconductor, and one can no longer observe the series of $S$-type peaks from
    the contribution of the levels just at the chemical potential.  For
    ferromagnets with high spin polarization, the current arises only from
    the resonant Andreev reflections by the spin-up electrons tunneling
    through the levels $0.5|\Delta_\cR|$  below the chemical potential.
    Thus the Andreev current spectrum ((Fig. 6(c3)) only consists of
    a series of peaks at the positions of the $T$-type peaks
    in the normal case. Whereas in the small polarization case ((Fig. 6(b3)), the
    Andreev reflections involving two neighboring levels and two
    of three levels contributing to the $T$-type peaks in the
    normal case can happen, and a series of
    resonant peaks with equal separation $0.25|\Delta_\cR|$ is observed.

    Schematic views of the above resonant Andreev reflection processes for electron transmission
    are given in Fig. 7(a). When the bias voltage is reversed, the current
    is contributed from hole transmissions, and the
    situation is now very similar
    to the normal case. The only difference is that the amplitudes
    of the peaks in the ferromagnetic case are suppressed due to the
    reduced DOS for the reflected electrons. The relation among the
    $S$-, $D$- and $T$-type current amplitudes still hold, for
    which a heuristic physical picture is given in the Fig.
    7(b). When the level spacings are not identical, more
    interesting and complicated resonant Andreev current
    patterns  can be expected.\cite{Sun}  However, we can still
    appreciate and analyze them from the intuitive pictures of Fig. 7 whatever the
    Andreev current spectra may be.

     \begin{figure}
     \epsfig{file=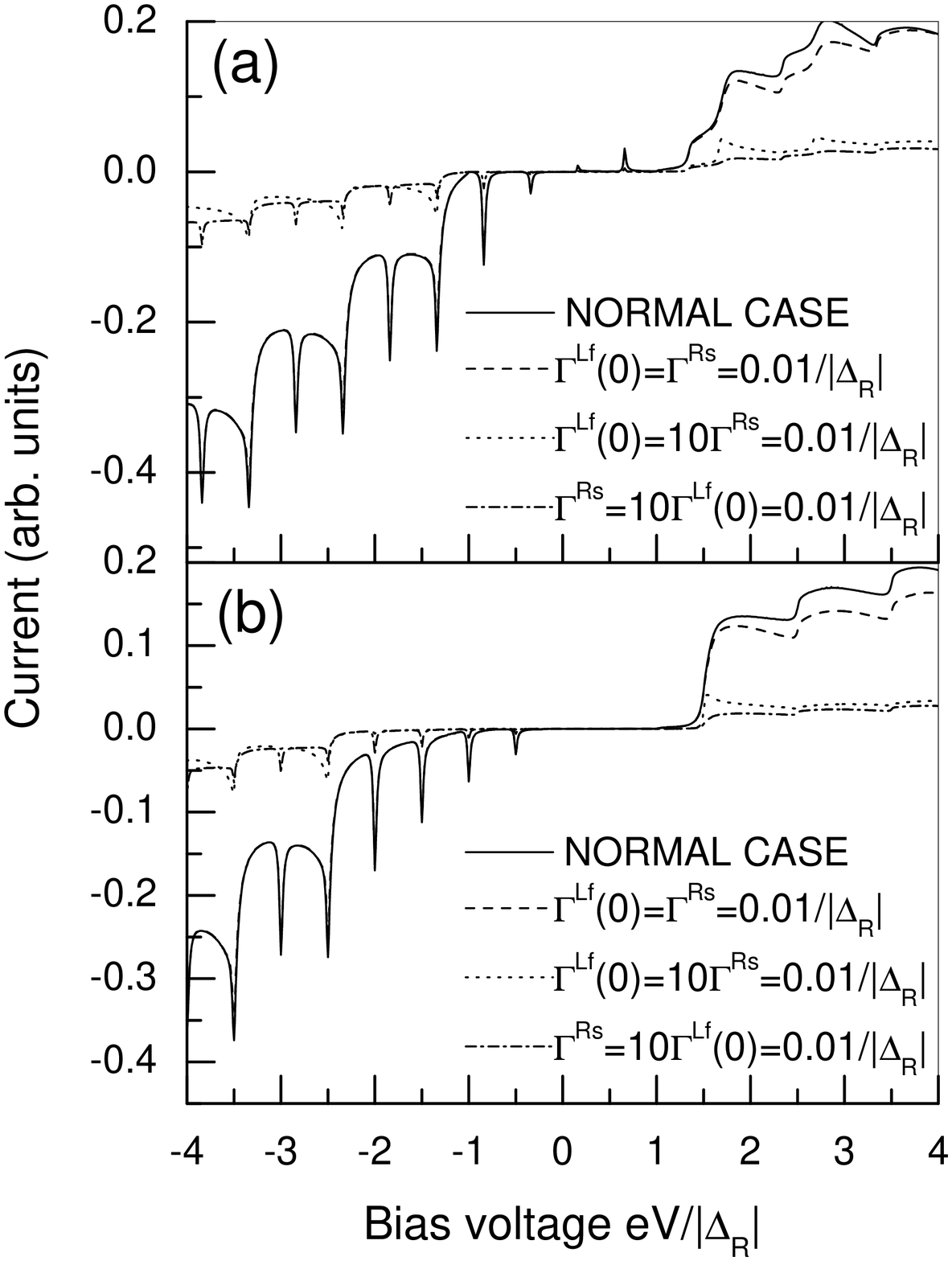, width=8cm}
      \caption{I-V characteristics
      of a $F/I/N/I/S$ resonant structure at temperature $0.1|\Delta_\cR|$
      for small level spacing $\Delta \varepsilon_n=$
      (a) $0.33|\Delta_\cR|$ below and (b)  $0.25|\Delta_\cR|$  above
      the chemical potential of the superconducting lead when $V=0$.
       The solid curves are the results for the $N/I/N/I/S$
       case with symmetric coupling $\Gamma^{\cL f}(0)=\Gamma^{\cR s}=0.01|\Delta_\cR|$, and
       the other curves for the $F/I/N/I/S$ structure with $h_\cL/W=0.5$.}
      \end{figure}

        \begin{figure}
        \epsfig{file=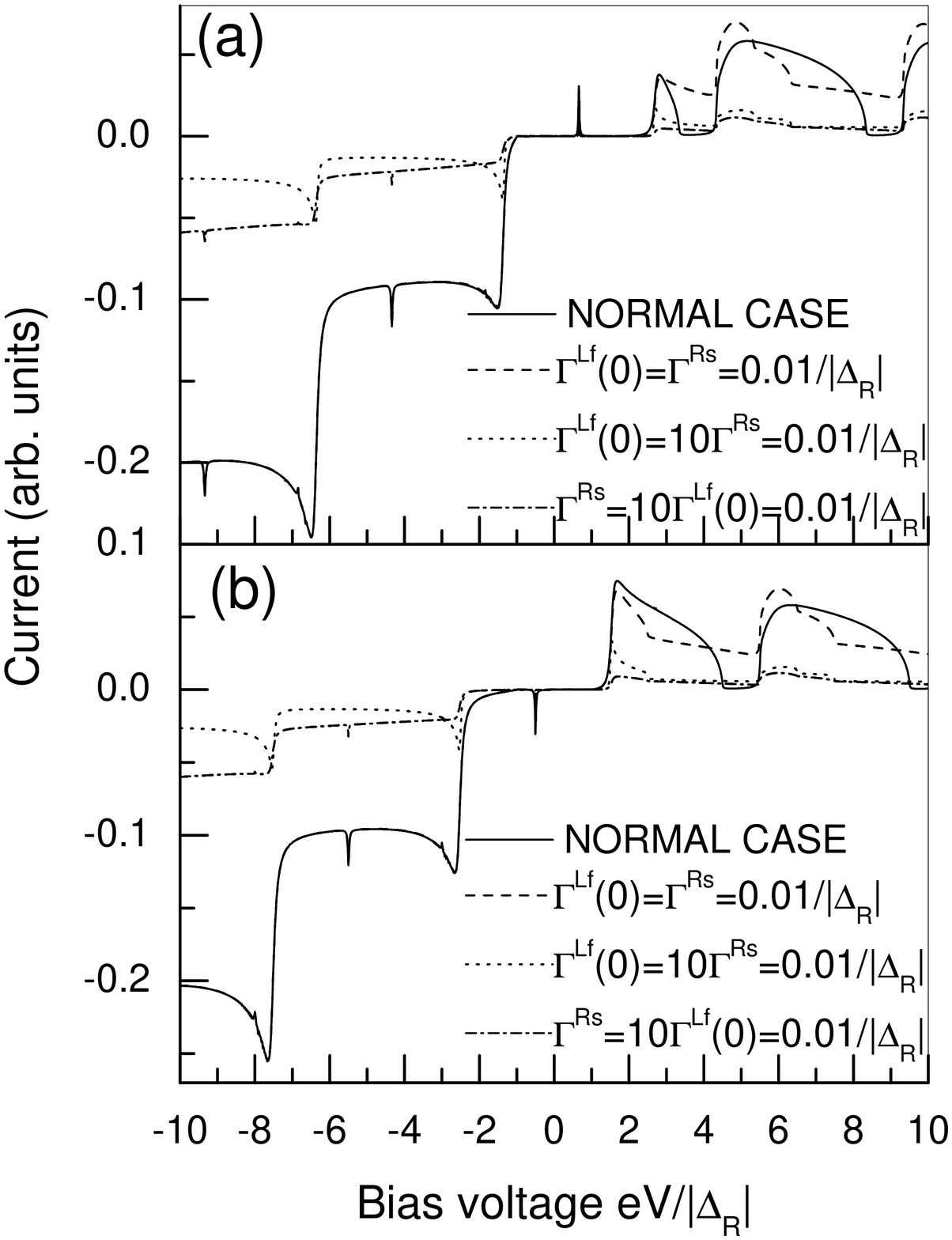, width=8cm}
         \caption{Same as Fig. 8, but for level spacing $\Delta \varepsilon_n=
         2.5 |\Delta_\cR|>W=2|\Delta_\cR|$ when the first level $\varepsilon_0$ lies
         (a) $0.33|\Delta_\cR|$ below and (b)  $0.25|\Delta_\cR|$  above
         the chemical potential of the superconducting lead when
         $V=0$.}
         \end{figure}

             The I-V characteristics of this kind of generic hybrid structure is
     also interesting.  It is known that the resonant Andreev reflection process also
     contributes to the current when the bias voltage $eV$ is greater than
     $|\Delta_\cR|$, the energy gap
      of the superconducting
     lead.  For this reason we consider cases in which the level spacing $\Delta \varepsilon_n$ can be
     either smaller or greater than the energy gap $|\Delta_\cR|$, as well as when the first
     level $\varepsilon_0$
     is either below or above the chemical potential of the
     superconductor when $V=0$.  Fig. (8a) and (8b) give the I-V
     curves for small level spacing $\Delta \varepsilon_n=0.5|\Delta_\cR|$
     when the first level $\varepsilon_0=-0.33|\Delta_\cR|$ lies
     below (Fig. 8a) and $\varepsilon_0=0.25|\Delta_\cR|$
     above (Fig. 8b) the chemical potential of the superconductor lead. As usual we approximate the variation
     of the energy levels with the bias voltage by
     $\varepsilon_n+0.5eV$. There is no substantial difference
    between the I-V characteristics  of the $F/I/N/I/S$ and
    $N/I/N/I/S$ systems when the level separation is small, as shown by the full and dashed
    lines in Fig. 8.  However, the current as a function of the bias
    voltage is strongly dependent on the symmetry between the
    couplings and the level configuration, especially when the applied bias is positive.
    The peaks in the I-V curves are originated from
    the resonant Andreev reflections, which emerge at some
    specific bias voltages when the resonant Andreev reflection
    condition is satisfied.  The irregular current plateaus come
    from the normal particle transmission. There are two types of
    current plateaus with different heights. Those with
    higher height are determined by particle tunneling through the
    level aligned with one of the edges of the superconducting energy gap at which the DOS is
    divergent, represented by a block arrow in Fig. 7, while the
    others are contributed from the levels  away from the edges.
    Such an analysis can be confirmed by the following simple estimation:
    If $\varepsilon_0=-0.33|\Delta_\cR|$, then at bias
    $-0.33|\Delta_\cR|+0.5eV=0$ and $0.17|\Delta_\cR|+0.5eV=0$, the resonant Andreev
    reflection condition is satisfied and current peaks emerge at
    $eV= 0.66|\Delta_\cR|$ and  $eV=-0.34|\Delta_\cR|$. For the
    first plateau to appear, we need $\varepsilon_n+0.5eV=\pm|\Delta_\cR|$
    and $eV\geq\pm|\Delta_\cR|$, i.e., $eV=1.66|\Delta_\cR|$ and
    $eV=-1.34|\Delta_\cR|$. The numerical results are consistent
    with this simple argument (Fig. 8(a)).

    In the case of
    negative bias $eV<-|\Delta_\cR|$, the levels are pushed down
    gradually inside the energy gap and  the resonant Andreev
    reflection condition is eventually satisfied at some bias voltages. Since both the normal particle tunneling
    and the resonant Andreev reflection contribute to the
    current, one observes a series of equally
    spaced resonant peaks superimposed onto the plateaus with
    identical widths. When the couplings become asymmetric (one
    of the couplings is weakened), both the Andreev current peaks and
    current plateaus are suppressed. This is because  the normal and Andreev current
    is proportional to the product of the couplings $\Gamma^{\cL f}(0)\Gamma^{\cR s}$, as can be found in the formulae
    (\ref{eq:current-fns-non}), (\ref{eq:andreev-transmission})
    and (\ref{eq:normal-transmission}). It has already been found
    analytically by Sun et al\cite{Sun} that the Andreev
    reflection probability is maximized for symmetric couplings in
    $N-QD-S$ structures and decreases rapidly with the increasing  coupling asymmetry.
    In the special asymmetric coupling case where $\Gamma^{\cR s}<<\Gamma^{\cL
    f}(0)$, the sharp DOS at the edges of the superconductor energy gap
     can be discerned clearly whenever particles
    transmit through a level aligned to  the edge (see the dotted lines in Fig. 8).
    This can be easily understood from the Breit-Wigner formula for resonant normal electron transmission at
    resonance $ \cT=4\Gamma^{\cL f}(0)\Gamma^{\cR s}\rho^{\cL f} \rho^{\cR
    s}/[\Gamma^{\cL f}(0)\rho^{\cL f}+\Gamma^{\cR s}\rho^{\cR
    s}]^2$.  The transmission probability $\cT$ depends on the
   ratio between the level-widths $\Gamma^{\cL f}(0)\rho^{\cL
    f}$ and  $\Gamma^{\cR s}\rho^{\cR s}$. At the edges of the gap, $\rho^{\cR s}$
    is divergent and thus the coupling constant should be small
    enough to balance these two level-widths to guarantee high transmission.
    Notice that the ferromagnetic feature can not be
   observed in the I-V curves of $F/I/N/I/S$ structures
    when the level spacing is small. It can only be displayed when the
    level separation is greater than the band-width of the
    ferromagnets. In Fig. 9 we present the results of I-V
    relation of this case  for different couplings and different
    level arrangements. As expected, the degree of spin
    polarization of the ferromagnetic lead is reflected in the I-V curves. In the
    normal case and when $eV>0$, the current first develops a
    resonant Andreev peak at $-0.33|\Delta_\cR|+0.5eV=0$ and exhibits
    usual resonant peaks in the double-barrier structure after a
    narrow peak with width determined from  $-0.33|\Delta_\cR|+0.5eV=|\Delta_\cR|$ and
    $-0.33|\Delta_\cR|+0.5eV=eV-W$. When the bias is reversed, the current
    displays a plateau from the position
    $-0.33|\Delta_\cR|+0.5eV=-|\Delta_\cR|$,
    superimposed by some weak Andreev peaks arising from the resonant Andreev
    reflections for $|\varepsilon|>|\Delta_\cR|$.  If
    $\varepsilon_0=0.25|\Delta_\cR|$ when $V=0$, one observes an I-V characteristics similar
    to that of a usual magnetic DBTS  in the
    case of positive bias $V>0$, and in the negative bias case a resonant Andreev
    peak at $0.25+0.5eV=0$ and a plateau-peak
    structure similar to the case $\varepsilon_0=-0.33|\Delta_\cR|$.
    For $F/I/N/I/S$ structures, the current develops both a
    resonant peak as well as a shoulder which are more prominent in
    the high bias case for $eV>0$, while they keep nearly the same
    as in the $N/I/N/I/S$ structure when the bias is negative.
    This picture is violated when the couplings become
    asymmetric.  As in the case of small level spacing, the sharp
    DOS at the edges of the energy gap is also prominent in the
    I-V characteristics when the coupling to the superconductor
    side $\Gamma^{\cR s}$ is much smaller than that to the
    ferromagnetic side $\Gamma^{\cL f}(0)$. The spin polarization
    of the ferromagnet can thus be measured when the couplings to
    the ferromagnetic and superconducting leads are
    symmetric, as suggested from the comparison of the
    dashed, dotted and dash-dotted lines in Fig. 9.  It is
    noted that the  I-V characteristics of
    $F/I/N/I/S$ resonant structures are qualitatively right,
    because the idea about the band-width of the superconductor is
    somewhat
    vague within its semiconductor model.\cite{Tinkham}

    In summary,  we have investigated in this subsection the Andreev current spectra
    and I-V relations of a $F/I/N/I/S$ resonant structure in
    detail.  Interesting dependence on the ferromagnetic spin polarization of
    the linear conductance is discussed in terms of analytic expressions,
    given by Eqs. (\ref{eq:conductance-fns1})
    and (\ref{eq:conductance-fns2}). Our results demonstrate that
    the peak structure of the Andreev current as a function of the gate voltage
    is determined by the applied bias (both the value and sign) and the degree of spin polarization,
    which differs substantially from the results under the
    wide-band approximation. The I-V characteristics, closely associated
    with the level arrangement, coupling symmetry, bias sign and
    spin polarization, can be employed to characterize the
    density of states (DOS) of both ferromagnets and
    superconductors by tuning the coupling strengths.

\subsection{S/I/N/I/S structures}

The discovery of the Josephson effect\cite{Josephson} has provoked
a lasting research interest in the properties of the DC-and
AC-Josephson current in mesoscopic $S/N/S$ junctions in thirty
years.\cite{Tinkham, Ohta}  When the width of the normal region is
smaller than the coherence length, electron pairs can coherently
tunnel from one superconductor to the other, inducing a
phase-dependent DC current even when the bias is zero. Early in
1963, Ambergaokar and Baratoff\cite{Ambergaokar} derived a useful
formula for the supercurrent in $S/I/S$ junctions with the help of
the Gor'kov Green functions. In the 1990's investigations on the
mesoscopic $S/N/S$ junctions became timely\cite{Ohta} due to the
advances in experimental techniques. In most of these work, the
scattering matrix method based on the Bogoliuboiv-de Gennes (BdG)
equation\cite{Gennes} is commonly used. In a $S/N/S$ junction, the
Andreev reflections at the $N/S$ boundaries confined the
quasi-particle inside the normal region, resulting in the bound
states sensitive to the superconducting phase difference of the
two superconductors.\cite{Kulik} Impurities inside the normal
region altering the quasi-particle wave
interference,\cite{Bagwell} and the asymmetry between the two
energy gaps can also modify the Josephson current.\cite{Chang}
Glazman and Matreev,\cite{Glazman} and Ishizaka et
al.\cite{Ishizaka} have studied the influence of the Coulomb
interactions on the Josephson current in $S-QD-S$ systems.
Research were also conducted on the DC Josephson current in
noninteracting symmetric $S-QD-S$ structures
 by Beenakker using the scattering matrix
 approach,\cite{Beenakker3}
 and by Lin's Group from the Keldysh NEGF method.\cite{Lin}    However,
 these investigations are restricted to the symmetric case-the
 same couplings and energy gaps.  Motivated by this limitation,  we
 investigate in this subsection the DC Josephson current through a general
 $S/I/N/I/S$ resonant structure, in order to reveal the dependence
 of the Josephson current on the  energy gaps.
 Results for the AC Josephson current will be reported elsewhere.

\begin{figure}
\epsfig{file=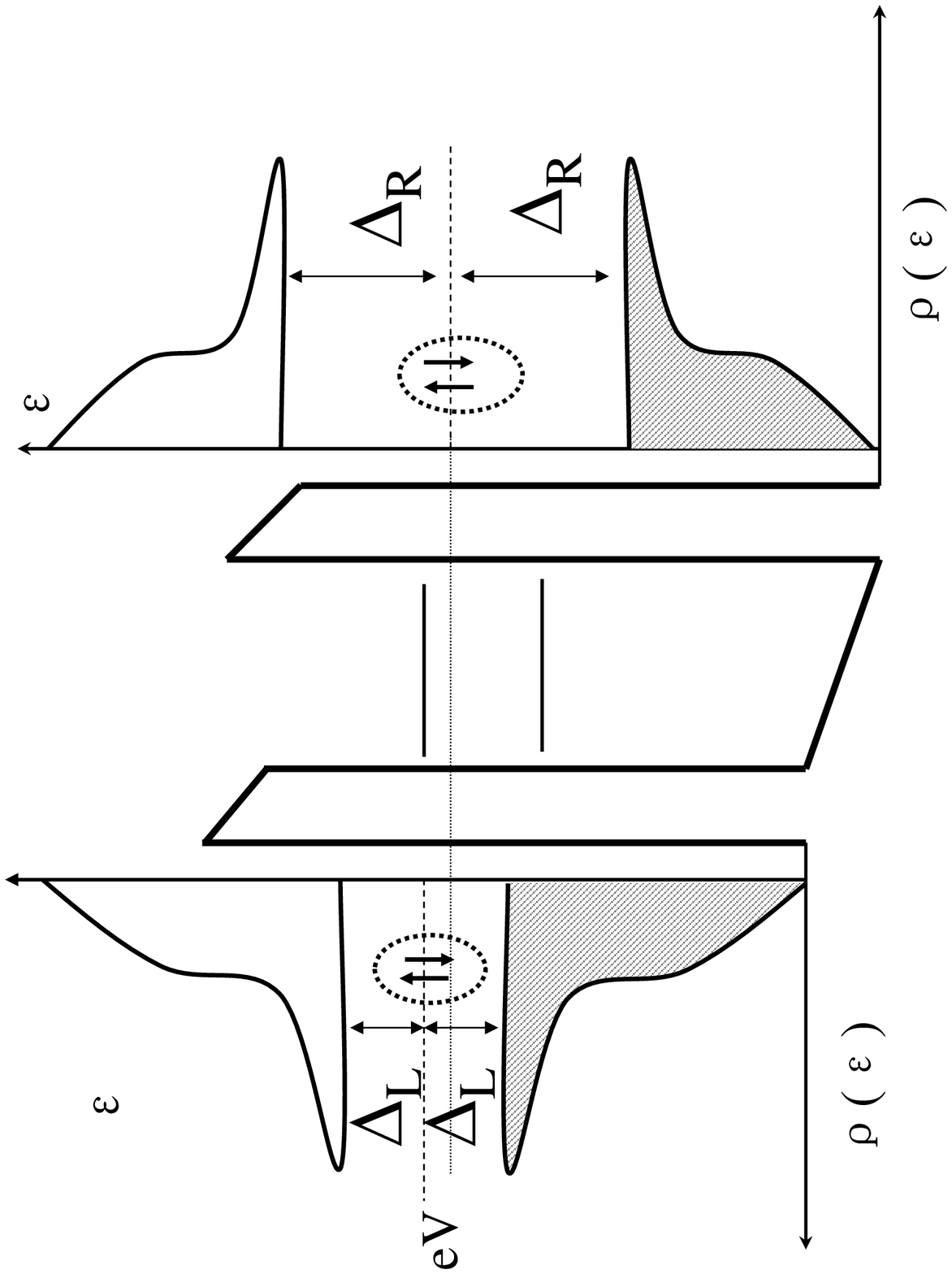, width=8cm} \caption{A schematic potential
profile for a biased DBTS attached to two
 superconducting leads with order parameters $\Delta_{\cL}$ and  $\Delta_{\cR } $.
 Here the hatched region represent occupied  electron states.}
\end{figure}

 Neglecting the interaction effects in the central region we
 obtain an unexpectedly simple form of the DC Josephson current
 formula for a general $S/I/N/I/S$ system
 from Eq. (\ref{eq:current-sns})
\begin{eqnarray}
\label{eq:current-sns-non}
 \cI_{sns}(\varphi_s) &=&\frac{e}{2h}\sum\limits^{i=1,3}\int d\varepsilon
{\rm Im}\Big\{f(\varepsilon)\Big(\big[\tilde{{\bf \Gamma}}^{\cL
s}_\varrho(\varepsilon)-{\bf \Gamma}^{\cR
s}_\varrho(\varepsilon)\big]\nonumber \\
&& \widetilde{{\bf G}}_{c,c}^{r}(\varepsilon)+\big[\tilde{{\bf
\Gamma}}^{\cL s}_{\varrho^*}(\varepsilon)-{\bf \Gamma}^{\cR
s}_{\varrho^*}(\varepsilon)\big]\widetilde{{\bf
G}}_{c,c}^{a}(\varepsilon) \Big)_{ii}\Big\},\nonumber \\
&=&-\frac{2e}{h}\sin{\varphi_s}\int_{-\infty}^{+\infty}
d\varepsilon \Gamma^{\cL s}\Gamma^{\cR
s}|\Delta_{\cL}\Delta_{\cR}|\frac{f(\varepsilon)}{\varepsilon^2}\nonumber\\&&
\hspace{3cm}{\rm Im}\big\{\frac{\varrho^{\cL s} \varrho^{\cR
s}}{\widetilde{G}_s^r} \big\}\nonumber\\
&=&-\frac{e}{\hbar}\sin{\varphi_s}
\sum\limits_{0<\varepsilon_p<|\Delta_\cL|} \tanh(\varepsilon_p/2k_BT)\nonumber\\
&& \lim\limits_{\varepsilon\rightarrow
\varepsilon_p}\frac{(\varepsilon-\varepsilon_p)\Gamma^{\cL s}
\Gamma^{\cR
s}|\Delta_{\cL}\Delta_{\cR}|}{\sqrt{(|\Delta_\cL|^2-\varepsilon^2)
(|\Delta_\cR|^2-\varepsilon^2)}\widetilde{G}_s^r(\varepsilon)}\nonumber\\&&-
\frac{2e}{h}\sin{\varphi_s}\Big[\overbrace{\int_{-|\Delta_\cR|}^{-|\Delta_\cL|}+
\int_{|\Delta_\cL|}^{|\Delta_\cR|}}^{2}+\nonumber\\
&&\overbrace{\int_{-\infty}^{-|\Delta_\cR|}+\int_{|\Delta_\cR|}^{+\infty}}^3\Big]
\Gamma^{\cL s}\Gamma^{\cR
s}|\Delta_{\cL}\Delta_{\cR}|\nonumber\\&&
\hspace{3cm}\frac{f(\varepsilon)}{\varepsilon^2}{\rm
Im}\big\{\frac{\varrho^{\cL s} \varrho^{\cR
s}}{\widetilde{G}_s^r} \big\}d\varepsilon \nonumber\\
&=&  \cI_1+\cI_2+\cI_3,
\end{eqnarray}
 where we
assume in general $|\Delta_\cL| \leq |\Delta_\cR|$, and
$\varepsilon_p$, the energies of the discrete Andreev bound
states, are the poles of the spectral function
$\widetilde{G}_{s}^r(\varepsilon)$
\begin{eqnarray}
\widetilde{G}_{s}^r(\varepsilon)&=&\Big\{\big(\sum\limits_{n}
\frac{1}{\varepsilon-\epsilon'_{n}+i0^+}\big)^{-1}+
\frac{i}{2}\big[\Gamma^{\cL s}\varrho^{\cL
s}(\varepsilon)\nonumber\\&& +\Gamma^{\cR s}\varrho^{\cR
s}(\varepsilon)\big]\Big\}\Big\{\big(\sum\limits_{n}
\frac{1}{\varepsilon+\epsilon'_{n}+i0^+}\big)^{-1}+\nonumber\\&&\hspace{-1cm}
\frac{i}{2}\big[\Gamma^{\cL s}\varrho^{\cL
s}(\varepsilon)+\Gamma^{\cR s}\varrho^{\cR
s}(\varepsilon)\big]\Big\}+\frac 14\Big[\frac{(\Gamma^{\cL
s}|\Delta_{\cL}|)^2}{\varepsilon^2-|\Delta_{\cL}|^2}\nonumber\\
&& +\frac{(\Gamma^{\cR s}|\Delta_{\cR}|)^2}{\varepsilon^2
-|\Delta_{\cR}|^2}+2\cos{\varphi_s}\Gamma^{\cL s}\Gamma^{\cR
s}\frac{|\Delta_{\cL}\Delta_{\cR}|}{\varepsilon^2}\nonumber \\
&&\hspace{2cm}\varrho^{\cL s}(\varepsilon)\varrho^{\cR s}(\varepsilon)\Big].
\end{eqnarray}
It is seen from Eq. (\ref{eq:current-sns-non}) that the DC
Josephson current  $\cI_{sns}$  in a general $S/I/N/I/S$ system
has contributions from three different scattering processes:
$\cI_1$ results from the resonant Josephson tunneling through the
discrete Andreev bound states given by
$\widetilde{G}^r_s(\varepsilon)=0$ within
$|\varepsilon|<|\Delta_\cL|$;
 $\cI_2$ from the quasi-particle escaping through  broadened levels
from the normal region to the weaker superconductor side, i.e.,
$|\Delta_\cL|\leq |\varepsilon|<|\Delta_\cR|$;  and $\cI_3$ from
quasi-particle tunneling from the normal region to both
superconductors $|\varepsilon|\geq|\Delta_\cR|$.\cite{Chang}  One
can consider Eq. (\ref{eq:current-sns-non}) as an extension to the
asymmetric case of the Beenakker's result for symmetric $S-QD-S$
systems.\cite{Beenakker3,Beenakker4}  In addition one can check
after simple algebra that
 the DC Josephson current formula (\ref{eq:current-sns-non}) for the general asymmetric $S/I/N/I/S$ system
 reduces to the known result for the symmetric
 case.\cite{Beenakker3,Lin,Beenakker4}

 \begin{figure}
  \epsfig{file=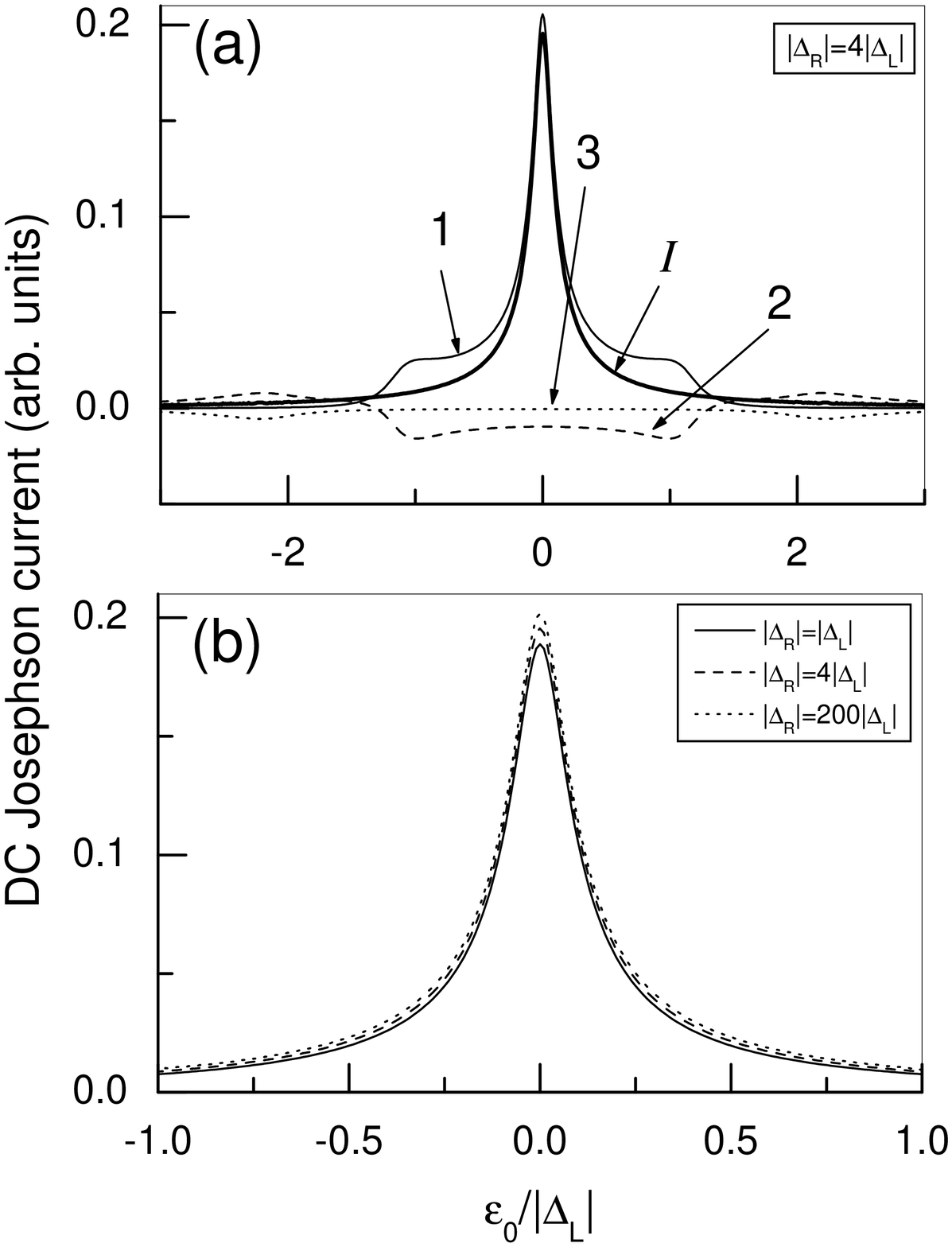, width=8cm}
  \caption{The DC Josephson current vs $\varepsilon_0$ for (a)
  different scattering processes, and (b) different energy gaps, where
  we take symmetric couplings as $\Gamma^{\cL s}=\Gamma^{\cR s}=0.01|\Delta_\cL|$. The superconducting
  phase difference is $\varphi_s=\pi/2$.}
  \end{figure}

  Now we consider the simplest situation in which there is only a
  single active
  level $\varepsilon_0$ in the central normal region.
  It is expected that the resonant Josephson scattering via the Andreev bound states $\varepsilon_p$
  will
  dominate the DC Josephson current.  In Fig. $11$ we plot this
  quantity
 at zero temperature, calculated from Eq.
  (\ref{eq:current-sns-non}) as a function of the single level energy $\varepsilon_0$  for symmetric couplings
  $\Gamma^{\cL s}=\Gamma^{\cR s}=0.01|\Delta_\cL|$. However, the superconducting energy gaps are allowed to be asymmetric.
  The superconducting phase difference is chosen as$\varphi_s=\pi/2$. The total
  current labelled by $I$ in Fig. 11(a) has a resonant peak when
  the single level is aligned with the chemical potential of the
  superconductor, i.e., $\varepsilon_0=0$,  resulting from the constructive
  interference between the forward Andreev state $+\varepsilon_p$
  and backward Andreev state $-\varepsilon_p$. Inspecting the
  current components for three different scattering regions $|\varepsilon|<|\Delta_\cL|$,
  $|\Delta_\cL|\leq |\varepsilon|<|\Delta_\cR|$
  and $|\varepsilon|\geq|\Delta_\cR|$,\cite{Chang}  labelled respectively by
  $1$, $2$ and $3$, one finds that the current component $1$ contributed by
  discrete Andreev levels makes the major contribution to the
  DC Josephson current. It possesses one peak at $\varepsilon_0=0$ plus two side peaks pinned at
  $\varepsilon_0=\pm|\Delta_\cL|$ which are offset by two peaks in the current
  component
  $\cI_2$, in which  two additional wider side peaks cancelled exactly by
  $\cI_3$.  The side peaks come from the abnormal superconductor
  DOS singularities at the edges of the energy gap(s), and the exact
  cancellation of these side peaks in the total Josephson current is
  due to the fact that both electron-like and hole-like
  excitations can escape through the active level from the
  superconductor to the central normal region. This is
  equivalent,  mathematically, to the vanishing residues of the spectral function
  $\widetilde{G}^r_s$.\cite{Lin}  In contrast to the symmetric case
  $|\Delta_\cL|=|\Delta_\cR|$, the DC Josephson current is
  slightly enhanced in the asymmetric case
  $|\Delta_\cL|<|\Delta_\cR|$, as demonstrated in Fig. 11(b). This result
  differs from  the usual $S/N/S$ structure, where the
  current is greatly enhanced.\cite{Chang} The reason is that the
  energy levels lie within $|\varepsilon|<|\Delta_\cL|$, and then
  resonant Josephson tunneling  dominates the DC
  current, as can be seen more clearly in the current-phase
  relation in Fig. 12. The significant enhancement of the DC
  Josephson current may be observed in a resonant $S/I/N/I/S$
  system with many levels located inside the region
  $|\varepsilon|<|\Delta_\cL|$, since in this case so many levels are
  active to
  contribute to the Josephson current.
   \begin{figure}
   \epsfig{file=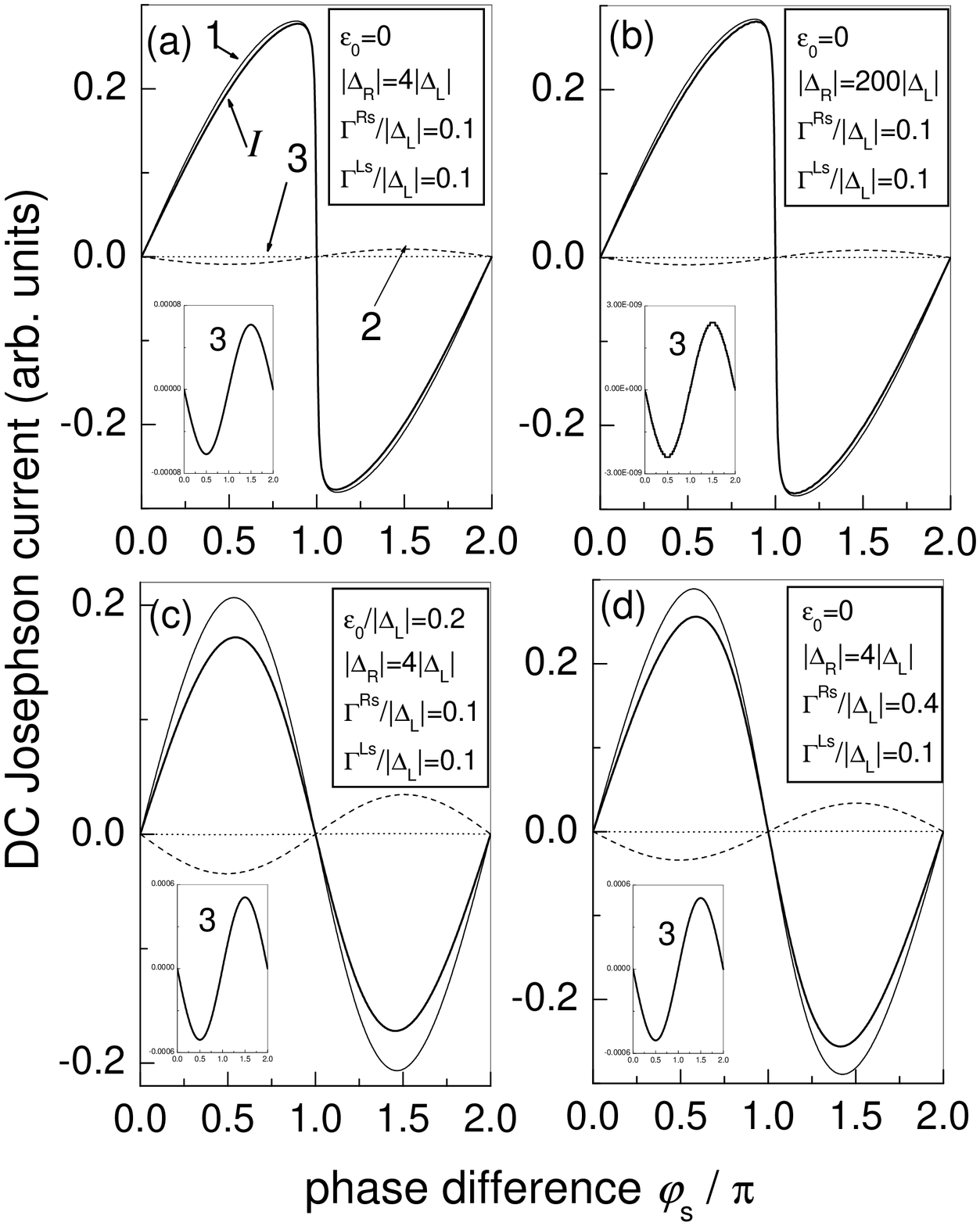, width=8cm}
   \caption{The current-phase relation for a resonant $S/I/N/I/S$
   structure, where the insets are enlargements of the current
   component
   $\cI_3$. Here the total DC Josephson current $\cI_{sns}$ and its components $\cI_1$, $\cI_2$ and $\cI_3$ are
   represented by the thick solid, thin solid, dashed and dotted lines, respectively. }
   \end{figure}
   Fig.12 shows the resonant DC Josephson current at zero temperature as a function of
   the superconducting phase difference $\varphi_s$ for general
   asymmetric $S/I/N/I/S$ systems.  When $\varepsilon_0=0$, i.e.
   the single level is exactly located at the position of the
   superconducting chemical potential,  the DC Josephson current
   $\cI_{sns}(\varphi_s)$ and its component $\cI_1$ vs the superconducting phase difference
   $\varphi_s$ is of the sawtooth shape (Fig. 12 (a) and (b)), no matter how big
   the difference between the two
   superconducting energy gaps is. The physical origin of
   the sharp discontinuity at $\varphi_s=\pi$ is the same as in
   the usual $S/N/S$ junction:\cite{Bagwell, Wees}  the Andreev
   levels $\varepsilon_p$ and $-\varepsilon_p$
   determined from the spectral function $\widetilde{G}^r_s(\varepsilon)=0$ at $\varphi_s=0$
  interchange their position at $\varphi_s=\pi$, producing a discontinuity in the DC
  Josephson current.  Unlike the usual $S/N/S$ structure, the
  supercurrent-phase relation has a weak dependence on the
  asymmetry between the superconducting energy gaps in resonant
  $S/I/N/I/S$ structures. This is because the main contribution to the DC Josephson
  current in resonant structures with a single level is mainly
  from the Andreev refection processes inside the
  region $|\varepsilon|<|\Delta_\cL|$, where the energy of the
  Andreev bound states has a trivial dependence on the energy gap
  $|\Delta_\cR|$, as can be seen from the spectral function
  associated with
  $\widetilde{G}^r_s(\varepsilon)$. When the single level moves
  away from the position of the chemical potential $\varepsilon_0=0.2|\Delta_\cL|$ (Fig. 12(c)), or
  the
  coupling to the superconductors becomes asymmetric $\Gamma^{\cR s}=4\Gamma^{\cL s}$ (Fig. 12(d)),
  the abrupt jump at  $\varphi_s=\pi$ is smoothed out and the current-phase
  approaches the sinusoidal relation. The transmission probability becomes smaller
  when the level moves away from the chemical potential or the
  elastic couplings becomes asymmetric. As a result the link between these two superconductors becomes
  wicker, and thus the current-phase relation $\cI(\varphi_s)\propto
  \sin\varphi_s$ is expected.\cite{Tinkham} As the single level moves away from
  the chemical potential or the couplings become unequal,
  the DC Josephson current is significantly suppressed,
  with the component $\cI_1$ from the discrete spectra
  suppressed while the component $\cI_2$ from the continuum spectra
   $|\Delta_\cL|<|\varepsilon|<|\Delta_\cR|$
   enhanced. The suppression of $\cI_1$ is due to the
   decrease in the resonant Josephson tunneling probability which
   is originated from the violation of the constructive
   interference between the wave-functions of the Andreev levels,
   while the enhancement of $\cI_2$ results from the fact that
   the Andreev levels are pushed towards the region $|\Delta_\cL|<|\varepsilon|<|\Delta_\cR|$
   and thus the leaky probability is increased.

   In summary, we have shown in this subsection that the DC Josephson current
   in an asymmetric-gap resonant $S/I/N/I/S$ structure with a single level is
   slightly enhanced in contrast to the symmetric-gap case. The
   current-phase relation is closely related to the position of
   the single level and the symmetry between the couplings to the
   two superconducting leads.

   \section{Concluding remarks}

We have developed a unified theory of electronic transport in a
general two-terminal hybrid nanosystem, in which each lead can be
either a ferromagnet or a superconductor. Within the Keldysh NEGF
formalism, the current is expressed in terms of the local
properties of the central interacting region, ${\bf G}^{r,a,</>}$
and the equilibrium distribution functions of the leads ${\bf
f}_\gamma$. The ferromagnetism and superconducting proximity are
treated on the same footing, incorporated into the tunneling
Hamiltonian and the self-energy matrices after introducing a
four-dimensional Nambu-spinor space and performing appropriate
Bogoliubov transformations. With the help of some unitary {\it
rotation} and {\it phase} matrices, one can demonstrate
analytically the gauge invariance  of the general current formula
(\ref{eq:current}), and simplify it to the Meri-Wingree forms for
specific structures.  For some quantities, such as the chemical
potential, magnetization orientation and the superconducting order
parameter phase, only their relative value appears explicitly in
the expressions of current. Moreover, resonant tunneling, strong
electron correlations (Coulomb blockade, Kondo resonance etc.),
ferromagnetism and superconductivity proximity effect can be
investigated in a unified transport theory without introduction of
any ad hoc assumptions. In addition, the energy- and
bias-voltage-dependence of the level-width functions and
distribution functions enters into the current formula in a strict
and natural manner, allowing us to explore the I-V characteristics
in the large bias limit. However, the disadvantage of applying
first the Bogoliubov transformation for the ferromagnetic lead
Hamiltonian is that we can only obtain the expressions for the sum
of the spin-up and spin-down current, while the information about
the spin components of the current is lost.

Applying the current formulae to the simplest DBTS where the
interactions are ignored some interesting transport properties are
revealed if one takes into consideration the finite energy band
structure of the ferromagnets. We have reported on the current
flowing through a non-interacting zero-dimensional central region,
thus the results obtained are qualitatively right for a $2D$
quantum well with the attached emitter and/or the collector being
ferromagnetic. In addition, we did not consider the spin-flip
process due to the interfacial scattering or the existence of
paramagnetic impurities inside the barrier. It is known that the
spin-flip process may reduce the magnitude of the tunnel
magnetoresistance of a $F/I/F$ junction.\cite{Zhangx}  Also, we
can expect that the Andreev current spectrum in $F/I/N/I/S$
structures will be modified to a great extent in the presence of
the spin-flip process, since the Andreev reflection may be
enhanced with the assistance of such process.

In $F-QD-F$, $F-QD-S$ or $S-QD-S$ systems,  electron-electron
interactions inside the QD are important and thus one should
consider the many-body correlation effect. One example is the
Kondo effect at low temperatures.\cite{Raikh} In such a
circumstance, one has to calculate the full Green functions of the
QD in the presence of electron-electron interactions, taking into
consideration the couplings between the QD and the leads. We are
aware of three recent preprints\cite{Martinek} on the Kondo
physics in $F-QD-F$ systems in which the wide-band approximation
is used.\cite{Martinek,Wang}  Such a simplification of the
ferromagnetic DOS may lead to even spurious results in $I-V$
characteristics. However. interesting and even unexpected Kondo
resonances in these systems may arise with the full consideration
of the finite energy band structure of the ferromagnetic leads.

\acknowledgments

  Z. Y. Zeng and B. Li have been supported in part by the Academic Research Fund of
 the National University of Singapore and DSTA of Singapore.
 F. Claro has been supported by  C\'atedra Presidencial en Ciencias of Chile and
  FONDECYT 1020829 of Chile.

\appendix
\section{Derivation of the Self-energy matrices ${\bf \Sigma}^{r,a,</>}_{\gamma f/s}$}
\label{selfenergy}

In this Appendix we derive  various kinds of self-energy matrices
for the elastic couplings between the central region and the
ferromagnetic and superconductor leads.

First,  we calculate the retarded/advanced self-energy matrix
${\bf \Sigma}_{\gamma f;nm}^{r/a}(t_1,t_2)$ arising from the
coupling between the central normal metal and the ferromagnetic
lead
\begin{eqnarray*}
 {\bf \Sigma}^{r/a}_{\gamma
f;nm}(t_1,t_2) &=&\sum_{k}{\bf V}^{\gamma f\dagger}_{kn}(t_1) {\bf
g}^{r/a}_{\gamma f k,\gamma f k}(t_1,t_2) {\bf
V}^{\gamma f}_{km}(t_2)\nonumber \\
&=&\sum\limits_k {\bf P}^\dagger(\mu_{\gamma \cC}t_1){\bf
V}^{\gamma f\dagger}_{kn}{\bf R}^{f\dagger}(\frac{\theta_{\gamma
f}}{2})\nonumber \\
 &&\hspace{0cm}{\bf
g}^{r/a}_{\gamma f,\gamma f}(t_1,t_2){\bf
R}^f(\frac{\theta_{\gamma f}}{2}){\bf V}^{\gamma f}_{km}{\bf
P}^\dagger(\mu_{\gamma \cC}t_2).
\end{eqnarray*}

The sum over momentum $k$ can be converted to energy integration,
i.e., $\sum\limits_k \rightarrow\int d\varepsilon_{k\sigma}
\rho_\sigma^{\gamma f} (\varepsilon_{k\sigma})$. Neglecting the
level shift term, we get
\begin{eqnarray}
{\bf \Sigma}^{r/a}_{\gamma f;nm}(t_1,t_2)&=&\mp \frac i2
\int \frac{d\varepsilon}{2\pi}
e^{-i\varepsilon(t_1-t_2)}{\bf P}^\dagger(\mu_{\gamma \cC} t_1)\nonumber \\
& & \hspace{0cm} {\bf
R}^{f\dagger}(\frac{\theta_{\gamma  f}}{2}){\bf \Gamma}^{\gamma
f}_{nm}(\varepsilon){\bf R}^f(\frac{\theta_{\gamma f}}{2}){\bf
P}(\mu_{\gamma \cC} t_2) \nonumber \\
&=&\mp \frac i2 \int \frac{d\varepsilon}{2\pi}
e^{-i\varepsilon(t_1-t_2)}{\bf R}^{f\dagger}(\frac{\theta_{\gamma  f}}{2})
\nonumber\\
 &&\hspace{1.3cm}  {\bf
\Gamma}^{\gamma f}_{nm}(\varepsilon\mp\mu_{\gamma \cC}){\bf
R}^f(\frac{\theta_{\gamma  f}}{2}),
\end{eqnarray}
in which
\begin{eqnarray}
\label{eq:linewidth-f0}
 &&{\bf \Gamma}^{\gamma f}_{nm}(\varepsilon\mp c)\nonumber \\
 &=&\left (\matrix{ \Gamma_{nm;\uparrow}^{\gamma
f}(\varepsilon-c) & 0  \cr
 0 &\Gamma_{nm;\downarrow}^{\gamma
f}(\varepsilon+c)
 \cr
 0 & 0\cr 0&0\cr}\right.\nonumber\\
 && \hspace{1.5cm} \left.\matrix{ 0 & 0 \cr
  0 & 0 \cr
\Gamma_{nm;\downarrow}^{\gamma
 f}(\varepsilon-c)&0\cr0&\Gamma_{nm;\uparrow}^{\gamma
f}(\varepsilon+c) \cr }\right), \\
 &&{\bf \Gamma}^{\gamma f}_{nm}(\varepsilon)
 ={\bf \Gamma}^{\gamma f}_{nm}(\varepsilon\mp 0),
\end{eqnarray}
with \begin{eqnarray}
 \Gamma_{nm;\sigma}^{\gamma
f}(\varepsilon)&=&2\pi \rho^{\gamma
f}_{\sigma}(\varepsilon)V^{\gamma f*}_{kn}V^{\gamma
f}_{km}.\nonumber
\end{eqnarray}

Similarly, after transforming the momentum sum
 $\sum_{k}$ into an
integral $\int d\varepsilon_{k} \rho^{\gamma
s}_N(\varepsilon_{k})$, where $\rho^{\gamma s}_N$ is the normal
state of the superconductor, we obtain the self-energy matrix due
to the coupling of the central region to the superconducting lead
\begin{eqnarray}
\label{eq:selfenergy-s}
{\bf \Sigma}^{r/a}_{\gamma s;nm}(t_1,t_2)
&=&\sum_{k}{\bf V}^{\gamma s\dagger}_{kn}(t_1) {\bf
g}^{r/a}_{\gamma s k,\gamma s k}(t_1,t_2) {\bf
V}^{\gamma s}_{km}(t_2)\nonumber \\
&=&\mp \frac i2\int \frac{d\varepsilon}{2\pi}
e^{-i\varepsilon(t_1-t_2)/\hbar} {\bf P}^\dagger(\mu_{\gamma \cC}t_1+\frac{\varphi_\gamma}{2})\nonumber
\\ &&   \hspace{1.5cm}
{\bf
\Gamma}^{\gamma s}_{\varrho/\varrho^*; nm}(\varepsilon) {\bf
P}(\mu_{\gamma \cC}t_2+\frac{\varphi_\gamma}{2}) \nonumber \\
&=&\mp \frac i2 \int \frac{d\varepsilon}{2\pi}
e^{-i\varepsilon(t_1-t_2)/\hbar}{\bf P}^\dagger(\mu_{\gamma \cC}t_1+\frac{\varphi_\gamma}{2}) \nonumber\\ && {\bf
\Gamma}^{\gamma s}_{\varrho/\varrho^*; nm}(\varepsilon\mp \mu_{\gamma \cC})
{\bf P}(\mu_{\gamma
\cC}t_1+\frac{\varphi_\gamma}{2})\\
&=&\mp \frac i2 \int \frac{d\varepsilon}{2\pi}
e^{-i\varepsilon(t_1-t_2)/\hbar}{\bf P}^\dagger(\mu_{\gamma
\cC}t_2+\frac{\varphi_\gamma}{2}) \nonumber\\ &&\hspace{-0.5cm}
[{\bf \Gamma}^{\gamma s}_{\varrho/\varrho^*; nm}]^T(\varepsilon\mp
\mu_{\gamma \cC}) {\bf P}(\mu_{\gamma
\cC}t_2+\frac{\varphi_\gamma}{2}),
\end{eqnarray}
where we have  defined the complex level-width matrix as
\begin{eqnarray}
 {\bf \Gamma}^{\gamma
s}_{\varrho;nm}(\varepsilon\mp c)&=&\Gamma^{\gamma s}_{nm}\nonumber\\
&&\hspace{-2cm} \left(\matrix{\varrho^{\gamma s}(\varepsilon-c)
  & -\frac{|\Delta_\gamma|}{\varepsilon+c}\varrho^{\gamma
s}(\varepsilon+c) \cr
-\frac{|\Delta_\gamma|}{\varepsilon-c}\varrho^{\gamma
s}(\varepsilon-c)& \varrho^{\gamma s}(\varepsilon+c) \cr
 0 & 0\cr 0&0\cr}\right.\nonumber\\
 &&  \left.\matrix{ 0 & 0 \cr
  0 & 0 \cr
 \varrho^{\gamma
s}(\varepsilon-c)&\frac{
|\Delta_\gamma|}{\varepsilon+c}\varrho^{\gamma
s}(\varepsilon+c)\cr
 \frac{
 |\Delta_\gamma|}{\varepsilon-c}\varrho^{\gamma
s}(\varepsilon-c)&\varrho^{\gamma s}(\varepsilon+c) \cr}\right),\nonumber
\\
 {\bf \Gamma}^{\gamma
s}_{\varrho^*;nm}(\varepsilon\mp c)&=& [{\bf \Gamma}^{\gamma
s}_{\varrho;nm}(\varepsilon\mp c)]^*,
\end{eqnarray}
with
\begin{eqnarray}
\Gamma^{\gamma s}_{nm}&=&2\pi \rho^{\gamma s}_N(0) V^{\gamma s*}_{kn} V^{\gamma s}_{km} ,\nonumber \\
\varrho^{\gamma s}(\varepsilon)&=&\frac{|\varepsilon|\vartheta
(|\varepsilon|-|\Delta_\gamma|)}
{\sqrt{\varepsilon^2-|\Delta_\gamma|^2}}-
i\frac{\varepsilon\vartheta(|\Delta_\gamma|-|\varepsilon|)}
{\sqrt{|\Delta_\gamma|^2-\varepsilon^2}}.
\end{eqnarray}
Here we have defined a complex superconducting DOS, extending to
the forbidden region in the usual BCS theory
$|\varepsilon|<|\Delta_\gamma|$ , inside which the Andreev
reflection processes arise, as shown in the
Blonder-Tinkham-Klapwijk theory.\cite{Blonder} When
$|\varepsilon|<|\Delta_\gamma|$, the quasi-particle density of
states $\varrho$ is purely imaginary, indicating evanescent states
in the gap which eventually decay into the pair condensate. The
quasi-particle density of states of the superconducting lead
$\gamma$ is defined as
\begin{eqnarray}
 \rho^{\gamma
s}(\varepsilon)&=&\frac{1}{\pi}{\rm Im}\hat{g}^{a}_{\gamma
s,\gamma s;11 }= \frac{|\varepsilon|
\vartheta(|\varepsilon|-|\Delta_\gamma|)}{\sqrt{\varepsilon^2-|\Delta_\gamma|^2}}\nonumber
\\
&=& {\rm Re}\varrho(\varepsilon).
\end{eqnarray}

In deriving Eq. (\ref{eq:selfenergy-s}) we have used the following
equalities
\begin{eqnarray}
\label{eq:super-green1}
\vartheta(\tau)
 \int_{-\infty}^{+\infty} &d\epsilon_{k}&  [\cos^2\theta_{\gamma sk}
e^{\pm i\sqrt{\varepsilon_{k}^2+|\Delta_\gamma|^2}\tau/\hbar}+\nonumber\\
&&\hspace{1.5cm} \sin^2 \theta_{\gamma sk} e^{\mp
i\sqrt{\epsilon_{k}^2+|\Delta_\gamma|^2}
\tau/\hbar}]\nonumber \\
&&\hspace{-2.2cm} = i\int \frac{d\varepsilon}{2 \pi} e^{-i\varepsilon\tau/\hbar}\int
_{-\infty}^{+\infty}d\varepsilon_{k}\frac{\varepsilon\pm
\varepsilon_{k}}{\varepsilon^2-\varepsilon^2_{k}-|\Delta_\gamma|^2} \nonumber\\
&&\hspace{-2.2cm}  = \int d\varepsilon e^{-i\varepsilon\tau/\hbar} \varrho^{\gamma
s}(\varepsilon),\nonumber\\
\label{eq:super-green2}
\vartheta(-\tau)
 \int_{-\infty}^{+\infty} &d\epsilon_{k}&  [\cos^2\theta_{\gamma sk}
e^{\pm i\sqrt{\varepsilon_{k}^2+|\Delta_\gamma|^2}\tau/\hbar}+\nonumber\\
&&\hspace{1.5cm} \sin^2 \theta_{\gamma sk} e^{\mp
i\sqrt{\epsilon_{k}^2+|\Delta_\gamma|^2}
\tau/\hbar}]\nonumber \\
&&\hspace{-2.2cm} =-i\int \frac{d\varepsilon}{2 \pi} e^{-i\varepsilon\tau/\hbar}\int
_{-\infty}^{+\infty}d\varepsilon_{k}\frac{\varepsilon\pm
\varepsilon_{k}}{\varepsilon^2-\varepsilon^2_{k}-|\Delta_\gamma|^2} \nonumber\\
&&\hspace{-2.2cm}  = \int d\varepsilon e^{-i\varepsilon\tau/\hbar} [\varrho^{\gamma
s}(\varepsilon)]^*,\nonumber\\
\label{eq:super-green3} \vartheta(\tau)  \int_{-\infty}^{+\infty}
&d\varepsilon_{k}&\frac{\sin (2 \theta_{\gamma sk})}{2}
(e^{-i\sqrt{\varepsilon_{k}^2+|\Delta_\gamma|^2}\tau/\hbar}-\nonumber\\
&&\hspace{3cm} e^{i\sqrt{\varepsilon_{k}^2+|\Delta_\gamma|^2}\tau/\hbar})\nonumber \\
&&\hspace{-2.2cm} = i\int \frac{d\varepsilon}{2\pi} e^{-i\varepsilon\tau/\hbar}
  \int_{-\infty}^{+\infty}
d\varepsilon_{k}\frac{|\Delta_\gamma|}{\varepsilon^2-\varepsilon^2_{k}-|\Delta_\gamma|^2}
\nonumber\\
&&\hspace{-2.2cm} = \int d\varepsilon e^{-i\varepsilon\tau/\hbar} \varrho^{\gamma
 s}(\varepsilon)\frac{|\Delta_\gamma|}{\varepsilon}, \nonumber\\
\label{eq:super-green4} \vartheta(-\tau)  \int_{-\infty}^{+\infty}
&d\varepsilon_{k}&\frac{\sin (2 \theta_{\gamma sk})}{2}
(e^{-i\sqrt{\varepsilon_{k}^2+|\Delta_\gamma|^2}\tau/\hbar}-\nonumber\\
&&\hspace{3cm} e^{i\sqrt{\varepsilon_{k}^2+|\Delta_\gamma|^2}\tau/\hbar})\nonumber \\
&&\hspace{-2.2cm} =-i\int \frac{d\varepsilon}{2\pi} e^{-i\varepsilon\tau/\hbar}
  \int_{-\infty}^{+\infty}
d\varepsilon_{k}\frac{|\Delta_\gamma|}{\varepsilon^2-\varepsilon^2_{k}-|\Delta_\gamma|^2}
\nonumber\\
 &&\hspace{-2.2cm} = \int d\varepsilon e^{-i\varepsilon\tau/\hbar} [\varrho^{\gamma
 s}(\varepsilon)]^*\frac{|\Delta_\gamma|}{\varepsilon}. \nonumber
\end{eqnarray}
Note that we have chosen different complex half-planes in the
contour integrations, in order to guarantee that the
retarded/advanced self-energy matrices satisfy ${\bf
\Sigma}^r_{\gamma s}(\varepsilon)=[{\bf \Sigma}^a_{\gamma
s}(\varepsilon)]^\dagger$.

The lesser/greater self-energy matrix defined by
\begin{eqnarray*}
{\bf \Sigma}_{\gamma f/s;nm}^{</>}(t_1,t_2)=\sum_k {\bf V}^{\gamma
s/s\dagger}_{kn}(t_1) {\bf g}^{ </>}_{\gamma f/s,\gamma
f/s}(t_1,t_2){\bf V}^{\gamma f/s}_{km}(t_2)
\end{eqnarray*}
 are obtained in a similar way
\begin{eqnarray}
{\bf \Sigma}^{</>}_{\gamma f;nm}(t_1,t_2)
&=&i\int\frac{d\varepsilon}{2\pi}
e^{-i\varepsilon(t_1-t_2)/\hbar}{\bf P}^\dagger(\mu_{\gamma \cC}
t_1){\bf R}^{f\dagger}(\frac{\theta_{\gamma  f}}{2}) \nonumber\\
&&\hspace{-1.0cm}{\bf \Gamma}^{\gamma f}_{nm}(\varepsilon){\bf
R}^f(\frac{\theta_{\gamma f}}{2})[{\bf
f}_{\gamma}(\varepsilon)-\frac12{\bf 1}\pm
\frac12{\bf 1}]{\bf P}(\mu_{\gamma \cC} t_2) \nonumber\\
&=&i\int \frac{d\varepsilon}{2\pi}
e^{-i\varepsilon(t_1-t_2)/\hbar}{\bf
R}^{f\dagger}(\frac{\theta_{\gamma  f}}{2})  \nonumber\\
&&\hspace{-2.2cm} {\bf \Gamma}^{\gamma f}_{nm}(\varepsilon\mp \mu_{\gamma \cC} )
 {\bf R}^f(\frac{\theta_{\gamma  f}}{2}) [{\bf
f}_{\gamma}(\varepsilon\mp \mu_{\gamma \cC} )-\frac12{\bf 1}\pm
\frac12{\bf 1}], \\
{\bf \Sigma}^{</>}_{\gamma s;nm}(t_1,t_2)&=& i \int
\frac{d\varepsilon}{2\pi} e^{-i\varepsilon(t_1-t_2)/\hbar}{\bf
P}^\dagger(\mu_{\gamma \cC} t_1+\frac{\varphi_\gamma}{2})\nonumber
\\ &&\hspace{-1cm}{\bf \Gamma}^{\gamma s}_{\rho;nm}(\varepsilon)[{\bf
f}_{\gamma}(\varepsilon)-\frac12{\bf 1}\pm
\frac12{\bf 1}]{\bf P}(\mu_{\gamma \cC} t_2+\frac{\varphi_\gamma}{2})\nonumber\\
&=& i \int \frac{d\varepsilon}{2\pi}
e^{-i\varepsilon(t_1-t_2)/\hbar}{\bf P}^\dagger(\mu_{\gamma \cC}
t_1+\frac{\varphi_\gamma}{2})\nonumber
\\
& &{\bf \Gamma}^{\gamma s}_{\rho;nm}(\varepsilon\mp\mu_{\gamma
\cC})[{\bf f}_{\gamma}(\varepsilon\mp\mu_{\gamma \cC})-\frac12{\bf
1}\pm
\frac12{\bf 1}]\nonumber\\
&&\hspace{2cm}{\bf P}(\mu_{\gamma \cC} t_1+\frac{\varphi_\gamma}{2})\nonumber\\
&=& i \int \frac{d\varepsilon}{2\pi}
e^{-i\varepsilon(t_1-t_2)/\hbar}{\bf P}^\dagger(\mu_{\gamma \cC}
t_2+\frac{\varphi_\gamma}{2})\nonumber
\\
& &[{\bf f}_{\gamma}(\varepsilon\mp\mu_{\gamma \cC})-\frac12{\bf
1}\pm \frac12{\bf 1}]
[{\bf \Gamma}^{\gamma  s}_{\rho;nm}(\varepsilon\mp\mu_{\gamma
\cC})]^T\nonumber\\
&&\hspace{2cm}{\bf P}(\mu_{\gamma \cC}
t_2+\frac{\varphi_\gamma}{2}),
\end{eqnarray}
where the real level-width matrix is defined as
\begin{eqnarray}
 {\bf \Gamma}^{\gamma
s}_{\rho;nm}(\varepsilon\mp c)&=&\Gamma^{\gamma s}_{nm}\nonumber\\
&&\hspace{-2cm} \left(\matrix{\rho^{\gamma s}(\varepsilon-c)
  & -\frac{|\Delta_\gamma|}{\varepsilon+c}\rho^{\gamma
s}(\varepsilon+c) \cr
-\frac{|\Delta_\gamma|}{\varepsilon-c}\rho^{\gamma
s}(\varepsilon-c)& \rho^{\gamma s}(\varepsilon+c) \cr
 0 & 0\cr 0&0\cr}\right.\nonumber\\
 && \hspace{0.1cm} \left.\matrix{ 0 & 0 \cr
  0 & 0 \cr
 \rho^{\gamma
s}(\varepsilon-c)&\frac{
|\Delta_\gamma|}{\varepsilon+c}\rho^{\gamma s}(\varepsilon+c)\cr
 \frac{
 |\Delta_\gamma|}{\varepsilon-c}\rho^{\gamma
s}(\varepsilon-c)&\rho^{\gamma s}(\varepsilon+c)
\cr}\right).\nonumber
\end{eqnarray}

The Fermi distribution matrix in the Numbu-spinor space takes the
following form
\begin{eqnarray}
&&{\bf f}_{\gamma}(\varepsilon\mp c)
=\nonumber \\
 &&\left (\matrix{ f(\varepsilon-c) & 0 & 0
  & 0 \cr
 0 &f(\varepsilon+c) & 0 & 0 \cr
 0 & 0&f(\varepsilon-c)  & 0 \cr
 0 & 0 & 0 & f(\varepsilon+c)  \cr}\right),\nonumber
 \end{eqnarray}
  \begin{equation}
{\bf f}_{\gamma}(\varepsilon)= {\bf f}_{\gamma}(\varepsilon\mp 0).
\end{equation}

\end{document}